\documentclass{aastex631}

\usepackage{graphicx}
\usepackage{xcolor}

\begin{document}

\title{SN 2017hcc and SN 2023usc --  a comparative spectroscopic study of type-IIn supernovae.}

\author[0009-0000-4635-4945]{V. Sethulakshmi}
\affiliation{Indian Institute of Astrophysics \\
Koramangala Block II \\
Bangalore 560034, Karnataka, India.}
\affiliation{Instituto de Estudios Astrofísicos, Universidad Diego Portales \\
Avenida Ejército Libertador 441, 
Santiago, Region Metropolitana, Chile.}
\affiliation{Millennium Institute of Astrophysics, Santiago, Chile.\\}

\author[0000-0003-1138-9747]{F. K. Sutaria}
\affiliation{Indian Institute of Astrophysics \\
Koramangala Block II \\
Bangalore 560034, Karnataka, India.}

\author[0009-0000-8186-5740]{R. Sharma}
\affiliation{Indian Institute of Astrophysics \\
Koramangala Block II \\
Bangalore 560034, Karnataka, India.}
\affiliation{Department of Physics, Florida State University \\
77 Chieftan Way, Tallahassee, Florida 32306, USA.}


\author[0000-0003-2404-0018]{A. Ray}
\affiliation{Homi Bhabha Center for Science Education \\ Tata Institute of Fundamental Research, \\
 Mumbai 400088, Maharashtra, India.}





\begin{abstract}

We report on a spectroscopic  study of the bright, nearby type-IIn supernova SN 2017hcc, and SN 2023usc using data obtained from the Himalayan Chandra Telescope (HCT). SN 2017hcc is well-studied event, and our sampling covers 7 epochs, starting from +14\,d post explosion, and continuing into the nebular stage, at +411\,d. The type-IIn event SN 2023usc was sampled over 5 epochs from +12\,d to +155\,d post explosion. The nearly featureless  (except H$\alpha$) late time (+62\,d onward) spectra of SN 2023usc, 
suggests a novel explosion route for this type-IIn event. Assuming a CSM model created by multi-epoch ejection of material from the pre-explosion progenitor, we present here a comparative study of both events with several other type-IIn / interacting supernovae in progenitors with persistent signatures of a CSM. We find that true narrow lines ($v \ll 1000$\,km\,s$^{-1}$) emerge in the early ($\sim$ +10\,d) spectra only in few events (SN 2017hcc, SN 2023usc and SN 2010jl) initially classified as type-IIn in our sample -- in most cases the line velocity hovers at $\sim 1000$\,km\,s$^{-1}$ even in the very early epochs. CSM line velocity being indicative of its extent and opacity, this suggests that progenitors with a highly extended CSMs, which are also optically transparent in their outer regions may be relatively rare.
\end{abstract}

\keywords{Supernovae -- Type-II Supernovae -- type-IIn -- SN 2017hcc -- SN 2023usc}

\section{Introduction} \label{sec:intro}

The violent end of the quiescent thermonuclear burning lifetime of massive stars ($M_{ZAMS} > 8 M_{\odot}$), via gravitational core-collapse, provides a unique opportunity to study not only the pre-explosion properties of the progenitors, but also to trace their post-main sequence evolutionary path. These evolutionary tracks in the HR diagram are often complicated by mass loss from the progenitor, either via a steady, metallicity dependent, wind, or by episodic ejection of matter from the progenitor, leading to the formation of a circumstellar medium (CSM). Evidence of CSM with complex morphologies, and varying densities, around core-collapse progenitors is seen across supernova sub-types, with the densest distribution inferred for the type-IIn supernovae \citep{Filipenko-97}. These events, whose spectra are characterized by persistent, narrow, emission lines with velocities of $\sim \rm{few} \times 100$\,km\,s$^{-1}$, are relatively rare, with the cosmic fractional rate of type-IIn to all core collapse supernovae (SNe) being $0.047\pm 0.009$ \citep{Cold-2023}. Fig.\,7 of the same study also sets the absolute peak magnitudes in the $\it i$ band in the range $\sim -16.5$ to $-21$, and as high as  $i \sim -22$ for the super-luminous type-IIn (SLSN-IIn). Likewise, their peak bolometric luminosities range from  $\sim \rm{few} \times 10^{42}$  to $\sim \rm{few} \times 10^{43}$ erg/s, comparable to  the more common SN-IIP at peak or greater \citep{Taddia-2013}. 

The progenitors of type-IIn supernovae cannot be indisputably  inferred post-explosion because of the difficulty in isolating the purely radioactive component in the tail of the bolometric light curve, from contribution of the still glowing CSM. Additionally, in the event that a progenitor can be located in archival data, assigning a spectral type becomes difficult since the object may not be in quiescence, e.g. as in the case of SN 2009ip (see Fig.\,3, \citealt{Smith_2010}), whose progenitor showed considerable variability for almost 4000\,d prior to first detected eruption in 2009. Nevertheless, both massive stars ($M \ge 25 M_{\odot}$) with LBV behavior \citep{Weis-2020} as well as thermonuclear explosions in low mass stars with very dense CSM (CSM-Ia supernovae, \citealt{Sharma-2023}) have been suggested. For example the progenitor of SN 2005gl was identified as an LBV from archival and post-explosion HST data (\citealp{Gal-Yam-2009,Dwarkadas-2011}).
A massive progenitor with a binary companion cannot be ruled out either \citep{Smith-2017}, as in the case of SN 1998S \citep{Fassia-2001}. Finally, ``ordinary'' core-collapse explosion in less massive stars embedded in a dusty CSM have also given rise to type-IIn SN, as in the case of SN 2008S \citep{Botticella-2009}. Using archival Spitzer/IRAC data, \citet{Preito-2008} identified a dust-enshrouded point source at the location of the SN 
with a most probable progenitor mass of $M \sim 10 M_{\odot}$. The host galaxies and immediate environment of SNe type-IIn also show considerable variation. \citet{Moriya-2023} find that events with a higher peak luminosities usually occur in low metallicity galaxies, or in younger star forming regions, consistent with the idea that progenitors of the brighter events must be very massive. However, no clear correlation was identified between the CSM density and environmental metallicity, suggesting that stellar winds from the massive progenitor may not be the sole precursor of the CSM. 
 Indeed, with such a range of progenitor properties, it becomes necessary to explore the commonalities and the differences in the (mainly) spectral characteristics of the events, to search for characteristics which may serve as a diagnostic of the progenitor properties, its environment and its evolutionary history. Here we present a comparative spectra study of two type-IIn  -- the long lived SN 2017hcc and the fast fading SN 2023usc -- using data obtained under our own program, and from archival sources. 

SN 2017hcc (=ATLAS17lsn), located at 00$^h$03$^m$50$^s$.280, --11$^{\circ}$28$'$28$''$.78 (J2000) was discovered \citep{di2017hcc} on 02 October 2017 (MJD 58028.57800), by the Asteroid Terrestrial-impact Last Alert System survey (ATLAS; \citealt{ATLAS}). 
It was associated with the FUV-bright, dwarf, spiral host-galaxy GALEXASC J000350.27-112827.3 \citep{Bianchi-2014}, which is cross-correlated with IrS WISEA J000350.27-112828.7.
The event has now been studied by several groups, across the electromagnetic spectrum. A study of optical/UV photometry and low resolution optical spectra  taken over 45 epochs from +6\,d to +1765\,d, as well NIR spectra taken over 5 epochs from +39\,d to +340\,d was presented by \citet{moran}. A combined spectroscopic study using low-resolution (MMT Blue channel and Binospec) spectra over 8 epochs (from 25\,d to 848\,d post explosion) and high-resolution MIKE Echelle spectra (taken on 24\,d, 282\,d and 351\,d post explosion) was 
done by \citet{smith}. The evolution of the supernova's intrinsic polarization during the 2 months post-explosion in V and R bands was characterized by \citet{kumar}, while \citet{mauerhan} carried out a multi-epoch spectropolarimetric study over 12 epochs spanning $\sim 1$\,yr  post-explosion. The latter study found that it maintained up to 6\% continuum polarization, significantly stronger in the blue part of the spectrum than in the red, for up to $\sim$29\,d post explosion, which was inferred as scattering from an aspherical distribution of dust grains in the CSM. The continued, low level, wavelength independent, polarization observed at later epochs is attributed to electron scattering at the interface of the SN ejecta and CSM.  Finally, a multiwavelength (IR/x-ray/radio) study was presented by \citet{chandra}.   All concluded that the progenitor was cocooned in a highly inhomogeneous CSM with an asymmetric morphology. Based on line asymmetry of H$\alpha$ and He I $\lambda$5876 lines, \citet{smith} suggested a CSM with a bipolar configuration, similar in geometry, but slower moving than that seen in $\eta$-Car, interacting with a near-spherical supernova ejecta. 

SN 2023usc (=Gaia24ajp, PS23kav, GOTO23awo and ZTF23abjhwem) located at 05$^h$21$^m$31$^s$.649, +00$^{\circ}$28$'$20$''$.73,  was discovered on 11 October 2023 (MJD 60228.8306)
by the ATLAS survey
 and was initially designated as ATLAS23tzi \citep{di2023usc}. The supernova is most likely associated with the host galaxy IrS  WISEA J052131.69+002817.9 \citep{Cutri}. A redshift of $z=0.06$ was estimated based on the shift in the core of the H$\alpha$ emission line in the classification spectra \citep{cl2023usc}. 

We present below in Section\,\ref{Obs-data}, the low-resolution optical spectra of both events, covering several epochs, as well as photometry of both sources at coeval epochs, and  present an analysis of our observations in Section\,\ref{ANALYSIS}. We carry out a comparative study with other events in the same class, observed at approximately the same epochs. While this  object has been spectroscopically well-sampled by other groups (notably \citealt{moran} and \citealt{smith}), we provide additional coverage at low to moderate resolution (R=1330 and R=2190) up to +411\,d.

The light curve of SN 2017hcc is seen to peak $57 \pm 2$\,d post explosion \citep{moran}, so our observations cover both the pre- and post- peak evolution of this event.  For SN 2023usc, our spectral coverage over 5 epochs covers the first 118\,d post explosion. A light curve provided by the ZTF, obtained via difference photometry, shows that the r- and g- band light curves peak at $\sim 33.5$\,d (MJD 60261.45) post discovery, and so our spectroscopic monitoring follows this object post its optical peak luminosity.  

We compare in Section\,\ref{Comparision}, the spectral characteristics, and Balmer line profiles, of these two supernovae with those other interacting supernovae at similar epochs. Our selection of interacting supernovae (with the distinctive narrow  lines in their early spectra events) for comparison is limited by the availability of spectra in the WISeREP repository \citep{WISEREP}. We have selected objects for which spectra is available for coeval epochs with SN 2017hcc. Finally, in Section\,\ref{Discussion},  we discuss the ejecta - CSM interaction characteristics of these events, in the context of our, and other archival observations.

\begin{deluxetable}{cccccccc} 
\tablecaption{Log of observations of SN 2017hcc and SN 2023usc, and of the standard stars used for flux calibration. All observations were done with the HCT/HFOSC instrument, and the epoch is the midpoint of observation.
\label{tab: Obs_log}}
\tablehead{\colhead{Sr.} & \colhead{Name of} & \colhead{Dispersion} & \colhead{MJD} & \colhead{Exposure time} & \colhead{Standard} & \colhead{MJD} & \colhead{Exposure time} \\
\colhead{ No.}& \colhead{object}& \colhead{element}& \colhead{}& \colhead{(s)}& \colhead{star}& \colhead{}& \colhead{(s)}}
\colnumbers
\startdata
1 & SN 2017hcc & Grism 7 & 58040.6838 & 1800 & Feige34 & 58040.9724 & 500\\
2 & SN 2017hcc & Grism 8 & 58040.7059 & 1800 & Feige34 & 58040.9796 & 500\\
3 & SN 2017hcc & Grism 7 & 58049.6918 & 1800 & HZ4 & 58049.8539 & 900\\
4 & SN 2017hcc & Grism 8 & 58049.7141 & 1800 & HZ4 & 58049.8658 & 900\\
5 & SN 2017hcc & Grism 7 & 58112.6562 & 1200 & Feige34 & 58112.8815 & 720\\
6 & SN 2017hcc & Grism 8 & 58112.6718 & 1200 & Feige34 & 58112.8916 & 720\\
7 & SN 2017hcc & Grism 7 & 58126.6133 & 1200 & Feige34 & 58126.4997 & 720\\
8 & SN 2017hcc & Grism 8 & 58126.6280 & 1080 & Feige34 & 58126.0086 & 600\\
9 & SN 2017hcc & Grism 7 & 58371.9421 & 2400 & Feige110 & 58400.8051 & 720\\
10 & SN 2017hcc & Grism 7 & 58400.7362 & 1800 & Feige110 & 58400.8051 & 720\\
11 & SN 2017hcc & Grism 8 & 58400.7649 & 3000 & Feige110 & 58400.8137 & 720\\
12 & SN 2017hcc & Grism 7 & 58437.6167 & 2700 & Feige34 & 58437.9502 & 600\\
13 & SN 2017hcc & Grism 8 & 58437.6822 & 2700 & Feige34 & 58437.9575 & 600\\
\hline 
14 & SN 2023usc & Grism 7 & 60264.3441 & 2100 & Feige110 & 60264.0442 & 600\\
15 & SN 2023usc & Grism 8 & 60264.3570 & 2100 & Feige110 & 60264.0567 & 600\\
16 & SN 2023usc & Grism 7 & 60289.2635 & 2400 & Feige110 & 60289.0305 & 720\\
17 & SN 2023usc & Grism 8 & 60289.2856 & 2400 & Feige110 & 60289.0587 & 720\\
18 & SN 2023usc & Grism 7 & 60320.1198 & 4800 & Feige34 & 60320.0476 & 720\\
19 & SN 2023usc & Grism 8 & 60320.1840 & 4800 & Feige34 & 60320.0573 & 720\\
20 & SN 2023usc & Grism 7 & 60355.1320 & 7200 & Feige34 & 60355.3819 & 600\\
21 & SN 2023usc & Grism 8 & 60355.2057 & 5400 & Feige34 & 60355.3902 & 600\\
22 & SN 2023usc & Grism 7 & 60382.1861 & 2695 & Feige34 & 60382.0826 & 600\\
\hline
\enddata
\end{deluxetable}

\begin{deluxetable}{ccccccccccccccc}
\tablecaption{Calibrated \textit{UBVRI} photometry (with uncertainties) of SN 2017hcc from the Himalayan Chandra Telescope (HCT) at Hanle. All epochs have been measured in days from the adopted explosion date of MJD 58027.4 .}
\label{hcc_phot}
\tablehead{\colhead{$U$epoch} & \colhead{$U$} & \colhead{$\sigma U$} & \colhead{$B$epoch}  & \colhead{$B$} & \colhead{$\sigma B$} & \colhead{$V$epoch} & \colhead{$V$} & \colhead{$\sigma V$} & \colhead{$R$epoch} & \colhead{$R$} & \colhead{$\sigma R$} & \colhead{$I$epoch} & \colhead{$I$} & \colhead{$\sigma I$} \\
\colhead{(d)} & \colhead{(mag)} & \colhead{(mag)} & \colhead{(d)} & \colhead{(mag)} & \colhead{(mag)} & \colhead{(d)} & \colhead{(mag)} & \colhead{(mag)} & \colhead{(d)} & \colhead{(mag)} & \colhead{(mag)} & \colhead{(d)} & \colhead{(mag)} & \colhead{(mag)}}
\colnumbers
\startdata
13.2615 & 13.87 & 0.06 & 13.2586 & 14.61 & 0.01 & 13.2563 & 14.74 & 0.02 & 13.2545 & 14.58 & 0.01 & 13.2527 & 14.54 & 0.03 \\
22.35 & 13.31 & 0.08 & 22.3468 & 14.02 & 0.03 & 22.3446 & 14.12 & 0.02 & 22.3428 & 14.02 & 0.02 & 22.3410 & 13.99 & 0.06 \\
52.3216 & 13.09 & 0.08 & 52.32 & 13.6773 & 0.02 & 52.3158 & 13.63 & 0.02 & 52.3122 & 13.47 & 0.02 & 52.3141 & 13.36 & 0.0 \\
--  & -- & -- & -- & -- & -- & 85.2838 & 13.97 & 0.03 & 85.2803 & 13.70 & 0.05 & 85.2821 & 13.48 & 0.05 \\
99.1926 & 14.66 & 0.06 & 99.1847 & 14.76 & 0.02 & 99.1818 & 14.27 & 0.02 & 99.1729 & 13.95 & 0.01 & 99.1793 & 13.73 & 0.04 \\
373.2656 & 18.11 & 0.07 & 373.2608 & 18.12 & 0.02 & 373.2344 & 17.77 & 0.04 & 373.2288 & 17.14 & 0.04 & 373.2310 & 17.02 & 0.02 \\
410.1678 & 18.33 & 0.07 & 410.1594 & 18.4843 & 0.019 & 410.1566 & 18.17 & 0.02 & 410.1530 & 17.59 & 0.03 & 410.1468 & 17.41 & 0.05 \\
\enddata
\end{deluxetable}

\begin{deluxetable}{ccc|cccccccccc}
\tablecaption{Calibrated \textit{ugriz} photometry (with uncertainties) of SN 2023usc from the Himalayan Chandra Telescope (HCT) at Hanle. All epochs have been measured in days from the adopted explosion date of MJD 60226.834.}
\label{usc_phot}
\tablehead{\colhead{Sr. No.} & \colhead{Date} & \colhead{Epoch} & \colhead{\textit{u}} & \colhead{$\sigma$\textit{u}} & \colhead{\textit{g}} & \colhead{$\sigma$\textit{g}} & \colhead{\textit{r}} & \colhead{$\sigma$\textit{r}} & \colhead{\textit{i}} & \colhead{$\sigma$\textit{i}} & \colhead{\textit{z}} & \colhead{$\sigma$\textit{z}} \\
\colhead{~} & \colhead{(MJD)} & \colhead{(days)} & \colhead{(mag)} & \colhead{(mag)} & \colhead{(mag)} & \colhead{(mag)} & \colhead{(mag)} & \colhead{(mag)} & \colhead{(mag)} & \colhead{(mag)} & \colhead{(mag)} & \colhead{(mag)}}
\colnumbers
\startdata
1 & 60263.88  & 37.04 & 17.16 & 0.05 & 16.76 & 0.01 & 16.54 & 0.01 & 16.48 & 0.01 & 16.43 & 0.02  \\ 
2 & 60288.73  & 61.90 & 17.81 & 0.06 & 17.09 & 0.01 & 16.78 & 0.02 & 16.64 & 0.02 & 16.54 & 0.02  \\ 
3 & 60319.58  & 92.75 & 18.68 & 0.07 & 17.89 & 0.01 & 17.40 & 0.02 & 17.28 & 0.02 & 17.14 & 0.02  \\ 
4 & 60354.58  & 127.74 & 18.82 & 0.08 & 18.30 & 0.02 & 17.83 & 0.02 & 17.68 & 0.02 & 17.50 & 0.02  \\ 
5 & 60381.62  & 154.79 & 18.84 & 0.06 & {\textendash} & {\textendash} & 17.85 & 0.03 & 17.70 & 0.02 & {\textendash} & {\textendash} \\
\enddata
\end{deluxetable}

\section{Data and Observations}
\label{Obs-data}
\subsection{From HCT}

The HCT/HFOSC grisms cover 3800\,{\AA} to 8350\,{\AA} for Grism 7 and from  5800\,{\AA} to 8350\,{\AA} for Grism 8 with a resolution $R=1330$ and $R=2190$ ($ \sim 5$\,{\AA} and $\sim 3$\,{\AA}  at H$\alpha$ $\lambda$6562.8) respectively.
 Our spectroscopic observations of SN 2017hcc, carried out with the HCT/HFOSC instrument, started from MJD  58040.728 (i.e.$\sim 14$\,d after first detection), and continued until +411\,d, at which the object was at $V = 18.25$ (from Table\,A.1, \citealt{moran}), after which it became too faint for this instrument. Our observations of SN 2023usc were initiated 35 d post discovery, and continued until +155\,d , up to the faintness limit of HFOSC. 
 The log of our observations is provided in Table\,{\ref{tab: Obs_log}}.
 
 The data was reduced using standard {\tt IRAF} packages \citep{iraf}. 
 Night sky emission lines  from [O I] $\lambda$5577, $\lambda$6300 and $\lambda$6363 were used to cross-check the wavelength calibration and small shifts were applied wherever necessary. Following \citet{moran} and \citet{smith}, the total extinction in the direction to SN 2017hcc is dominated by that from the Milky Way, and so we use the maximum value of $E(B - V) = 0.016$  to deredden the spectra. Additionally, telluric band correction were found to be necessary for SN 2023usc (at $z=0.061$), to correct for the prominent telluric feature at $\lambda$6756, which overlapped with H$\alpha$ in its spectra. 

The extinction for SN 2023usc was estimated using the spectrum taken 37\,d post explosion. We used the relation between the extinction and equivalent width of Na I D1/D2 lines, estimated by \citet{Poznanski}. We identified and deblended the Milky way NaD feature, leading to an equivalent width of 0.1472\,{\AA} for the  5889.950\,{\AA} (D$_1$) feature, and 0.118\,{\AA} for the 5895.9\,{\AA} (D$_2$) feature. Thus, for the Milky way, the former leads to $E(B-V)=0.04^{+0.02}_{-0.01}$
while the latter results in
$E(B-V)=0.022^{+0.009}_{-0.006}$. In the same spectrum, we identified the host galaxy's NaI D$_1$/D$_2$ features at  6244.4\,{\AA} and 6247\,{\AA}, with equivalent widths 0.0543 and 0.0494\,{\AA} respectively. Using the same relations, this leads to $E(B-V)=0.024^{+0.011}_{-0.007}$(for $D_1$) and 
$E(B-V)=0.016^{+0.004}_{-0.006}$, assuming $R_V=3.1$ \citep{cardelli89} for the host galaxy as well, leading to a maximum extinction in the direction of the host galaxy as $E(B-V)=0.09$. 


Additionally, we also note that, based on the estimates of dust reddening, obtained using stellar colors and spectra from the Sloan Digital Sky Survey \citep{schlafly}, and assuming $R_V=3.1$, $E(B-V)_{Milky Way}=0.12$ in the direction of SN 2023usc. Thus, we adopt a maximum value of $E(B-V)=0.155$ for this work.

\subsection{From archival data} 

For events other than SN 2017hcc and SN 2023usc, data were taken from Transient Name Server (TNS; \citealp{TNS}) and Weizmann Interactive Supernova Data Repository (\citealp{wiserepp,WISEREP}).
When comparing the profile of the H$\alpha$ feature across events, we used the range from 6000\,{\AA} to 7000\,{\AA}, resolved the continuum from the H$\alpha$ emission feature in each case, and then normalized each spectrum with respect to its continuum. Some of the WISeREP data was provided with telluric correction, while others were not. In the latter case, if no telluric standard was available, we have proceeded without this correction. However, instances where the H$\alpha$ line was influenced by the telluric line are mentioned in their respective sections.

 \subsection{Photometric reduction}
 
 Imaging observations of SN 2017hcc were obtained with the HCT/HFOSC on seven nights between $\sim 13$\,d and $\sim 410$\,d post-explosion, in the Bessel $UBVRI$ filters. For SN 2023usc, we obtained five epochs of data in the Sloan $ugriz$ filters, covering the period from $\sim 37$\,d to $\sim 155$\,d post-explosion.  The images were reduced using standard \texttt{IRAF} routines for CCD aperture photometry, and the calibrated  magnitudes obtained for SN 2017hcc and SN 2023usc are reported in tables\,\ref{hcc_phot} and\,\ref{usc_phot} respectively. 

The HCT observations of both supernovae were carried out within the first $\sim 500$\,d after the explosion date, when the supernovae were still substantially brighter than their host galaxies, and so we did not perform template subtraction for these events. 
We did, however, perform PSF-fitting photometry using \texttt{IRAF}/\texttt{daophot} on a few images, the results  which were consistent with our aperture photometry to within 0.1 mag. Moreover, our photometry of SN 2023usc is consistent with the r and i band forced photometry obtained from LASAIR-ZTF. 



\begin{figure*}
\centering

 \begin{minipage}{0.45\textwidth}
   \centering
   \includegraphics[scale=0.25, trim=80 0 0 0, clip]{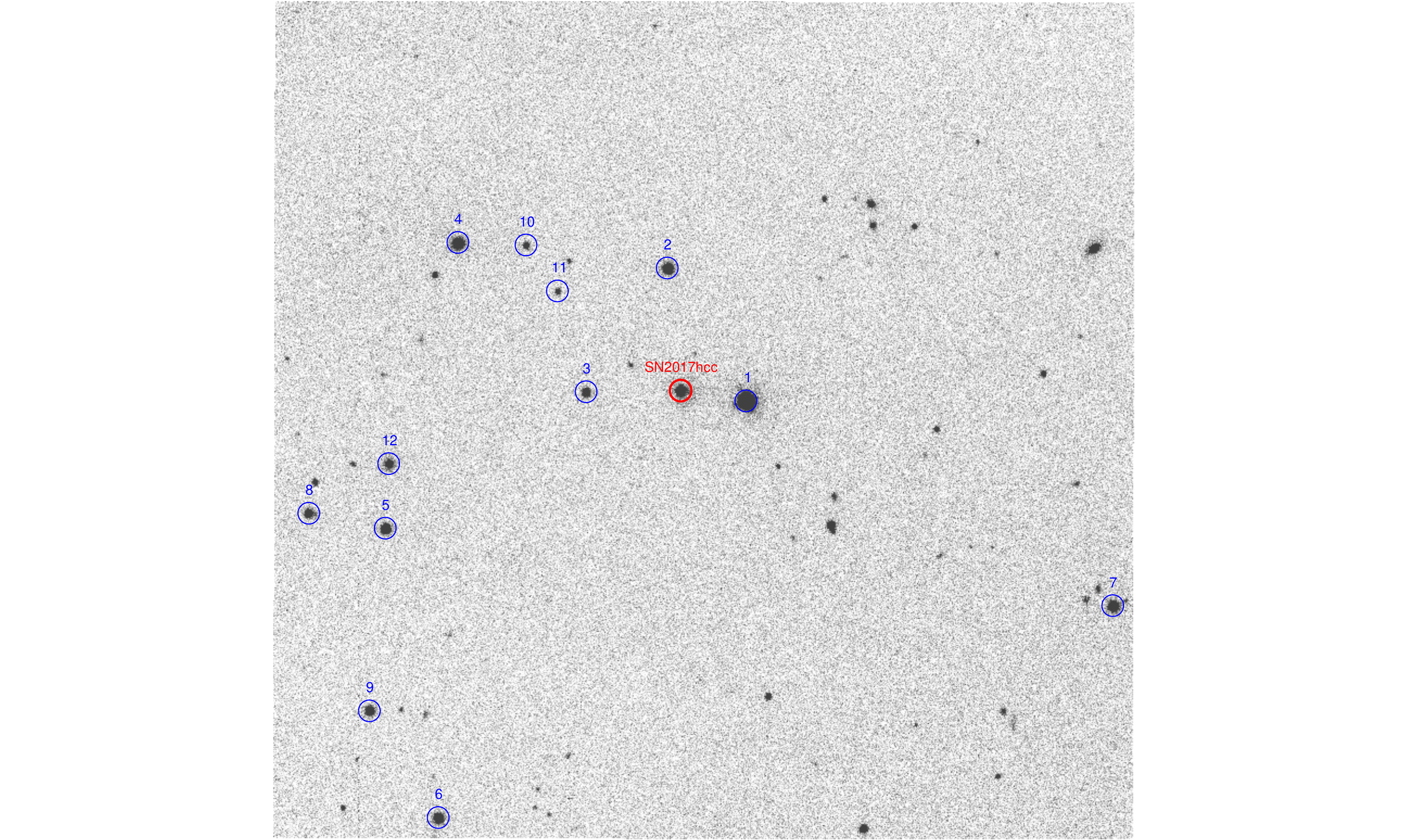}
   \caption{Bessel-$R$ band field of view of SN 2017hcc, with the supernova circled in red, and the selected secondary calibration stars circled in blue.}
   \label{hcc_FoV}
 \end{minipage}
 \hfill
 \begin{minipage}{0.45\textwidth}
   \centering
   \includegraphics[scale=0.25, trim=80 0 0 0, clip]{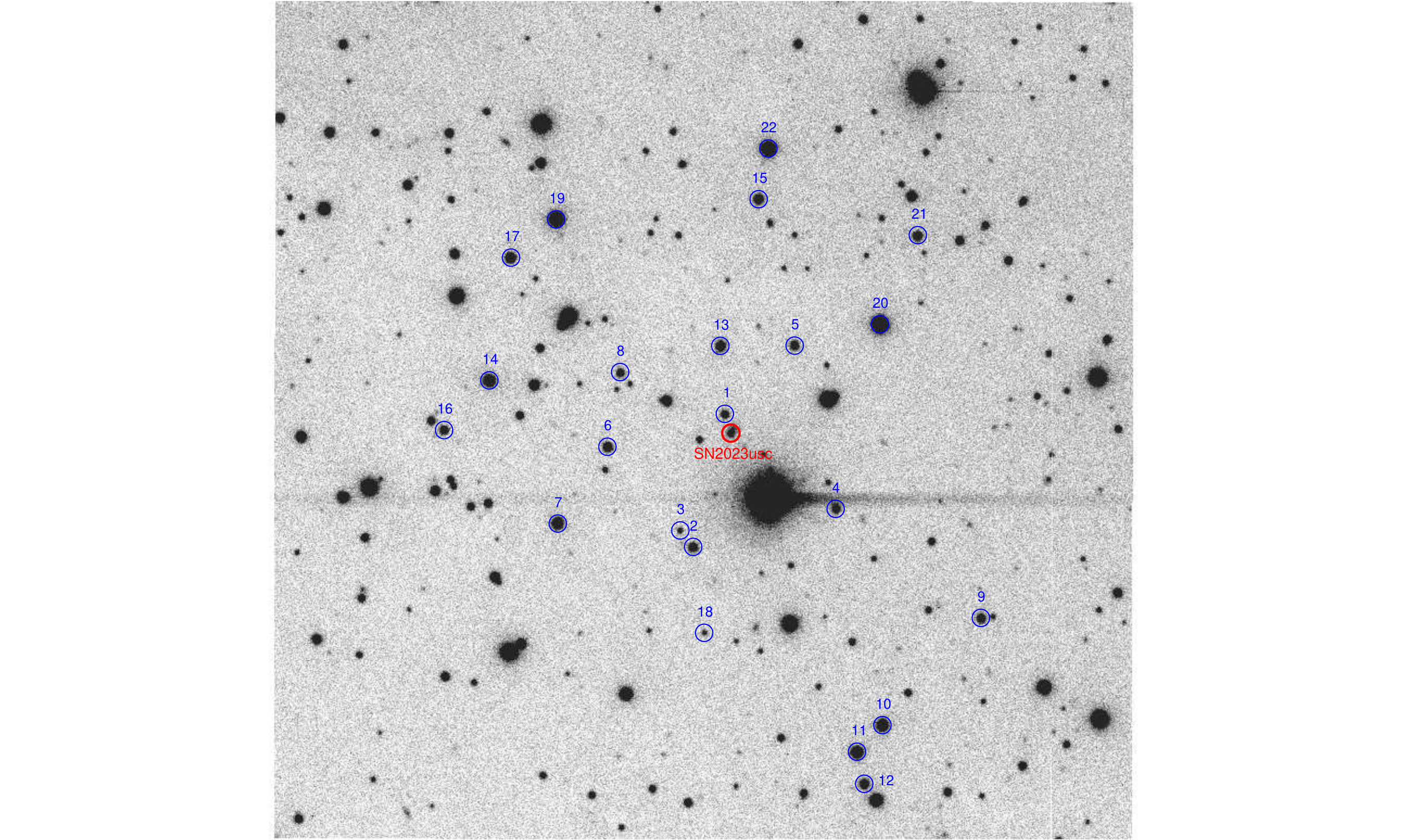}
   \caption{Sloan-$r$ band field of view of SN 2023usc, with the supernova circled in red, and the selected secondary calibration stars circled in blue.}
  \label{usc_FoV}
 \end{minipage}

\end{figure*}

\begin{figure*}
    \centering
    \begin{minipage}{0.45\textwidth}
        \centering
       \includegraphics[scale=0.5]{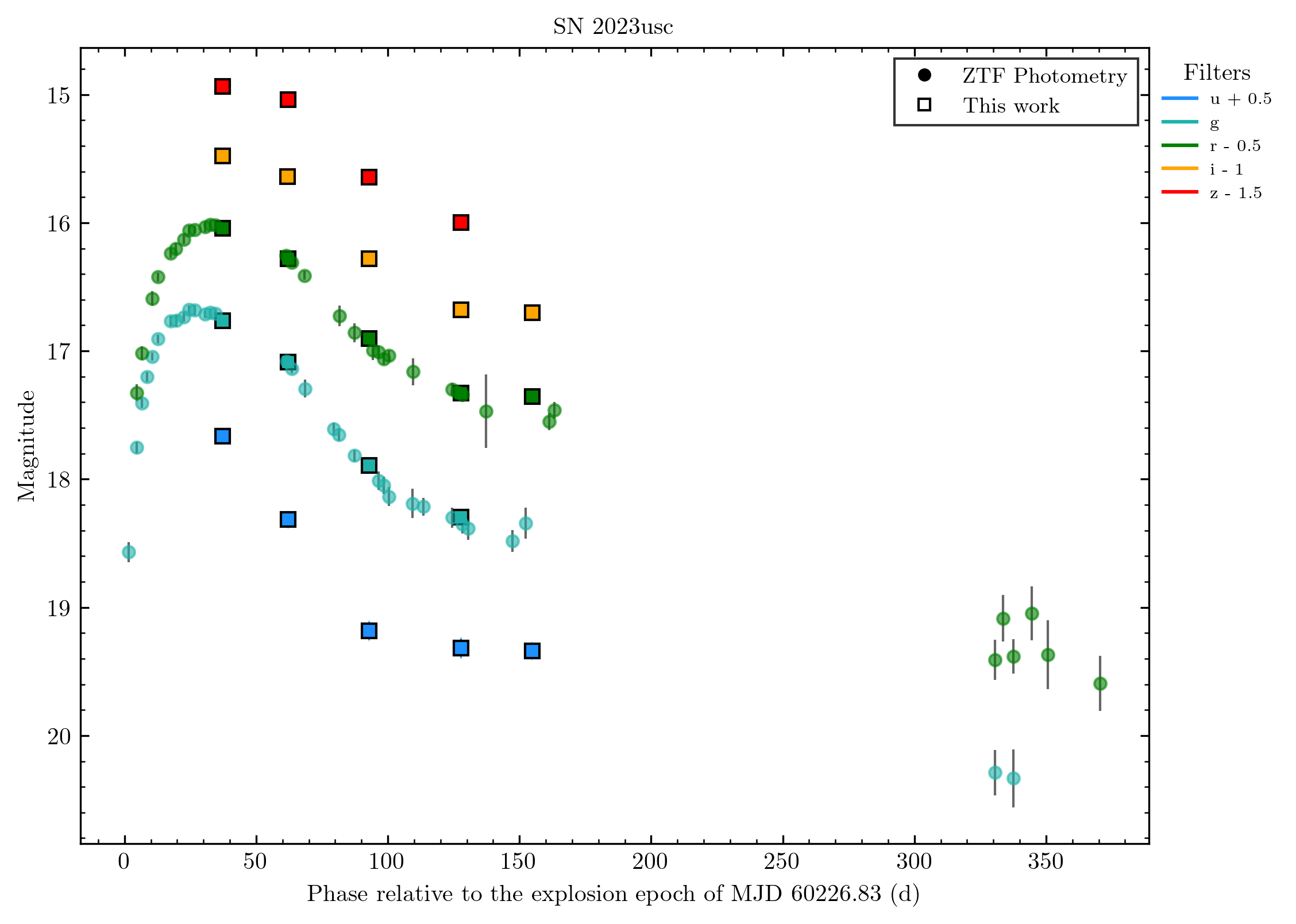}
 \caption{Multi-band optical light curves of SN 2023usc. Our observations have been overlaid on photometry from \citet{lasair} for comparison, and the magnitudes in various filters have been offset by different constants for ease of viewing.}
  \label{usc_lc}
    \end{minipage}
    \hfill
    \begin{minipage}{0.45\textwidth}
        \centering
        \includegraphics[scale=1.1]{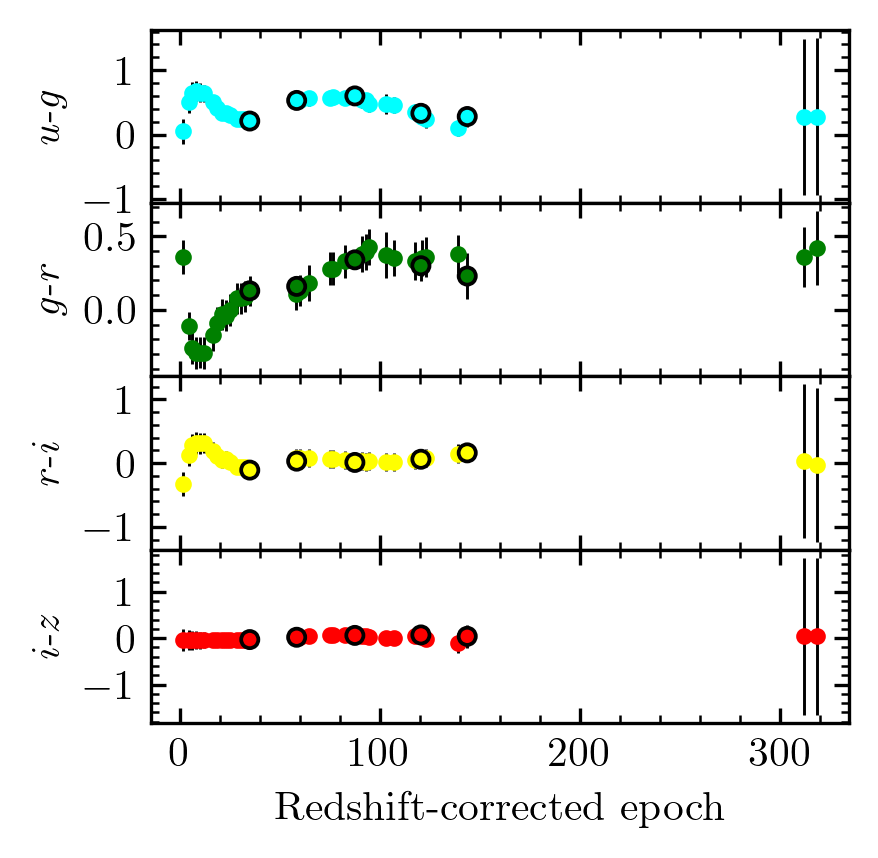}
    \caption{Evolution of the (top to bottom) $u-g$, $g-r$, $r-i$, and $i-z$  photometric colors of SN 2023usc with redshift-corrected epoch ($z=0.06$). Epochs on which HCT photometry was obtained are marked with black circles. The magnitudes in various bands have been corrected for interstellar reddening taking $E(B-V) = 0.155$, and missing magnitudes have been interpolated with respect to the $g$-band.}
       \label{usc_colors}
    \end{minipage}
\end{figure*}

The photometric calibrations of both supernovae were carried out using selected field stars. For SN 2023usc, photometric calibration was carried out using 
field stars (marked in Fig.\,\ref{usc_FoV}) from the SDSS DR16 \citep{ahumada202016th} catalog, applying a zero-point transformation. For SN 2017hcc, suitable local standards were unavailable, so we used stars in the Landolt field PG0231+051 \citep{landolt1992ubvri}, observed with the HFOSC on 2017-10-14 UT (MJD 58040.7), to derive standard magnitudes for local field stars (shown in Fig.\,\ref{hcc_FoV}). We employed transformation equations with color terms to account for filter transmission differences between the instrumental and standard systems. The supernova instrumental magnitudes were finally calibrated against the local standards using a zero-point transformation. Figures \ref{hcc_lc} and \ref{usc_lc} contain the multi-band light curve plots for SN 2017hcc and SN 2023usc respectively.

\subsection{Bolometric light curves}

 The multi-band magnitudes of SN 2017hcc and SN 2023usc were corrected for distance/redshift using a distance modulus of $\mu = 34.31$ for the former and  $\mu = 37.16$ for the latter. Here the redshift for SN 2023usc is taken as $z = 0.06$.  Using the extinction law given by \citet{cardelli89} and assuming $R_V = 3.1$, we corrected the supernova magnitudes for interstellar reddening using $A_V = 0.091$ for SN 2017hcc \citep{moran} and a host galaxy + Milky Way $E(B-V)$ of $0.155$ \citep{schlafly} for SN 2023usc. 

\begin{figure*}
    \centering
    \begin{minipage}{0.45\textwidth}
        \centering
       \includegraphics[scale=0.5]{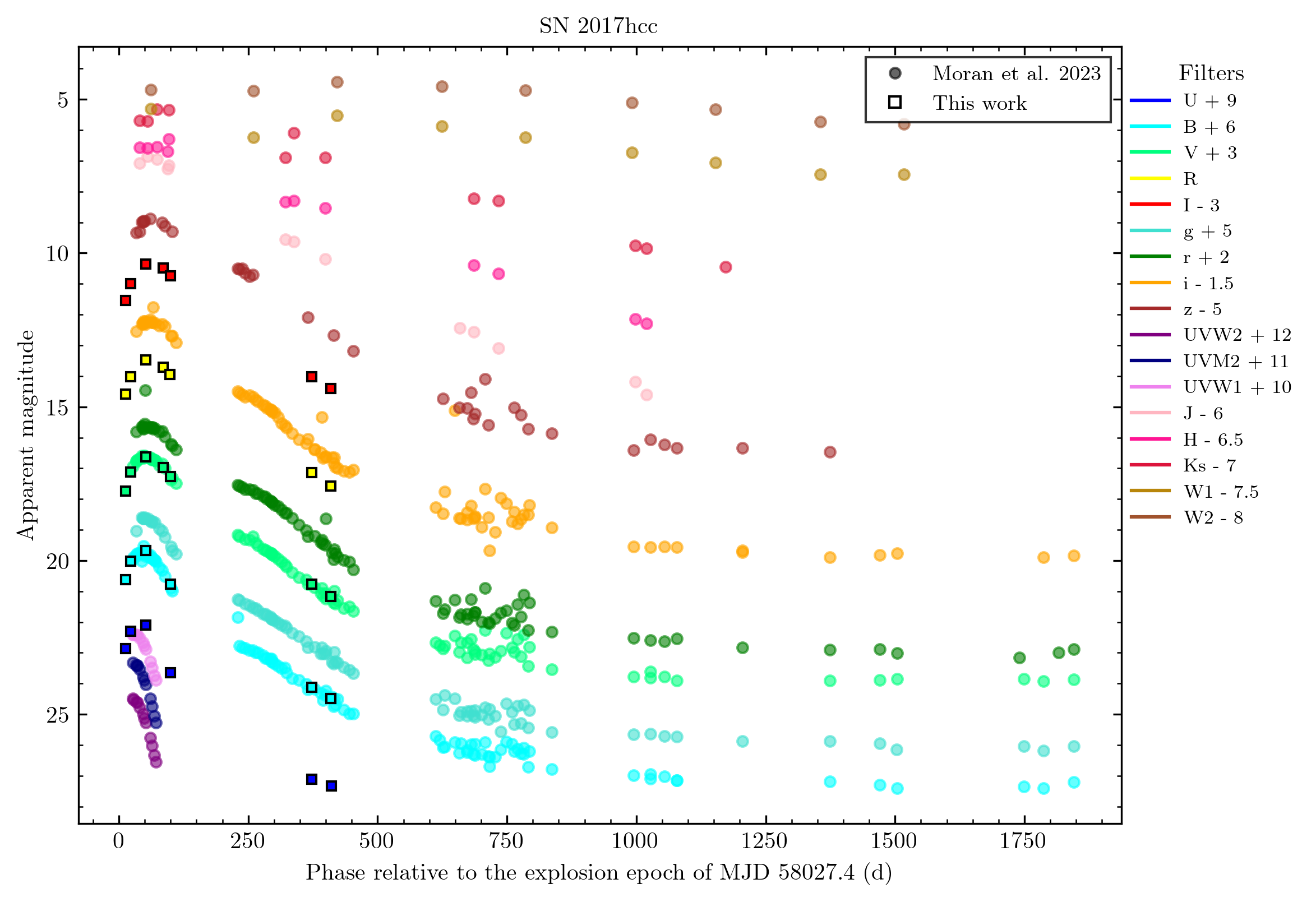}
 \caption{Multi-band optical light curves of SN 2017hcc. Our observations have been overlaid on photometry from \citet{moran} for comparison, and the magnitudes in various filters have been offset by different constants for ease of viewing.}
  \label{hcc_lc}
    \end{minipage}
    \hfill
    \begin{minipage}{0.45\textwidth}
        \centering
        \includegraphics[scale=0.45]{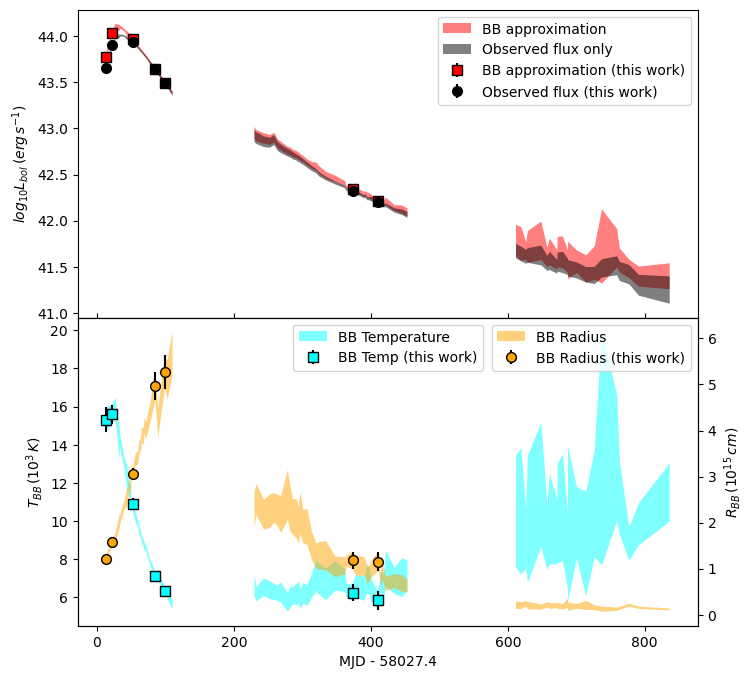}
    \caption{Top panel: Pseudo-bolometric and bolometric light curves of SN 2017hcc. The plot includes luminosities from direct integration and blackbody approximations. Bottom panel: Temperature and radius evolution estimates from the blackbody approximation. Data from this study has been plotted along with luminosities derived from \citet{moran}. The shaded regions indicate the uncertainties.}
       \label{hcc_bol}
    \end{minipage}
\end{figure*}

The pseudo-bolometric and bolometric light curves of the supernovae were generated using \texttt{SUPERBOL} \citep{nicholl2018superbol}. For SN 2017hcc, we combined our distance- and reddening-corrected HCT $UBVRI$ photometry with multi-band observations from \citet{moran}, which include the \textit{Swift}/UVOT filters $UVW2$, $UVM2$, $UVW1$, and the $BgVrizJHK_s$ bands. As per \citet{nicholl2018superbol}, we performed polynomial interpolations and constant-color extrapolations to map the magnitudes across various filters to a common set of epochs, for which we adopted the $V$ band as reference. In accordance with \citet{moran}, the ultraviolet \textit{Swift} bands were only included in the first $\sim 150$\,d post-explosion, after which their contribution to the total luminosity fell off rapidly. These bands were fitted with a quadratic polynomial. At later times, luminosities were computed using only the $BgVrizJHK_s$ bands. Pseudo-bolometric light curves were obtained via trapezoidal integration over the observed bands at each epoch after converting magnitudes to spectral luminosities ($L_\lambda$). To derive bolometric luminosities, we applied blackbody corrections to the pseudo-bolometric curves through SED fitting (see Fig.\,\ref{hcc_bol}).

\begin{figure*}
    \centering
    \includegraphics[scale=0.55]{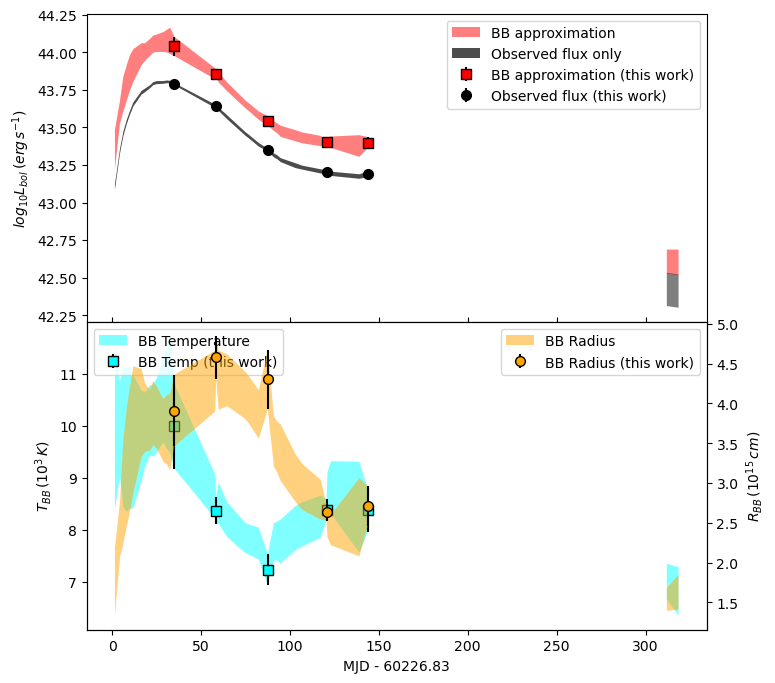}
    \caption{Top panel: Pseudo-bolometric and bolometric light curves of SN 2023usc. The plot includes luminosities from direct integration and blackbody approximations. Bottom panel: Temperature and radius evolution estimates from the blackbody approximation. Data from this study has been plotted along with luminosities derived from \citet{lasair}. The shaded regions indicate the uncertainties.}
    \label{usc_bol}
\end{figure*}

To construct the bolometric light curves for SN 2023usc, we used our HCT $ugriz$ photometry together with $g$- and $r$-band magnitudes from the \citet{lasair} archive, adopting the $g$-band as reference. Unlike SN 2017hcc, the absence of ultraviolet data precluded the use of separate ``early" and ``late" seasons. Given the more limited wavelength coverage, the logarithmic blackbody-corrected luminosities exceed the corresponding pseudo-bolometric values by $\sim2\%$ in the first $\sim 150$\,d; i.e., $(\log_{10}{L_{bb}}-\log_{10}{L_{obs}})/\log_{10}{L_{obs}} \approx 0.02$ (see Fig.\,\ref{usc_bol}).

\newpage
 
\section{ANALYSIS}
\label{ANALYSIS}

\subsection{SN 2017hcc}

\citet{moran} estimated the explosion epoch of SN 2017hcc at  MJD $58027.4 \pm 1.0$, based on the mid-point of the epoch of last non-detection and first detection. We use this value here for the sake of comparison with other works \citep{smith,Prieto-2017}.  The optical light curves of the long lived SN 2017hcc have been extensively discussed by \citet{moran}, and we find that our additional $U B V R I$ photometry, combined with their reported Swift and (terrestrial) optical as well IR photometry in the $UVW2$, $UVM2$, $UVW1$ and $UBgcVroizJHKs$ bands consistently reproduces the bolometric light curves in their paper, with the complete datasets shown in Figs.\,\ref{hcc_lc} and\,\ref{hcc_bol}. Below, we discuss the spectroscopic evolution of this event, and in further sections, do a comparative study with other events in the same class.

\begin{figure*}
    \centering
    \includegraphics[scale=0.55, trim=3cm 1cm 1cm 0cm ]{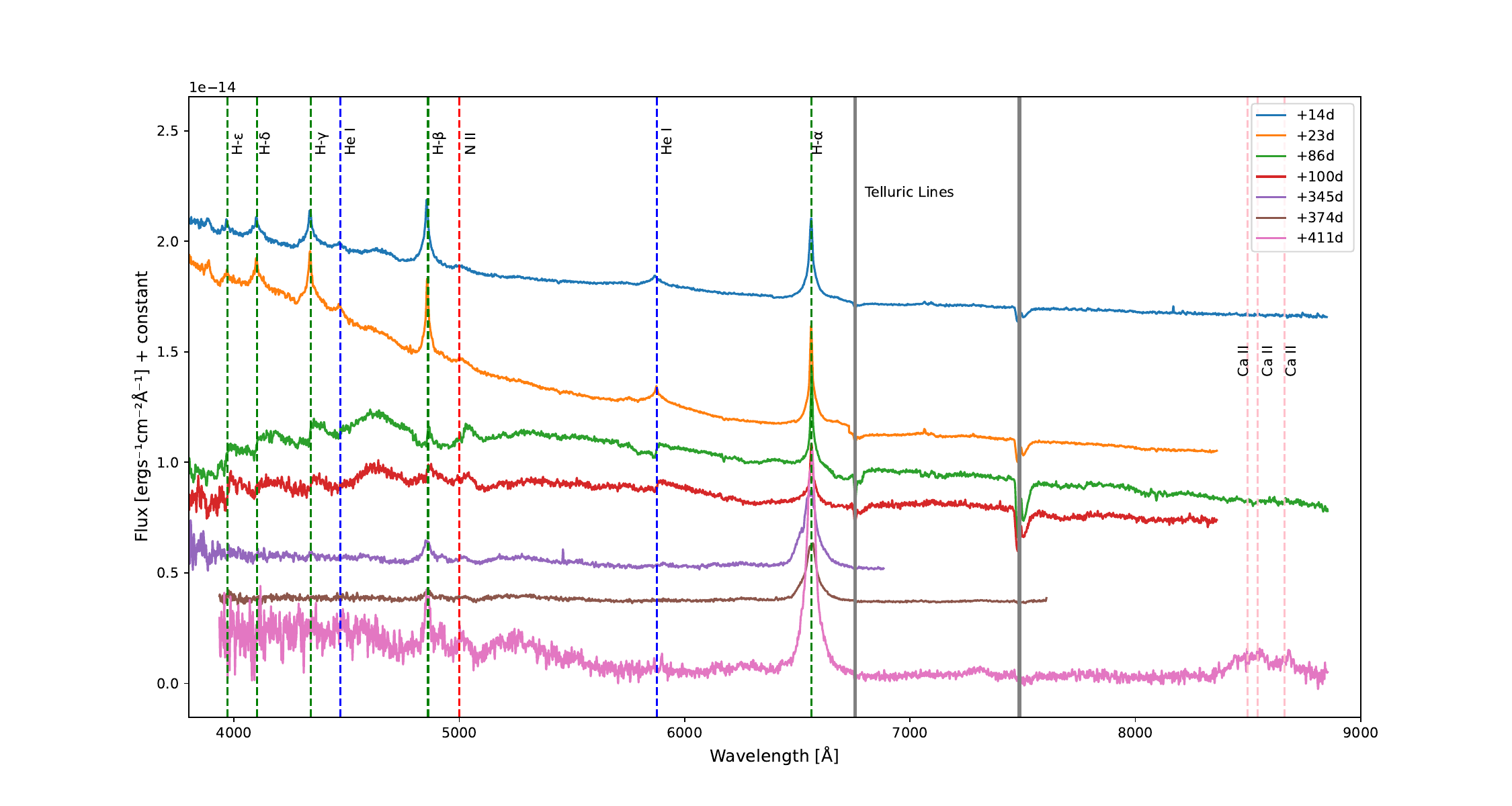}
    \caption{Multi-epoch (+14 to +411\,d post explosion) spectra of SN 2017hcc, taken with the HCT. The spectra have been shifted to the rest-frame of host galaxy, but no telluric correction has been applied. The fluxes have been scaled on the y-axis.}
    \label{hccspec7epoch}
\end{figure*}

Our spectra, redshift corrected to the rest frame of the host galaxy, and dereddened, are shown in Fig.\,\ref{hccspec7epoch}. We identify broad lines of H$\alpha$, H$\beta$, H$\gamma$, H$\delta$, H$\epsilon$ and He I. We also identify Ca II lines in the last epoch of observation of +411\,d, although they are first detected at +104\,d (see Fig.\,7, \citealt{moran}).
The NaI D$_1$/D$_2$ lines associated with Milky Way absorption at 5890/5896\,{\AA} are well separated from the host galaxy's own absorption at 5989/5995\,{\AA}. We thus estimate a redshift $z=0.0168$ for the host galaxy, comparable to the values used by \citet{smith}. As expected, the early time spectra reveals traces of a hot continuum, and is dominated by H-Balmer lines, as well as He I 5876\,{\AA}. The progenitors of Type-IIn supernovae have substantial mass-loss, and therefore it is to be expected that they would be embedded in a dusty halo, at least some of which should have been sublimated, and flash ionized, by the SN shock. The very early time spectra of even some type-IIP supernovae, with progenitor masses $< 18 M_{\odot}$ shows the C IV doublet (5801/5811\,{\AA}) in their spectra, as in the case of SN 2023ixf \citep{Sutaria-2023-I, Sutaria-2023-II, Bostroem+2023_SN2023ixf}, which can be attributed to carbon grains. This feature, however, is absent in SN 2017hcc, possibly because the earliest spectrum of SN 2017hcc was taken 6\,d post explosion (see Fig.\,7, \citealt{moran}), by which epoch the flash ionized features had faded away.

 The H$\alpha$ line at any epoch can be resolved into multiple components, including a narrow emission from the CSM, as well multiple broad components which can be attributed to the various stages of shocked ejecta emission and ejecta-CSM interaction. 
Here, we refer to velocities in the range $\sim 100 ~~{\rm to}~~  1000$\,km\,s$^{-1}$ as “narrow" lines. The intermediate-width component, ranging from $\sim 1000  ~~{\rm to}~~ \sim 5000$\,km\,s$^{-1}$ primarily emerges due to Thomson scattering  \citep{chugai} deep within the CSM while the “broad" lines with $v \gg 5000$\,km\,s$^{-1}$, can be attributed to the supernova the ejecta. 
  
\begin{figure*}
\centering
 \includegraphics[scale=0.5]{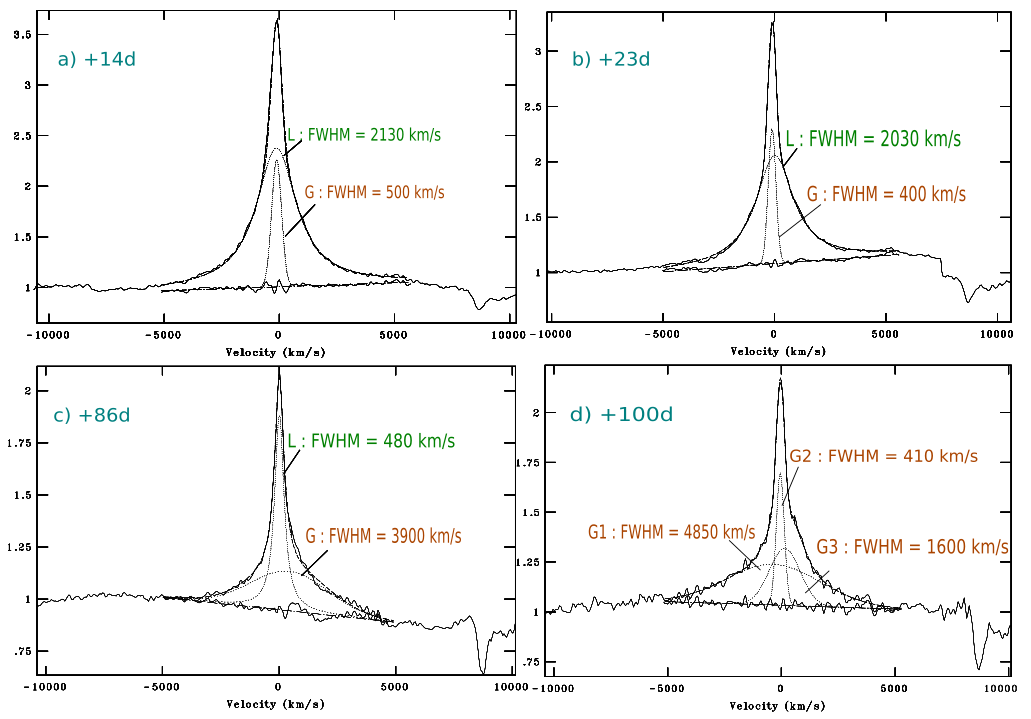}
 \caption{Resolving the H$\alpha$ line profile of SN 2017hcc into multiple components, from +14\,d to +100\,d post explosion. All spectra have been boxcar smoothed with a kernel of width 3, and  $v=0$\,km\,s$^{-1}$ is set at 6562.8\,{\AA}. The normalized flux is on the y-axis. The best fit profiles are marked as L (Lorentzian) or G (Gaussian), as the case may be.}
  \label{hcc-hal1}
\end{figure*}

\begin{figure*}
 \includegraphics[scale=0.6, angle=180]{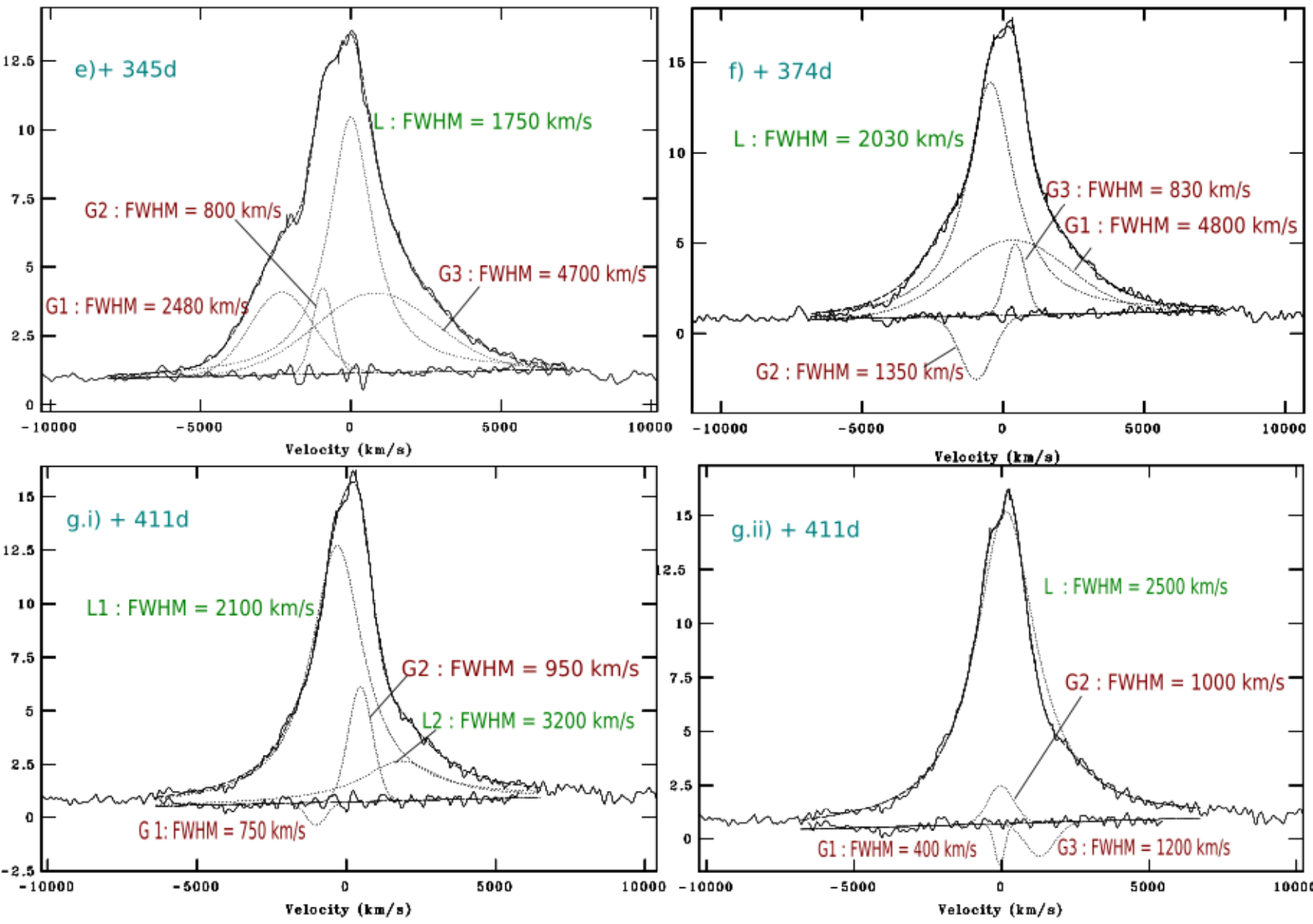}
 \caption{Evolution of the H$\alpha$ line profile of SN 2017hcc, on later epochs from +345\,d to +411\,d post explosion. The continuum was extracted over selected, line-free regions in the 6000\,{\AA} to 7000\,{\AA} band and the normalized flux is on the y-axis. All spectra have been boxcar smoothed by a kernel of width 3\,{\AA}, and $v=0$\,km\,s$^{-1}$ is set at 6562.8\,{\AA}
 (\tt{IRAF/splot}).
 }
  \label{hcc-hal2}
\end{figure*}

\begin{figure*}
    \centering
    \begin{minipage}{0.45\textwidth}
        \centering
       \includegraphics[scale=0.32, trim=4cm 0cm 0cm 1cm]{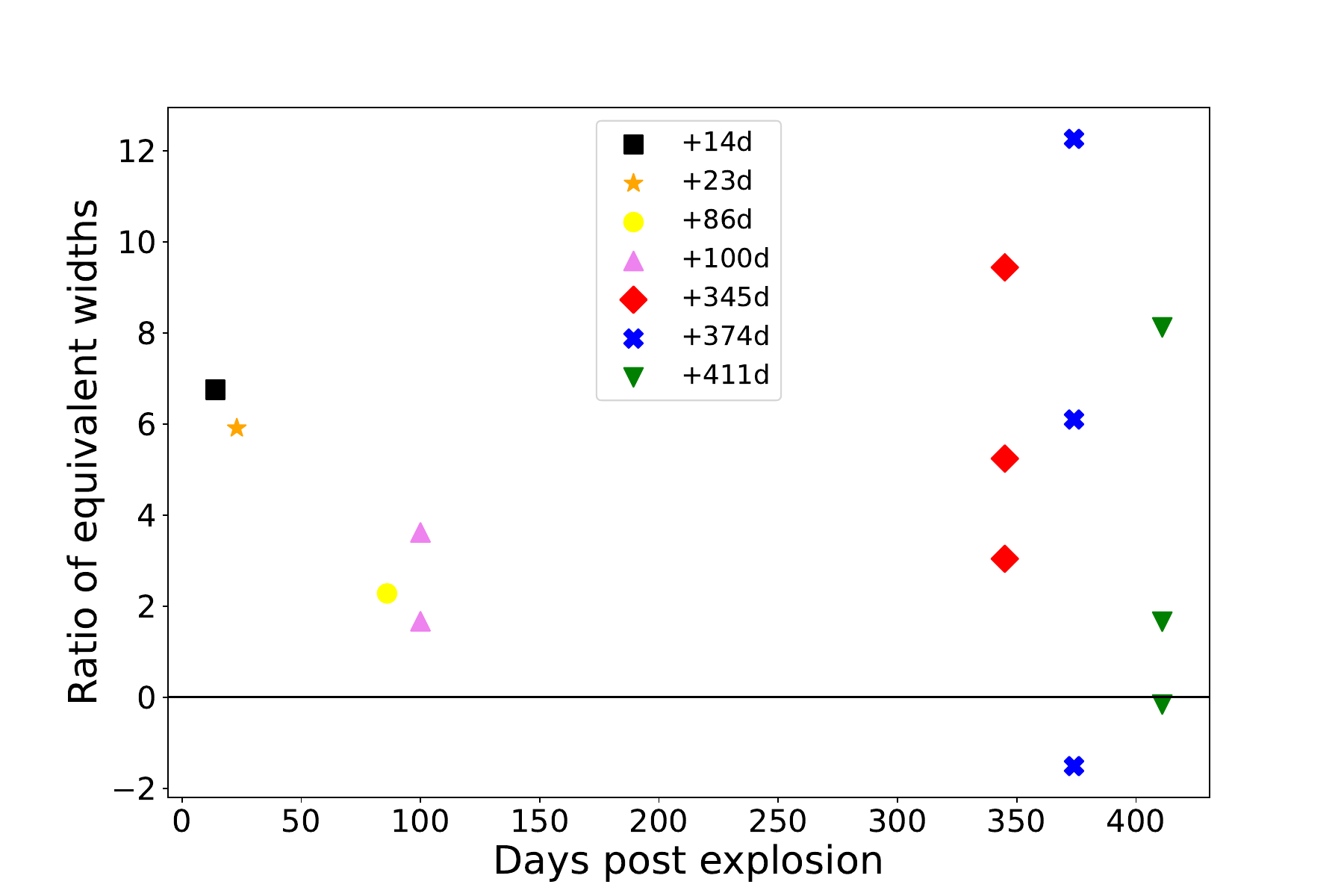}
 \caption{Ratio of equivalent width of the broad component of H$\alpha$ with respect to narrow and intermediate (wherever present) component on different epochs of SN 2017hcc. }
  \label{equivalent}
    \end{minipage}
    \hfill
    \begin{minipage}{0.45\textwidth}
        \centering
        \includegraphics[scale=0.28, trim=4cm 0cm 2cm 1cm]{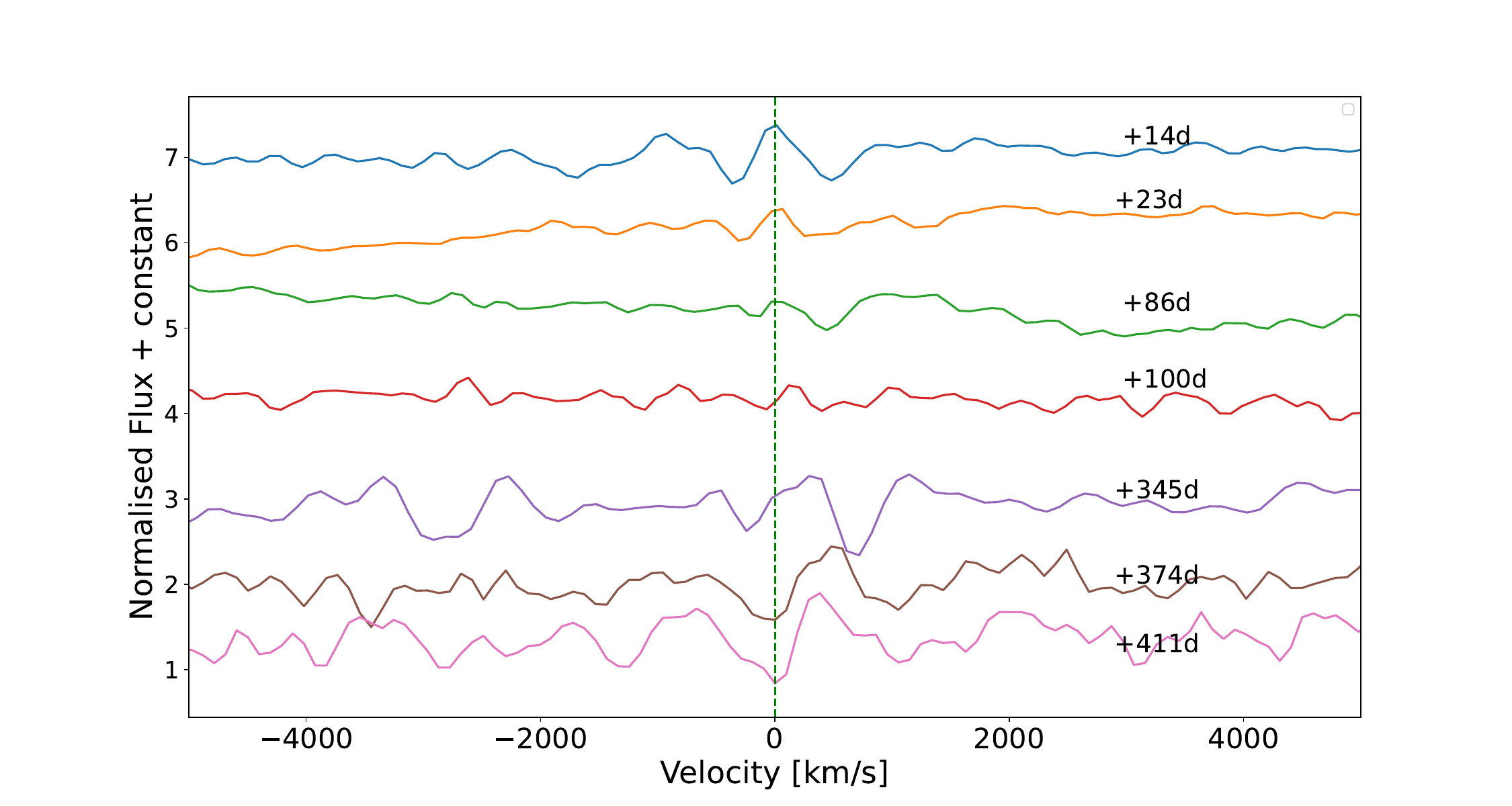}
    \caption{The fit residuals obtained after removing the narrow, intermediate and broad components of SN 2017hcc, mentioned in table.\,\ref{hcc}, and shown in Fig.\,\ref{hcc-hal1} and\,\ref{hcc-hal2}.}
       \label{pcygni hcc}
    \end{minipage}
\end{figure*}

 In the discussion below, we set $v=0$\,km\,s$^{-1}$ at the rest wavelength of H$\alpha$ (6562.81\,{\AA}) and deblend the line using the {\tt IRAF/splot} package, and Gaussian or Lorentzian profiles. Our aim is to deblend and remove the line, and to examine the the residual spectra for presence of a true P-Cygni structure. While supernovae line profiles are the result of multiple distortions introduced by the changing local opacities and thermodynamic conditions in the rapidly expanding material, we do note that a purely Gaussian line profile is typically associated with Doppler broadening (a purely thermal effect), while a purely Lorentzian profile is associated with collisional (pressure) broadening.  
  Using the dereddened spectra shifted to the rest frame, we provide, in Table\,\ref{hcc}, the line characteristics, consisting of the full width at half maximum (FWHM) velocity $v$ in km\,s$^{-1}$, centroid $\lambda_0$ in\,{\AA} and the goodness-of-fit statistic. We also provide a fit uncertainty to the FWHM. For reference, we note that the peak of the H$\alpha$ line is at ~600 $\sigma$ over the instrumental and sky background on +411\,d. The fitted profiles for 14\,d to 100\,d post explosion, and 345\,d to 411\,d post explosion, are shown in Figs.\,\ref{hcc-hal1}  and\,\ref{hcc-hal2} respectively, and the residuals are shown in Fig.\,\ref{pcygni hcc}. We note that neither a pure Lorentzian nor a Gaussian is a perfect fit to the line components, but changing the fit profiles is found to worsen the fit statistic, often considerably.

\begin{figure*}
 \includegraphics[scale=0.6]{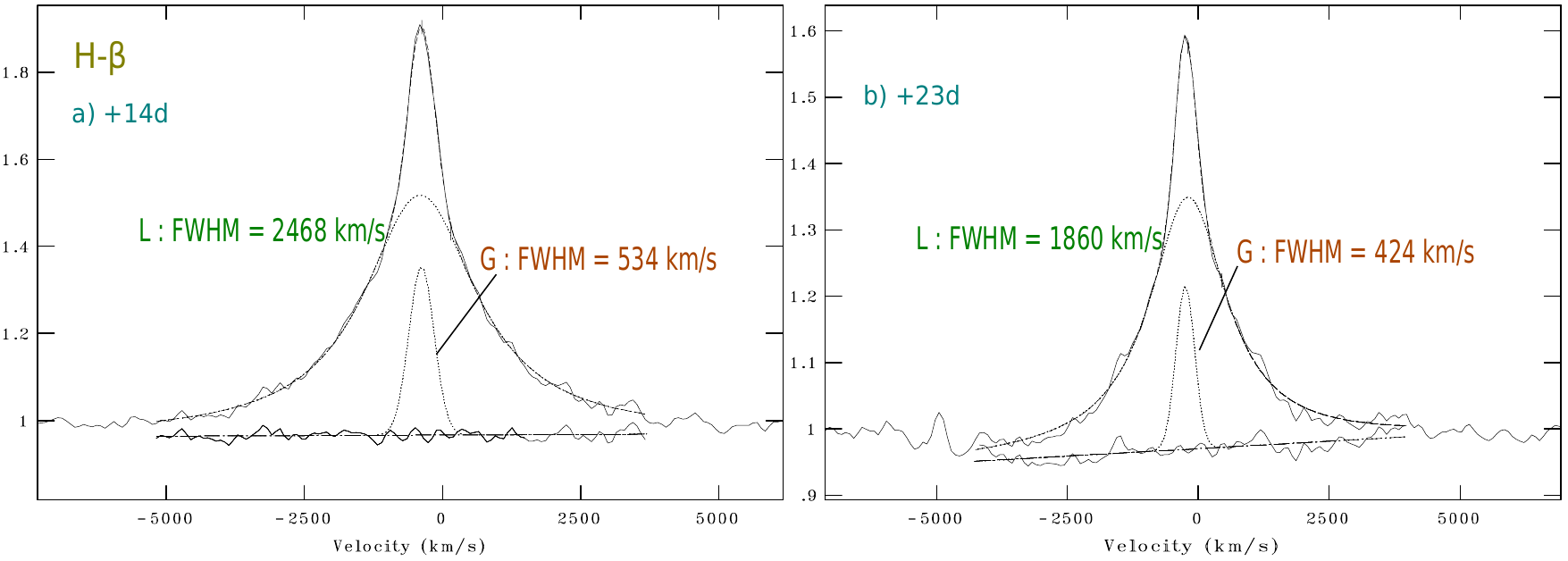}
\caption{Evolution of the H-$\beta$ line profile of SN 2017hcc, on earlier epochs of +14\,d and +23\,d post explosion. $v=0$\,km\,s$^{-1}$ is set 4861\,{\AA} for both events. The normalized flux is on the y-axis. }
  \label{hccbeta}
\end{figure*}

\begin{figure*}
 \includegraphics[scale=0.6]{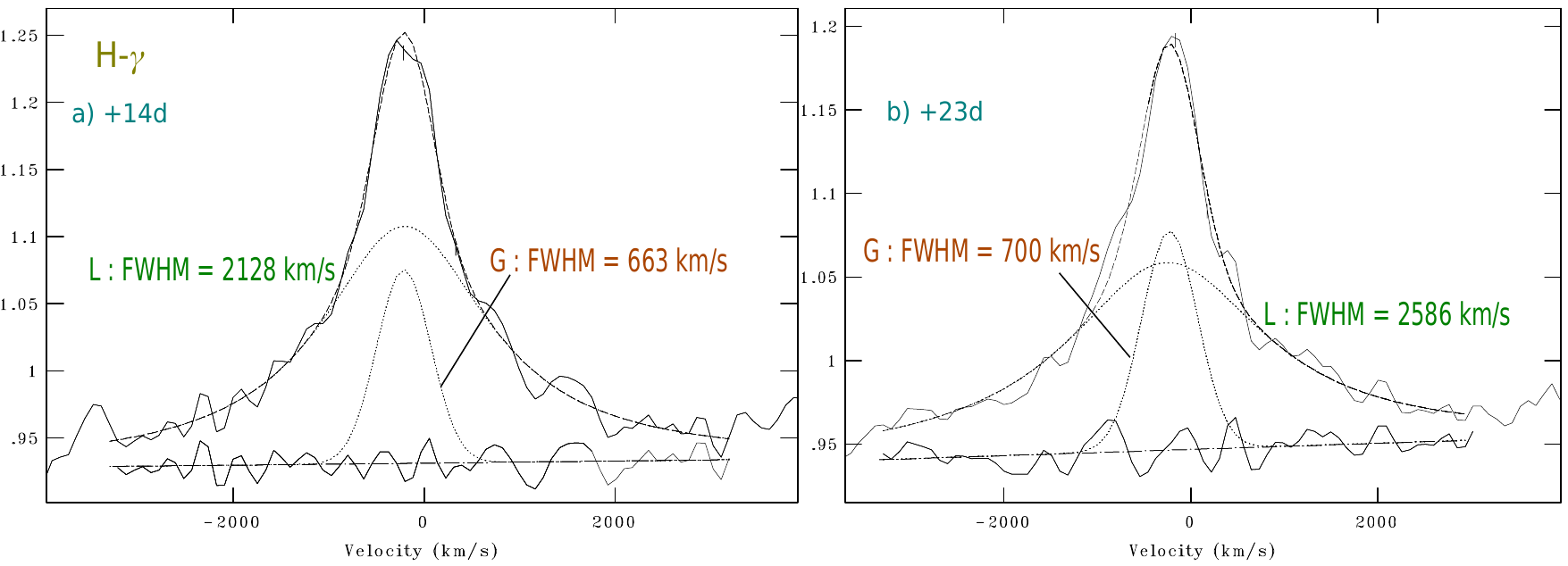}
 \caption{Evolution of the H-$\gamma$ line profile of SN 2017hcc, on earlier epochs of +14\,d and +23\,d post explosion. $v=0$\,km\,s$^{-1}$ is set 4340\,{\AA} for both events and the normalized flux is on the y-axis. The dotted lines represent the individual line components, as well as the  modeled spectral line.}
  \label{hccgamma}
\end{figure*}

The spectral evolution of SN 2017hcc demonstrates a clear transition from simple to increasingly complex line profiles, reflecting the changing dynamics of the ejecta–CSM interaction. During the early epochs (+14\,d, +23\,d, and +86\,d), the spectra can be characterized by a narrow component ($\sim$450\,km\,s$^{-1}$) and an intermediate component, primarily originating from the outer layers of the CSM (Fig\ref{hcc-hal1}). The H$\beta$ and H$\gamma$ lines exhibit similar behavior, reinforcing this interpretation(see in Fig\ref{hccbeta} and Fig\ref{hccgamma}). By around +100\,d, the H$\alpha$ profile appears blue-shifted and is best represented by a combination of narrow and intermediate-width components, accompanied by a weak P-Cygni absorption that gradually shifts redward, consistent with an expanding shell. Beyond this phase, the line profiles develop multiple components, likely arising from the expanding ejecta and its interaction with innermost CSM. In the later stages (+345\,d, +374\,d, and +411\,d), the initially narrow CSM emission becomes broader, while the intermediate and broader components associated with the ejecta become more prominent(Fig.\ref{hcc-hal2}). By +411\,d, the profile requires two alternative fits that both adequately reproduce the observed features, as summarized in Table\ref{hcc} and illustrated in Fig\ref{hcc-hal2}g. The overall progression from narrow, symmetric emission to complex, multi-component structures highlights the strengthening and eventual fragmentation of the ejecta–CSM interface over time.

We note here that \citet{smith} present a high-resolution MIKE-Echelle spectra of SN 2017hcc on near-coeval epochs of +24\,d and +351\,d,  as well as low-resolution MMT spectra on +25\,d, +88\,d and +340\,d. 
While they use only two components to resolve the H$\alpha$ profile in MMT spectra, they too note that at earlier epochs, the wider velocity component has a FWHM in the range 2000\,km\,s$^{-1}$. 
Moreover, their Gaussian fits to the H$\alpha$ profile on +351\,d (see Fig.\,6, \citealt{smith}) also shows a strong, red-shifted flux deficit, the extent of which they find to be both wavelength and epoch dependent. Finally, their MIKE Echelle spectra shows a P-Cygni component with the absorption trough centered at $\sim$ -51\,km\,s$^{-1}$ consistently on three epochs: 24\,d, 281\,d and 354\,d post-explosion, and is attributed to an overlying, pre-shocked, slow-moving, CSM material. The detection of such features is very unlikely in our low resolution spectrum, and we therefore regard it as an indication of the lines not being true Gaussians or Lorentzians. 
%

In Fig.\,\ref{equivalent} we plot the ratio of the equivalent width of the intermediate and broad H$\alpha$ component to the narrow component at different epochs. A consistent trend of decreasing ratio is seen in the earlier epochs. Up to +100\,d, the narrow component is clearly distinguishable from the intermediate component ($1000-2000$\,km\,s$^{-1}$) with an average velocity of $\sim 450$\,km\,s$^{-1}$. In the ``canonical" picture, the early emission from type-IIn emerges from a slow moving CSM as well as an ionized, low-density CSM-halo \citep{chugai}. The narrow line is the direct emission from the outermost layers of the CSM, while the intermediate component is generated via multiple scattering of photons in that expanding medium. 
In the case of SN2017hcc, on +14\,d, we find that the narrow component contributes 52.5\% of the total emergent flux. Using the expression in \citet{chugai} Sec.2.1, we estimate the line-of-sight Thomson scattering opacity for this component at $\tau_T= 1.004 $. This increases to 1.895 by +23\,d, as the narrow component's contribution declines to 44.9\% of the total flux. However from 345\,d onward, the narrow emission component widens, and has the velocity 800 to 950\,km\,s$^{-1}$. While this component still dominates the total flux, the increasing velocity suggests that either the fast moving SN ejecta has now collided with the CSM, thus imparting momentum to it, or that there are multiple shells to the CSM, corresponding to different epochs of ejection from the pre-supernova progenitor, and that an outer, faster moving layer has now encountered the supernova shock.

\begin{figure*}
    \centering
    \begin{minipage}{0.45\textwidth}
        \centering
        \includegraphics[scale=0.38, trim=3.5cm 0 4cm 2cm]{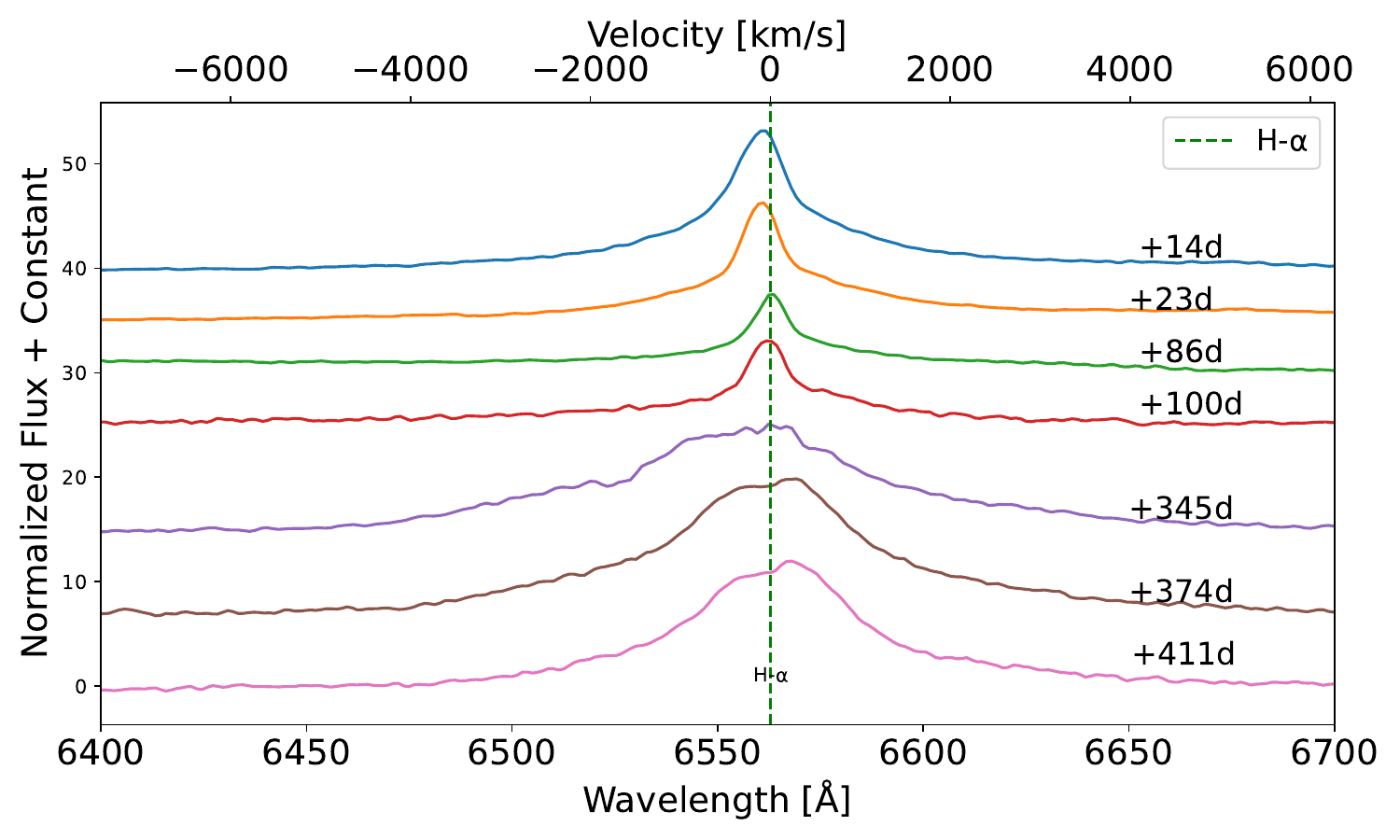}
    \caption{The variation of the H$\alpha$ over our 7 observation epochs of SN 2017hcc, from +14\,d to +411\,d post-explosion, are shown here. The spectra have been normalized to the local continuum. The profile's peak is seen to “flatten" +100\,d onward.}
    \label{hal7}
    \end{minipage}
    \hfill
    \begin{minipage}{0.45\textwidth}
        \centering
        \includegraphics[scale=0.25, trim=4.0cm 0cm 3cm 0cm]{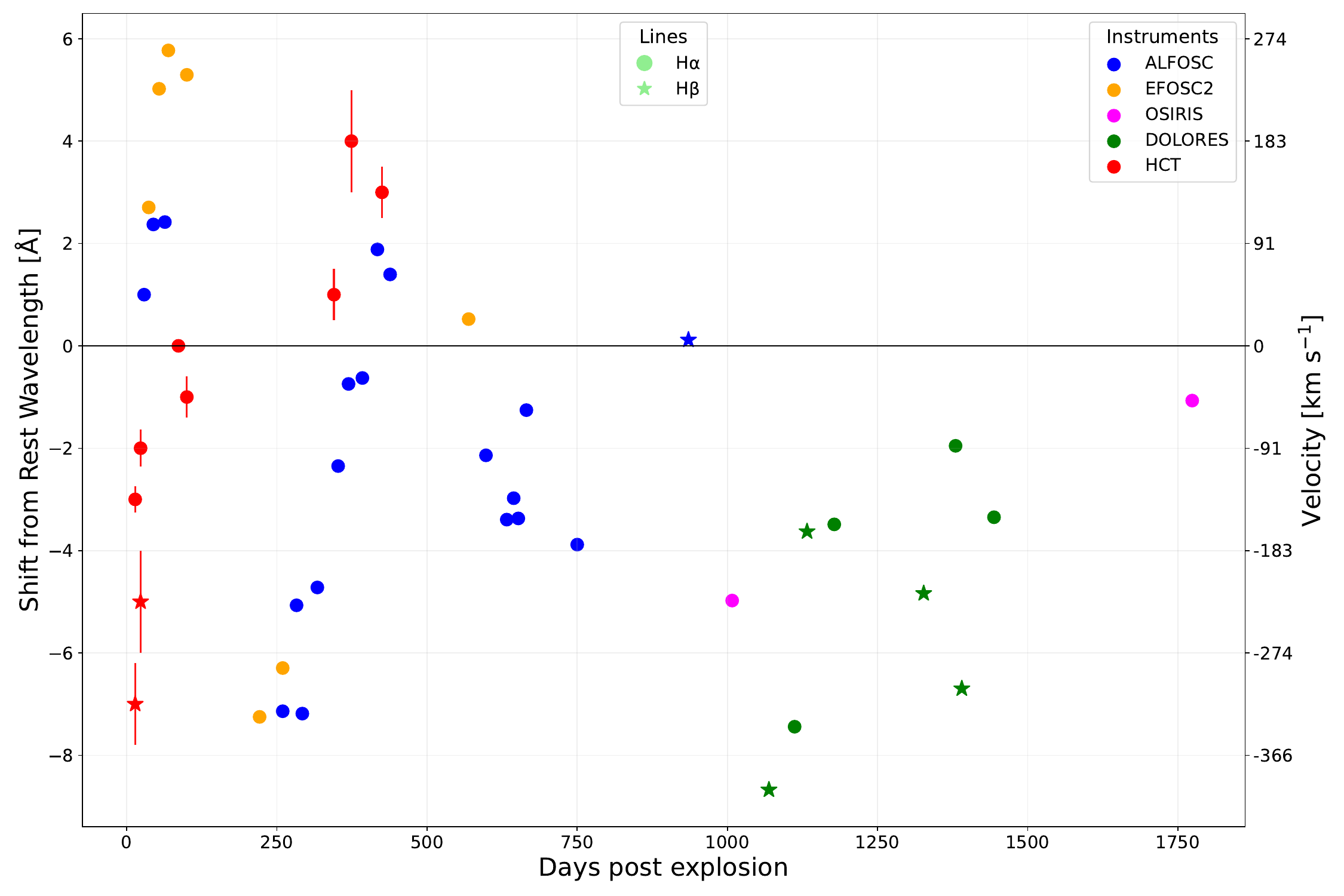}
     \caption{Peak of the H$\alpha$ line (narrow component) is plotted as a function of the epoch of observation. The centroids for spectra observed with NOT/ALFOSC, EFOSC2, OSIRIS and DOLORES instruments have been taken from Fig.\,17, \protect\citep{moran}, while that for the HCT are from this work. Following \protect\citep{moran}, the spectral calibration has been tied to the position of the [O I] $\lambda$6300.31 sky line on the HCT/HFOSC CCD. 
     }
     \label{shift}
    \end{minipage}
\end{figure*}

We now examine the shift in the H$\alpha$ line center over multiple epochs. The centroids of the ``narrow'' emission component typically falls closest to rest wavelength. SN 2017hcc exhibits a drift in the peak centroid (relative to the rest wavelength of H$\alpha$), with time -- a feature that is also seen in the more frequently sampled spectra presented by \citet{moran}. In Fig.\,\ref{shift} we use the data from Fig.\,13 of \citet{moran} and include our own results to show the complete trend in the “oscillations" from blue shift to red shift in the centroid. Other interacting supernovae (e.g. SN 2010jl; \citealt{fransson}) also show such a drift. The initial blue shift of the lines is variously attributed to obscuration of the red component by pre-existing dust, or from the dust formed in the cool dense shell \citep{smith}. Alternatively, acceleration of asymmetric pre-shocked CSM along the line of sight \citep{fransson} can also result in a blue shifted line. The early blue- to red- shift could also be explained by emission from an optically thick, shocked, CSM, which gradually thins by 100\,d, permitting transmission of the red-shifted low-velocity component into the line-of-sight. However, it is difficult explain the subsequent repetition of this trend from 250\,d to 450\,d, and again, to a lesser extent, beyond 500\,d, in  the single shell CSM model. If dust grain condensation blocks the red-shifted component, this would imply dust formation in a very short time frame. Instead, it may be possible that the blue-shifted lines are emitted from a cool, dense shell created between the forward and the reverse shock, formed by the interaction of the SN ejecta with a dense CSM \citep{dessart-2015}.  We note that the cool dense shell (CDS) forms at $\sim$78\,d for SN 1998S, as seen from the boxy profiles in the optical spectrum (Fig.\,7, \citealt{Fassia-2001}), although SN 2010jl (also a SNSN-IIn) showed no such feature for up to +1100\,d \citep{2010fransson}. Since the “boxiness" of the line profile depends on the density profile of the CSM,
this suggests that the progenitor of SN 2010jl may be significantly different from that of SN 1998S and SN 2017hcc.

After +80\,d, the spectrum profile begins to exhibit the presence of heavy metals, especially the Fe group in the 4500\,{\AA} to 5500\,{\AA} region. These are likely from the now optically thinning supernova ejecta. The Ca II $\lambda$$\lambda$$\lambda$ 8498,8542,8662 triplet is seen in our +411\,d spectrum, consistent with that observed by \citet{moran}.
 The Helium lines in SN 2017hcc are notably prominent and broad. At +14\,d the Helium $\lambda 5876$ line can be resolved into three Gaussians components: (a) a red-shifted one with FWHM 3280\,km\,s$^{-1}$ centered at 5840\,{\AA}, (b) a blue-shifted one with FWHM 3050\,km\,s$^{-1}$ centered at 5885\,{\AA} and (c) a narrow component with FWHM 860\,km\,s$^{-1}$ blue-shifted by 8\,{\AA} to 5867\,{\AA}. However, by +23\,d, this becomes a two Gaussian component feature, both at 5872\,{\AA}, but one with FWHM 500\,km\,s$^{-1}$ and the other with 2080\,km\,s$^{-1}$.
 
\begin{deluxetable}{ccccccccccc} 
\tablecaption{Multiple velocity components of Balmer lines in the spectra of SN 2017hcc. $\lambda_0$ is the component's centroid in \AA,  the velocities $v_1$, $v_2$, $v_3$, $v_4$ and their associated uncertainties are also in km\,s$^{-1}$.\label{hcc}}
\tablehead{\colhead{Spec.date (MJD)} & \colhead{ Epoch} & \colhead{ $v_1$} & \colhead{ Error}& \colhead{$v_2$} & \colhead{ Error} & \colhead{$v_3$} & \colhead{ Error}& \colhead{$v_4$} & \colhead{ Error}&\colhead{r.m.s. of fit} \\
 \colhead{(Date)}& \colhead{d}& \colhead{($\lambda_0$ {\AA})}& \colhead{$\pm$}& \colhead{($\lambda_0$  {\AA})}& \colhead{$\pm$}& \colhead{($\lambda_0$  {\AA})}& \colhead{$\pm$}& \colhead{($\lambda_0$ {\AA})}& \colhead{$\pm$} &\colhead{}} 
\colnumbers
\startdata
\multicolumn{7}{c}{H$\alpha$} \\
\hline
   58040.6948 &   +14   & 2130 &234 & 500  &40&- & - &-&- & 0.019\\ 
    (2017-10-14)         &          &(6560) && (6560)    &  & \\
  58049.5226 &  +23 & 2030  &374 & 400 &30&-& -&-&-& 0.016\\
     (2017-10-23)        &        & (6563) &&(6560)   &  &      \\
 58112.6640  &  +86 & &-      & 480 &28  & 3900 &100& - &- &0.018\\
    (2017-12-25)        &        & &        &(6562)& &(6566) & &&& \\
58126.6207  & +100 & 4850   &1138& 410 &31& 1600 & 155 &-&-& 0.015 \\
     (2018-01-08)       &        & (6555)  &&(6562)&&(6566)  & & \\
58371.9421  & +345 & 2480 &236  &1750&117 & 800&68 & 4700&280 & 0.085\\
     (2018-09-10)       &       & (6512)  && (6563) && (6542) &&(6580)\\
58400.7506 & +374 & 1350 & 156 & 830 &89& 2030 & 542&4880 &860& 0.121\\
    (2018-10-09)        &       & (6542) && (6572) && (6553) && (6569)\\
58437.6494 & +411 & 2100 &275  &950 & 175& 3200 &1084& 750&25&0.231\\
(2018-11-15)            &       &(6556) &&(6573 )&&(6602)& &(6540) \\
\hline
\multicolumn{7}{c}{H$\beta$} \\          
\hline
58040.6948 &  +14   & 2468 &368  & 534&136 & - & - &-&- & 0.012\\ 
  (2017-10-14)        &          & (4854) && (4854)& &  & & & \\
58049.5226 &  +23 & 1860   &140& 424 &35& - & - &-&-&0.0106\\
    (2017-10-23)        &       & (4857) && (4857)& & & & &\\    
\hline            
\multicolumn{7}{c}{H$\gamma$} \\      
\hline
     58040.6948  & +14   & 2128  &163&  663&56 &-&-& - & - & 0.0102\\ 
            (2017-10-14)      &       & (4337) & &(4336)&  &   & &&      \\
     58049.5226  &  +23 & 2586 & 175& 700  &89&-&-& -  & - & 0.008\\
        (2017-10-23)         &       & (4336) &&(4336)&    & &&  & \\      
\hline  
\enddata
\tablecomments{Multiple velocity components obtained by fitting different Gaussian and Lorentzian profiles. The FWHM velocity $v_4$ is an absorption component except on +345\,d, where it is an emission component.}
\end{deluxetable}



\subsection{SN 2023usc (= ZTF23abjhwem)}
 SN 2023usc was first detected \citep{di2023usc} on 11 October 2023(MJD 60228.3307). The last date of non-detection being 08 October 2023(MJD 60225.3373), we set the date of explosion at the mid-point epoch of  MJD 60226.8340 (09 October 2023, 20:01). The event was classified as a type-IIn supernova by \citet{cl2023usc} and was faint at discovery with $o=18.66$ \citep{di2023usc}. The $g$- and $r$-band light curves (taken 
 from ZTF/LASAIR\footnote{https://alerce.online/object/ZTF23abjhwem}, \citealp{alerce2023} and also shown in Fig.\,\ref{usc_lc}) peak at $r=16.519 \pm 0.042$ on  MJD 60261.4539. Our photometric sampling of this event is logged in Table\,\ref{usc_phot}. Additionally, we have also used the $g$- and $r$-band ZTF data to construct the bolometric light curves shown in Fig.\,\ref{usc_bol}.
 
\subsubsection{Photometric evolution of SN 2023usc:}

  We now examine the photometric evolution of SN 2023usc, which was tracked for +150\,d with the HCT in $ugriz$  bands and for 345\,d in the  ZTF $g$- and $r$- bands. The $r$- band light curve peaks $32.6$\,d post detection (on MJD 60261.4539) and the post-maximum it declines at an average rate of $\sim 0.02$ mag d$^{-1}$, while the rate of post-maximum decline in the $g$-band was $0.03$ mag d$^{-1}$.



 Fig.\,\ref{usc_colors} shows the extinction-corrected broadband color evolution of SN 2023usc as a function of rest-frame days since explosion. The colors were obtained from polynomial interpolations of various filter magnitudes at the epochs where we had $g$-band photometry. The $u-g$ color  is generally blue-to-neutral at early times. During the first $\sim 10$\,d it shows a sharp redward jump, reaching values of $\sim 0.7-0.8$ mag, after which it gradually shifts towards the blue over the next $\sim 30$\,d. The $u-g$ then rises redward again and eventually flattens out. The $g-r$ color displays the most pronounced evolution. In the first $\sim 10$\,d it exhibits a rapid blueward excursion before reddening steadily to $g-r \approx 0.3-0.5$ mag by $\sim 50-120$\,d, after which it remains approximately constant. This early blueward trend followed by monotonic reddening represents the dominant color change in SN 2023usc. The $r-i$ color shows a similar trend as the $u-g$ with a sharp redward rise during the first few days, and returns to near zero mag by $\sim 50$\,d. Thereafter it remains nearly flat and close to neutral throughout the covered epochs. Finally, $i-z$ is essentially constant throughout, indicating little color change in the reddest optical bands.  

Taken together, the colors indicate that the strongest evolution occurs during the first few weeks after explosion, with early excursions in $u-g$, $g-r$, and $r-i$, followed by a phase of much slower evolution dominated by the reddening of $g-r$. Since there is a substantial gap in our data between $150$ and $300$\,d, we cannot comment on the behavior of the colors during that interval. In addition, the final two measurements have large uncertainties, so they cannot be used to reliably constrain the late-time color evolution. 

Compared with SN 2017hcc at equivalent epochs (season one in \citealt{moran}), SN 2023usc shows a markedly quicker settling to near-neutral red colors and no obvious, strong long-lived reddening in $r-i$ or $i-z$ over the same early epochs. The strongest similarity is that both objects exhibit a prominent color shift of $g-r$, being initially blue followed by reddening over the first $\sim 100$\,d before flattening. The $r-i$ color of SN 2023usc is also consistently bluer than that of SN 2017hcc, especially between $40$\,d and $150$\,d after explosion. The bluest reported color of SN 2017hcc, the $B-V$, evolved steadily towards the red in the first season, in contrast to SN 2023usc's $u-g$, which has a redward trend very early on but becomes bluer at later times. In the redder colors, SN 2017hcc exhibited a strong redward evolution in $V-K_s$, reflecting the development of an NIR excess, whereas SN 2023usc shows almost no change in $i-z$ and remains close to zero throughout.

We note that \citet{Salamaso-2025} present optical pseudo-bolometric light curves (trapezoidal integration over $gri$ bands) for several strongly interacting SNe II, including SN 2022qml and SN 2021gci, which provide a useful comparison sample. Although our reconstructed pseudo-bolometric luminosity for SN 2023usc peaks at higher luminosities than the examples in \citet{Salamaso-2025} -- a difference that may partly reflect an overestimation of the line of sight extinction on our end -- the decline rate of these objects, $\Delta \log_{10} L_{obs}/\Delta t \approx 0.005$ d$^{-1}$, is similar to that of SN 2023usc, especially in the first $\sim 100$\,d.

\subsubsection{Spectral evolution of SN 2023usc:}
We used the publicly available spectrum obtained +12\,d after the date of discovery of SN 2023usc \citep{di2023usc} and compare it with that of SN 2017hcc taken +14\,d post-explosion with the HCT. The first HCT observation of SN 2023usc was on +37\,d, and we compare this with the +23\,d spectrum of SN 2017hcc. The velocity components for the H$\alpha$ lines are tabulated in Table\,\ref{usc-1}. For reference we note that the peak of H$\alpha$ does not exceed 60$\sigma$ above the 
background in any observation because of the intrinsic faintness of the object.

\begin{figure*}
    \centering
    \includegraphics[scale=0.471]{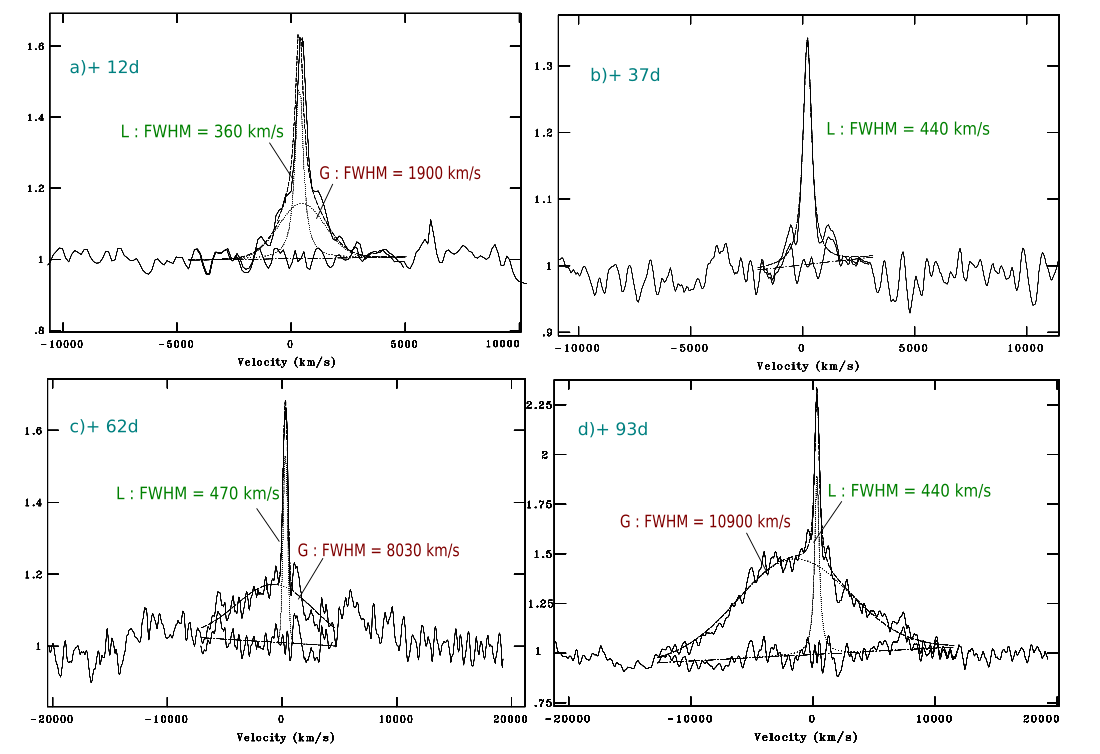} \\
    \includegraphics[scale=0.47]{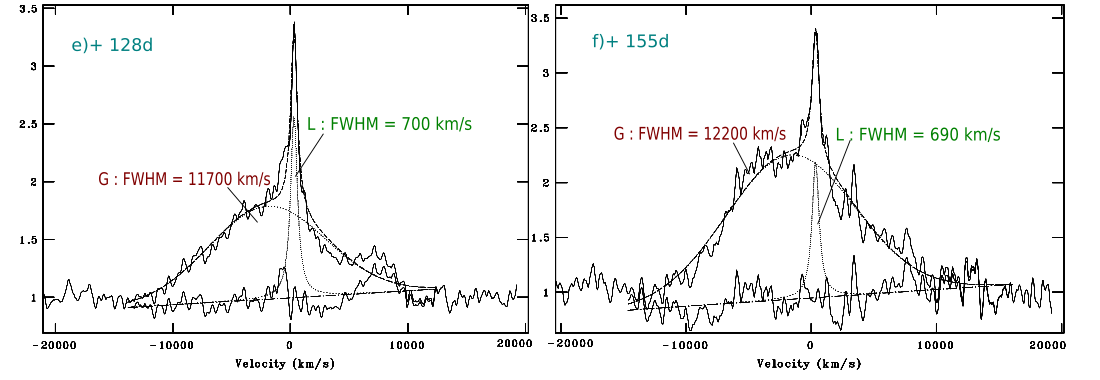}
    \caption{Evolution of the H$\alpha$ line profile of SN 2023usc, from +12\,d to +155\,d post explosion. $v=0$\,km\,s$^{-1}$ is set 6562.8\,{\AA} for all epochs and the normalized flux is on the y-axis. The continuum was extracted over selected, line-free regions in the 6000\,{\AA} to 7000\,{\AA} band.}
    \label{uschal}
\end{figure*}

 The H$\alpha$ profile on +12\,d was best fitted by a narrow Lorentzian profile of FWHM 360\,km\,s$^{-1}$ centered at $\lambda$6570 and an intermediate Gaussian of FWHM 1900\,km\,s$^{-1}$centered at $\lambda$6573 (Fig.\,\ref{uschal}). The presence of Galaxy emission line [N II] $\lambda$6583 appears to be prominent in the residuals. 
 However, because of the overall low S/N for the spectrum, any P-Cygni like residual is contaminated by noise and cannot be quantified.

It is difficult to isolate a continuum on either side of H$\alpha$ in the +37\,d spectrum because of the emergence of strong absorption features in the range $\lambda$6500 to $\lambda$6540 (-2500\,km\,s$^{-1}$ to -1150\,km\,s$^{-1}$ see Fig.\,\ref{uschal}b).  Moreover, at a red-shift of $z=0.06$, the broad H$\alpha$ line falls on top of the telluric band at $\lambda$6870, and this affects our analysis of SN 2023usc greatly compared to other events in this paper. Even after the telluric correction, we are unable to fit a broad or an intermediate component to the H$\alpha$ at this epoch. Taking the continuum at $\pm$ 2000\,km\,s$^{-1}$, the H$\alpha$ feature is fitted by a  single red-shifted Lorentzian of FWHM 440\,km\,s$^{-1}$ centered at $\lambda$6567. Host galaxy [NII] emission lines are possibly seen at $\lambda$6548.05 and $\lambda$6583.46.

From the +62\,d onwards, the H$\alpha$ profile shows a broad emission line, possibly from the ejecta. The narrow Lorentzian component, with FWHM 470\,km\,s$^{-1}$, is centered at $\lambda$6569, while the broad Gaussian component of 8030\,km\,s$^{-1}$ is centered at $\lambda$6550.

On +93\,d H$\alpha$ is fitted with a broad Gaussian component centered at $\lambda$6528, with FWHM 10900\,km\,s$^{-1}$, and a narrow Lorentzian with FWHM 440\,km\,s$^{-1}$ centered at $\lambda$6570.
The same spectral trend appears in the spectra at subsequent epochs (see Fig.\,\ref{uschal}d), but after this epoch, we start to see a ``broadening" of the narrow component. 
On +128\,d, H$\alpha$ resolves into a Gaussian component centered at $\lambda$6520, with FWHM 11700\,km\,s$^{-1}$, and a narrow Lorentzian component of FWHM 700\,km\,s$^{-1}$ on $\lambda$6570.
On +155\,d, the Gaussian component centered at $\lambda$ 6526, becomes much broadened, with FWHM 12200\,km\,s$^{-1}$, while the narrow Lorentzian component remains at  FWHM 690\,km\,s$^{-1}$, centered at $\lambda$6570.

We note that, unlike in the case of SN 2017hcc, there is no drift in the center of the H$\alpha$ line. The centroid of the narrow component remains consistently red-shifted relative to the rest wavelength (6562.8\,{\AA}), which suggests  strong blue-shifted absorption, possibly  due to a dense CSM along the line-of-sight, or due to a residual contribution from the rotation of the host galaxy.
The H$\beta$ profile is contaminated with noise, and cannot be easily distinguished from the continuum. Presence of [N II] were noted on +12\,d and later days (+93\,d,+128\,d), but the overall low S/N precludes the detection of other features. The spectrum is almost featureless in the later epochs, with the absence of other Balmer lines, except H$\alpha$.

\begin{figure*}
    \centering
    \includegraphics[scale=0.58, trim=3cm 1.5cm 2cm 2cm]{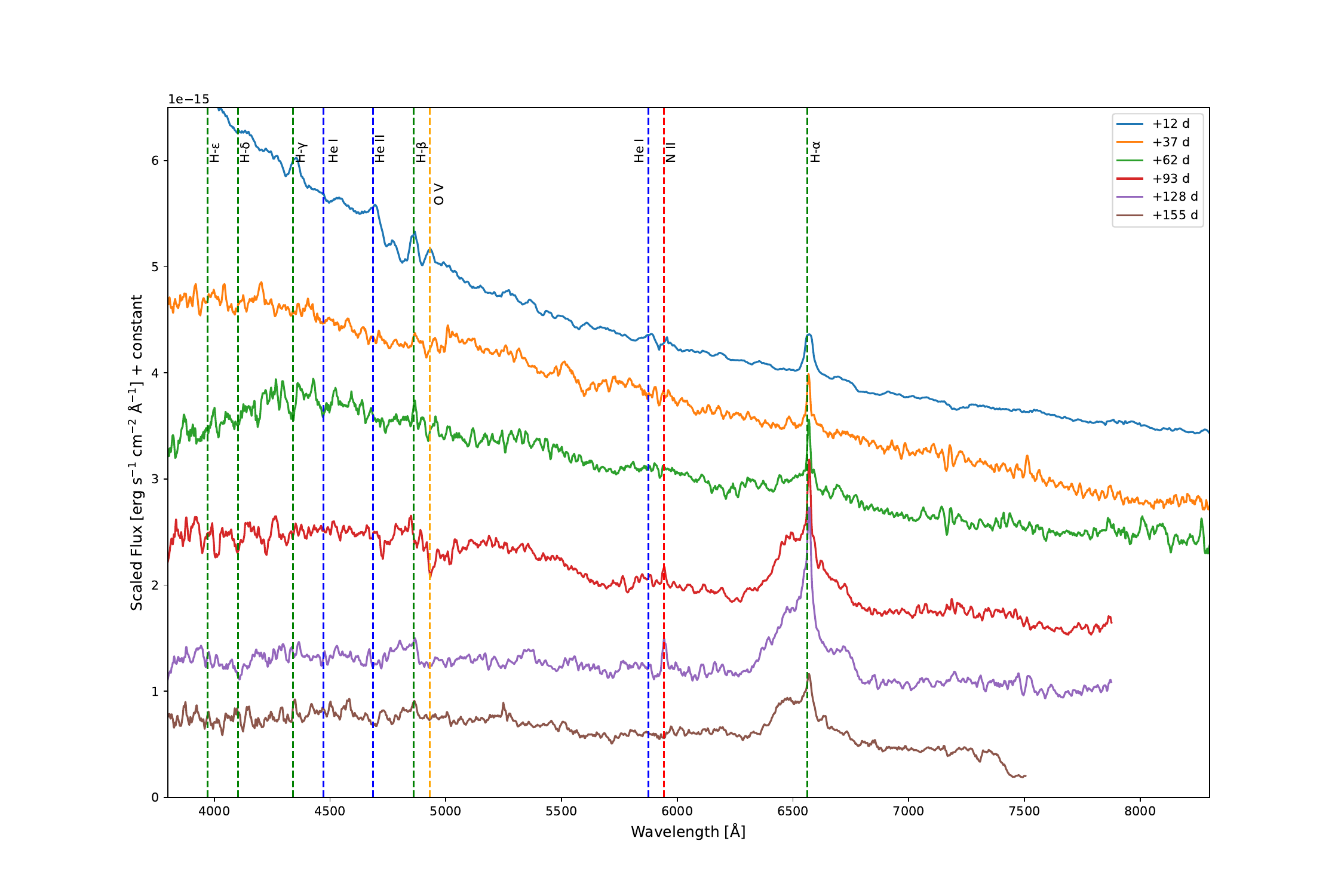}
    \caption{Multi-epoch (12\,d to 155\,d post explosion) spectra of SN 2023usc. The +12\,d spectrum is from UH88 / SNIFS (SCAT), while the rest of the spectra are from the HCT.
  The spectra have been shifted to the host-galaxy rest frame. Telluric correction has been applied, and the fluxes are presented on additive and multiplicative scales. All spectra have been boxcar smoothed with a kernel of width 12.}
    \label{uscspec5}
\end{figure*}

\begin{deluxetable}{ccccc} 
\tablecaption{Multi component velocity profiles of the H$\alpha$ line in the spectra of SN 2023usc. $v=0$\,km\,s$^{-1}$ corresponds to H$\alpha$ rest wavelength 6562.8\,{\AA}. $v_1$ and $v_2$  are the velocity components into which the line can be resolved with associated uncertainties estimated from profile fitting.
\label{usc-1}}
\tablehead{ \colhead{Spec.date (MJD)} & \colhead{ Epoch} & \colhead{ $v_1$ (km\,s$^{-1}$)} & \colhead{$v_2$ (km\,s$^{-1}$)}  & \colhead{r.m.s. of fit} \\
 \colhead{(Date)}& \colhead{}& \colhead{(centroid\,{\AA})}& \colhead{(centroid\,{\AA})}&  \colhead{}}
\colnumbers
\startdata
60239.5229 &   +12\,d & 1900$\pm$120 &360$\pm$42  &0.019\\ 
  (2023-10-22)                 &       &(6573) &(6570 ) \\ 
60264.3505 & +37\,d   & -     &  440$\pm$34  &0.018\\
    (2023-11-16)        &       &       &(6567) \\
60289.2745 &  +62\,d  & 8030$\pm$346 &  470$\pm$28  &0.032\\
  (2023-12-11)                      &       &( 6550)& (6569) \\ 
60320.1519 &  +93\,d &10900$\pm$1214  & 440$\pm$25   &0.041\\
  (2024-01-11)                     &       & (6528)&( 6570)   \\ 
60355.1688 &   +128\,d&  11700$\pm$1534& 700$\pm$45   &0.095\\
  (2024-02-15)                      &       &( 6520)&(6570)\\
60382.1861 &   +155\,d& 12200$\pm$2441 &690$\pm$48   &0.121 \\
   (2024-03-13)          &       &(6526) &(6570 ) \\
\enddata
\tablecomments{Multi component velocity profiles of the H$\alpha$ line in the spectra of SN 2023usc, at different epochs. The components (Lorentzian or Gaussian) are as marked in Fig.\,\ref{uschal}.  The continuum was extracted over selected, line-free regions in the 6000\,{\AA} to 7000\,{\AA} band.}
\end{deluxetable}

 \section{Analysis and comparison}
\label{Comparision}
 Our sampling of SN 2017hcc being somewhat sparse, we selected interacting supernovae for which near-contemporaneous spectra could be obtained for comparison purposes from the WISeREP archive. A comparison of the velocity components at +14\,d, +23\,d and +100\,d post-explosion is provided in Table.\,\ref{compare1}, \ref{compare2} and \ref{compare3} respectively. We aim to compare the spectroscopic properties of interacting supernovae both in the early as well as in the late stages.  H$\alpha$ being the most prominent spectral feature we make a detailed analysis of the multiple velocity components of the H$\alpha$ as the supernovae evolve. Additionally, the light curves of SN-IIn show considerable diversity, with slow rise times. For example, SN 2017hcc peaks on +57$\pm$ 2\,d in the ATLAS $o-$ band \citep{moran}, while  SN 2015da peaks on +100$\pm$ 5\,d in the R-band \citep{daTartaglia}.  However, as seen in Fig.\,\ref{usc_lc}, SN 2023usc peaks much earlier, at +37\,d in the ZTF $r-$.
We note from the SVO filter service\footnote{http://svo2.cab.inta-csic.es/theory/fps/} \citep{svophotometry,SVOversion1,SVOfilterprofile} that the band passes of the R- ($5462 - 8580$\,{\AA}, $\lambda_{ref}=6383$\,{\AA}), Atlas $o-$ ($5582 - 8249$\,{\AA}, $\lambda_{ref}=6750$\,{\AA})  and the ZTF $r-$ ($5498 - 7394$\,{\AA}, $\lambda_{ref}=6417$\,{\AA}) bands are approximately comparable at the blue end of the filter.


 Typically,  the H$\alpha$ profiles of the long lived type-IIn become “boxy" by $\sim100$\,d, as in the case of SN 2017hcc and SN 1998S \citep{Fassia-2001}. This flat-topped line profile is conventionally attributed in the literature to emission from a homologously expanding cool dense shell formed between the forward and reverse shocks (\citealp{DESSART2022,moran}). 
 Following \citet{chugai} and assuming a double shelled structure of the CSM consisting of a low-density halo with a higher density core, the pre-light curve peak H$\alpha$ profile can be modeled  being dominated by emission from the CSM. However, the H$\alpha$ profile at $\sim$100\,d and beyond also has some contribution from the interaction of the shocked ejecta with the CSM. 

\begin{deluxetable}{cccccccc} 
\tablecaption{Comparison of multiple velocity components with other SN type lln on +14\,d of SN 2017hcc.\label{compare1}}
\tablehead{\colhead{Object} & \colhead{Spec.date (MJD)} & \colhead{ Telescope/}  & \colhead{ Epoch}& \colhead{ $v_1$ (km\,s$^{-1}$)} & \colhead{$v_2$(km\,s$^{-1}$)} & \colhead{$v_3$(km\,s$^{-1}$)} &  \colhead{r.m.s. of fit} \\
\colhead{(red-shift)}& \colhead{(UTC)}& \colhead{Instrument}& \colhead{}& \colhead{(centroid\,{\AA})}& \colhead{(centroid\,{\AA})}& \colhead{(centroid\,{\AA})}& \colhead{}}
\colnumbers
\startdata
SN 2017hcc & 58040.6948&  HCT /HFOSC   & +14\,d &  2130 $\pm$234 &   500 $\pm$40 &   -&  0.019\\
(z=0.0168) &  	(2017-10-14)         &               &      &(6560) &(6560)    &            \\ 
SN 2023usc&60239.5229 &  UH88 / SNIFS & +12\,d & 1900$\pm$120 &360$\pm$ 42    &-   &  0.019  \\
(z=0.06)      &  (2023-10-22 )         &               &      &( 6573)&(6570 )    &            \\ 
SN 2022zi &	59602.1992& ESO-NTT / EFOSC2-NTT &+08\,d   &- &1450$\pm$256 &   - &  0.032 \\
(z=0.035)       &  (2022-01-23)         &               &      &       &(6564 ) &          &\\ 
SN 2021foa&59295.0000 &Ekar / AFOSC   &      +13\,d    &-       & 1070$\pm$114& - &0.03\\
(z=0.008386)        &    (2021-03-22)       &      &      &       &( 6565) &          \\
SN 2020cui&	58900.1541 & ESO-NTT / EFOSC2-NTT   &+12\,d &1056 $\pm$56 & 3555$\pm$589& - &  0.046\\
(z=0.025)        & (2020-02-21)          &                        &     &(6560)&( 6560)  &      \\
SN 2018hfg& 58401.84 & BAO-2.16m / Cassegrain &+05\,d & 700 $\pm$48 &2463$\pm$140 &-&0.029  \\
(z=0.023863)       &  (2018-10-10)         &                        &     &(6557)&(6563) &      \\
SN 2018khh& 58482.0512&  ESO-NTT / EFOSC2-NTT  &+12\,d &-& 660$\pm$86& 3300$\pm$550&0.015      \\
(z=0.023)       & (2018-12-30)          &                &     &      &(6564) &(6579) \\
SN 2016aiy& 57442.1189& ESO-NTT / EFOSC2-NTT   & +11\,d& -&882$\pm$145 & 3038 $\pm$347&  0.011\\
(z=0.01)     &    (2016-02-24)       &                        &        &   &( 6563) & (6567) \\
SN 2015da & 57041.8769&  Lijiang-2.4m / YFOSC & +11\,d&-&  1590 $\pm$560 &    -&0.067\\
(z=0.0.0066)       &  (2015-01-19)         &               &     &     &( 6556)     &     \\
\hline   
\enddata
\tablecomments{A comparison of deblended line velocities of SN 2017hcc, with other type-IIn events, at 14\,d post-explosion. $v=0$\,km\,s$^{-1}$ corresponds to H$\alpha$ rest wavelength 6562.8\,{\AA}. $v_1$,$v_2$ and $v_3$ are the velocity components into which the line can be resolved with associated uncertainties estimated from profile fitting. $v_2$ is the component whose centroid is closest to 6562.8\,{\AA}. All spectra except those of SN 2023usc and SN 2017hcc are taken from WISeREP \citep{WISEREP}.}
\end{deluxetable}

\subsection{\texorpdfstring{SN 2023hpd ($=$ PS23cup, ZTF23aajjzbm, ATLAS23knp)}{SN 2023hpd (= PS23cup, ZTF23aajjzbm, ATLAS23knp)}}

SN 2023hpd, discovered \citep{di2023hpd} on 02 May 2023(MJD 60066.5930)  at $z=0.0198$, showed clear signatures of a type-IIn supernova in its early spectra \citep{cl2023hpd}, which is dominated by narrow, Balmer emission features. We used the +29\,d post-explosion spectrum, shifted it to its rest-frame, and normalized it relative to the continuum.
Then it is compared with SN 2017hcc at +23\,d. The H$\alpha$ profile of SN 2023hpd in the $\pm 5000$\,km\,s$^{-1}$ region ($v=0$ at 6562.8\,{\AA}) is poorly fitted by a two component model, and we need to add a third Gaussian component to improve the fit. The best-fitted, lowest velocity component is a Gaussian of FWHM 1200\,km\,s$^{-1}$, centered at 6560\,{\AA}. Additionally, a blue shifted Gaussian with FWHM of $\sim$1300\,km\,s$^{-1}$ centered at 6523\,{\AA}, and a red shifted Gaussian with FWHM of 3090\,km\,s$^{-1}$ centered at 6582\,{\AA} are also present (see Fig.\,\ref{23}a).

We note that, at this epoch, SN 2017hcc required only two components: a narrow component directly emitted from the transparent outer layers of the CSM, and an intermediate component generated deeper within it via Thomson scattering (\citealp{chugai}). The early detection of a third component at a slightly higher velocity in SN 2023hpd suggests early CSM-ejecta interaction, indicating CSM formation close to the explosion epoch, i.e. mass ejection from the pre-SN star just prior to explosion.

\begin{figure*}
 \includegraphics[scale=0.54, trim=3cm 1cm 1cm .8cm]{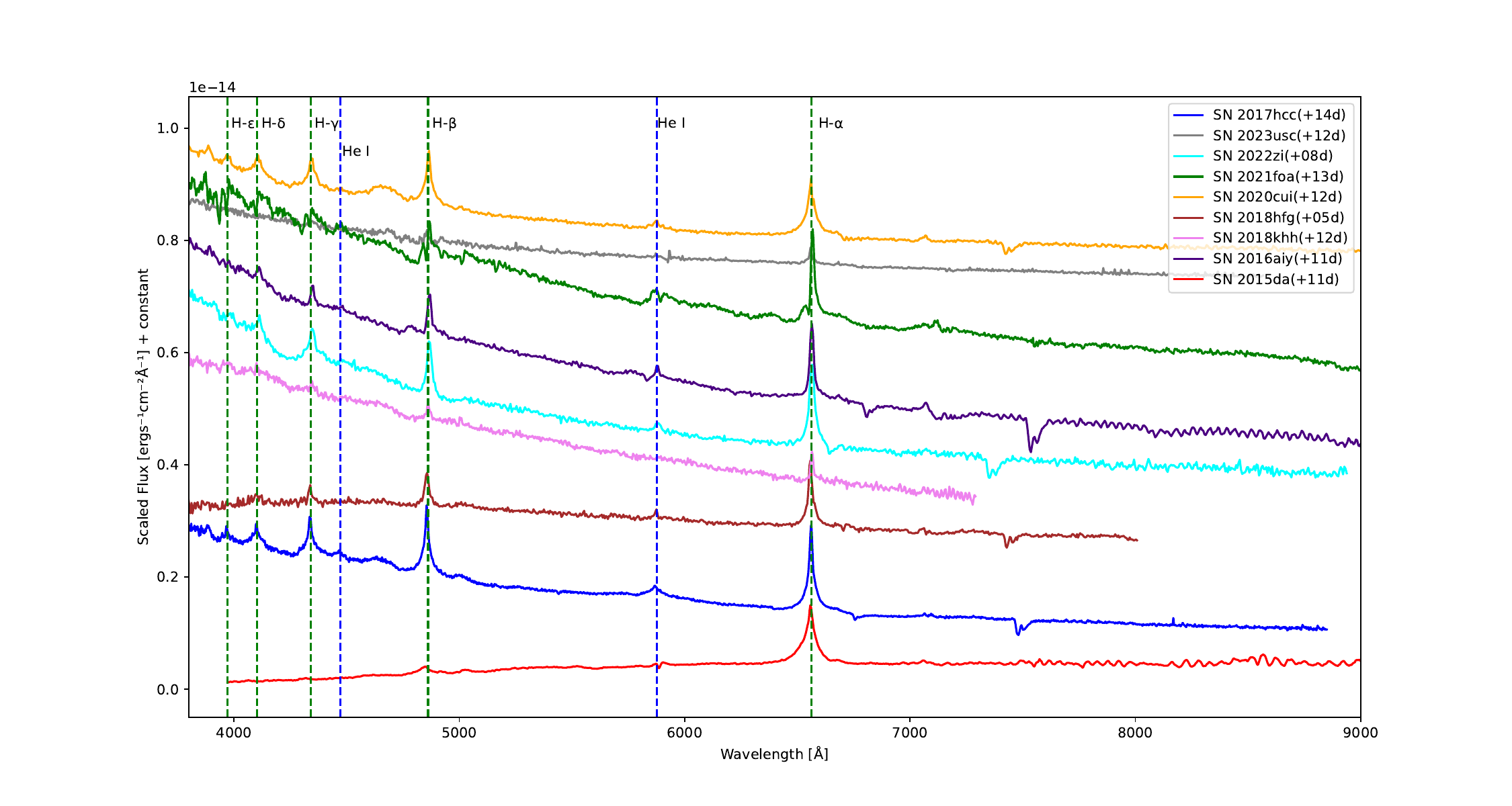}

\caption{\bf Comparison of other type-IIn supernovae at early epochs  with SN 2017hcc on +14\,d. The spectra have been scaled, and all have been shifted to the rest frame (for references to the spectra see the text).}

  \label{SNlln14 d}
\end{figure*}

\begin{figure*}
\centering
 \includegraphics[scale=0.55, trim=3cm 1cm 2cm 1cm]{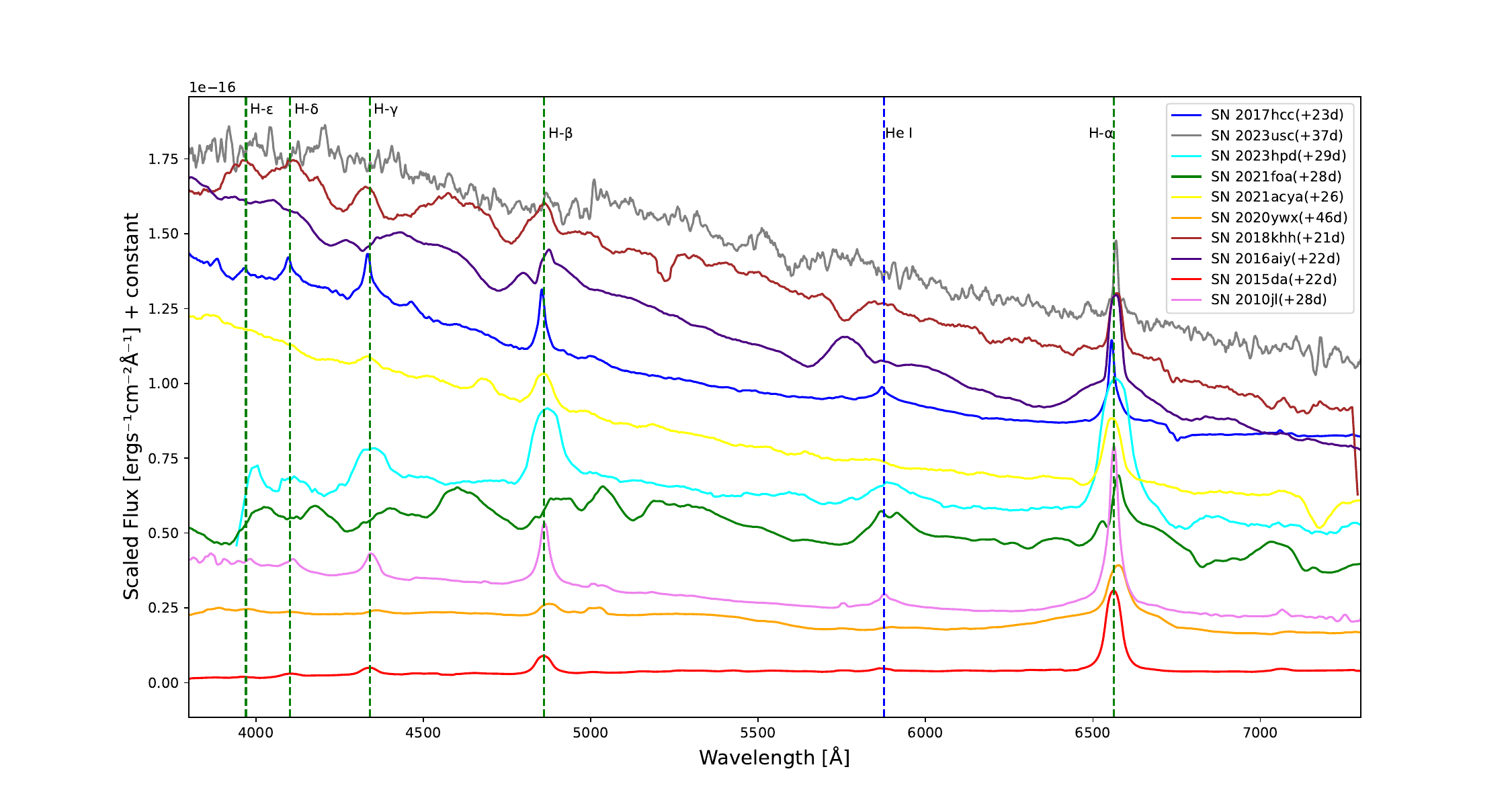}

 \caption{Spectral comparison of SN 2017hcc and SN 2023usc with other type-IIn supernovae on $\sim$+23\,d post explosion. The spectra have been scaled,  and all have been shifted to the rest frame (for references to the spectra see the text). }
  \label{SNlln23 d}
\end{figure*}
\begin{figure*}
\centering
 \includegraphics[scale=0.59,  trim=4cm 0 5cm 0]{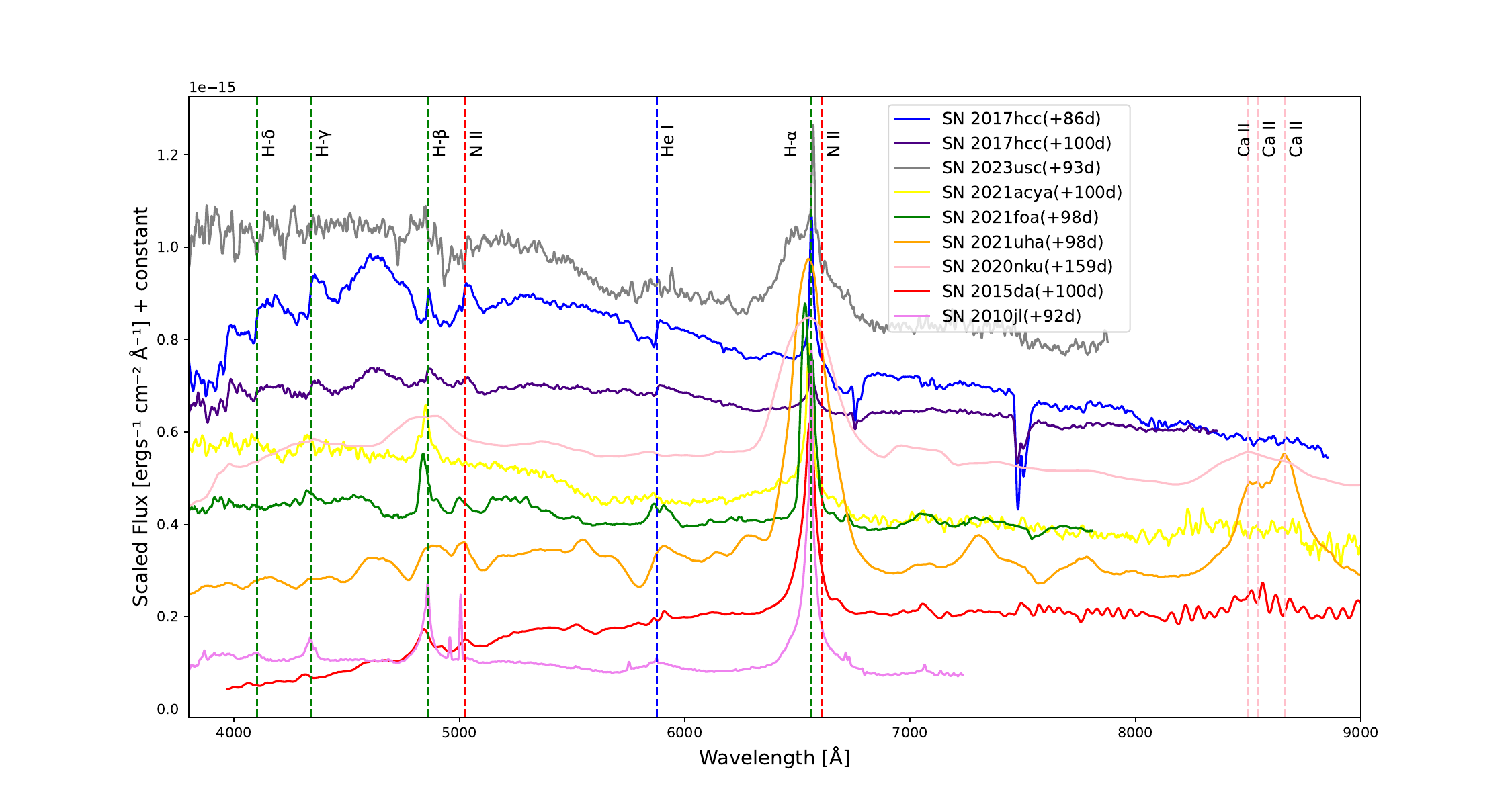} 
  \caption{Spectral comparison of SN 2017hcc and SN 2023usc with  other type-IIn events at $\sim $ +100\,d. The spectra have been scaled, and all have been shifted to the rest frame (for references to the data please see the text). }
  \label{SNlln100 d}
\end{figure*}

\subsection{\texorpdfstring{SN 2022zi ($=$ ATLAS22cbh, Gaia17 dkh)}{SN 2022zi (= ATLAS22cbh, Gaia17 dkh)}}

 SN 2022zi, discovered  \citep {di2022zi} on 17 January 2022 (MJD 59596.8890), at $z=0.035$ was classified as a young type-IIn supernova \citep{cl2022zi}, with the earliest publicly available spectrum at +8\,d post-discovery. 
  The date of last non-detection being 13 January 2022 (MJD 59592.8950), the mid-point between this epoch and the date of detection  -- 15 January 2022 (MJD 59594.892) -- was set as the explosion date.
 The rest frame shifted spectrum is continuum dominated, with characteristic and strong H$\alpha$, H$\beta$, H$\gamma$, H$\delta$,  H$\epsilon$ and He lines, confirming its similarity with  early stages of type IIn. A single, intermediate velocity, Lorentzian profile fits the H$\alpha$ region well, with FWHM$=1450$\,km\,s$^{-1}$ centered at 6564\,{\AA}, while the H$\beta$ is well fitted by a single, intermediate velocity Lorentzian of 1500\,km\,s$^{-1}$ centered at the rest wavelength (4861\,{\AA}). We note that there is no narrow velocity component even at this early stage, although we cannot rule out the possibility of a tenuous presence immediately post-explosion. This implies that the CSM is sufficiently dense to permit only transmission of Thomson scattered photons by +8\,d.

\subsection{\texorpdfstring{SN 2021uha ($=$ Gaia21 dno)}{SN 2021uha (= Gaia21 dno)}}
SN 2021uha \citep{cl2021uha} at $z=0.0079$ was associated with the nearby galaxy ESO 499-26. The last non-detection was on 17 July 2021(MJD 59412.9850), and the first detection was on 28 July 2021(MJD 559423.9150) using g-band Gaia photometric data \citep{di2021uha}.  We therefore set a mean date of 23 July 2021 (MJD 59418.4500)as the explosion epoch. The only publicly available spectrum of this object was taken on MJD 59517.34, i.e. +98.39\,d, and we compare it with that of SN 2017hcc at +100\,d.

In Fig.\,\ref{100}c, the H$\alpha$ line was fitted with four components: a blue shifted Gaussian component of width 5350\,km\,s$^{-1}$ centered at  $\lambda$6518, a red shifted Gaussian component of 5630\,km\,s$^{-1}$ centered at  $\lambda$6655, a Lorentzian profile with a FWHM of 3900\,km\,s$^{-1}$ centered at  $\lambda$6572, followed by an intermediate absorption component of Gaussian width 2900\,km\,s$^{-1}$ centered at  $\lambda$6570.  The post explosion epoch of 100\,d being the onset of interaction between the CSM and ejecta in the case of SN 2017hcc, the need for 4 components in the case of SN 2021uha suggests that a similar scenario may be playing out in this event as well. However, we note that  SN 2021uha also shows strong, broad, emission lines of  Ca II  $\lambda$$\lambda$$\lambda$8498,8542,8662 and [Ca II] $\lambda$$\lambda$7291,7323, at this epoch. Metal lines are indicative of emission from an optically thin ejecta, and not seen in either SN 2017hcc nor SN 2010jl at this epoch. Additionally, the light curve of SN 2017hcc gives absolute magnitude of -20.83 in the Gaia G-band(AB magnitude filters) on +59\,d. 
However, SN 2021uha has an absolute magnitude of -17.86 in the same band implying it is intrinsically fainter than SN 2017hcc. This suggests that SN 2021uha may actually be a Type II or Type IIP event with a possibly tenuous but complex CSM, as in the case of SN 2023ixf.

Notably, SN 2021uha shows a prominent split in H$\alpha$ at +98\,d, while SN 2017hcc did not exhibit this feature in their early epochs (showed in +345\,d).

\subsection{\texorpdfstring{SN 2021acya ($=$ ATLAS21blfc)}{SN 2021acya (= ATLAS21blfc)}}

SN 2021acya  was found in LEDA 742271 galaxy \citep{di2021acya} at $z=0.06203$.  A photometric and spectroscopic evolution of this object, with spectral sampling over 8 epochs ranging from +24\,d to +462\,d has been carried out by \citet{Salamaso-2025}. Following them, we set the mean date of explosion as 30 October 2021 (MJD 59517)$\pm 1$. The pseudo-bolometric light curve of this event (see Fig.\,1, \citealt{Salamaso-2025} and the discussion therein), reveals it to be a long-lived, strongly interacting one, and so here we use the spectra publicly available to us at the time of submission, and compare it with that of SN 2017hcc at similar epochs. 

On the +26\,d, the blue-shifted, trailing edge of the H$\alpha$ line is blended with the telluric feature. No telluric correction could be done, since no telluric standard spectra are available for this observation.  We note that the H$\alpha$ profile is well fitted by single, intermediate, Lorentzian velocity component (FWHM 1586\,km\,s$^{-1}$ centered at $\lambda$6562) with absorption troughs at -517\,km\,s$^{-1}$ and at 710\,km\,s$^{-1}$ in the fit residual (Fig.\,\ref{23}b). Moreover, no narrow ($ \ll 1000$ km $s^{-1}$) feature is present. A small emission feature of He II is present on $\lambda$4685.  At this early stage, most of the intermediate line emission being due to Thomson scattering of photons in the CSM, this implies a CSM morphology which is very dense and close to the progenitor, with little to no CSM-halo. Such a CSM could have been formed close to the explosion epoch, giving it little time to disperse before the terminal explosion, as in the case of SN 2019bxq, which showed an LBV-like eruption $\sim 10 $yr prior to the explosion \citep{Reguitti}.
 

 We note here that \citet{Salamaso-2025} estimated a rather large mass-loss rate for the progenitor of SN 2021acya, at  $0.06$ -- $0.8$ M$_{\odot}$ yr$^{-1}$ over a duration of $34$ to $37$ yr, leading to a CSM of $18$ -- $19$ M$_{\odot}$. This implies a very massive, LBV, progenitor with episodic mass ejections, possibly that of an entire shell.

On +100\,d, the H$\alpha$ profile is fitted by two components -- a blue-shifted Gaussian of FWHM 3840\,km\,s$^{-1}$ centered at $\lambda$6542, and a Lorentzian centered at $\lambda$6560 of FWHM 1013\,km\,s$^{-1}$. The Ca II NIR triplet  $\lambda \lambda \lambda 8498,8542,8662$ also emerges.

\subsection{\texorpdfstring{SN 2021foa ($=$ ASASSN-21 dg)}{SN 2021foa (= ASASSN-21 dg)}}

\begin{figure*}
 \includegraphics[scale=0.38]{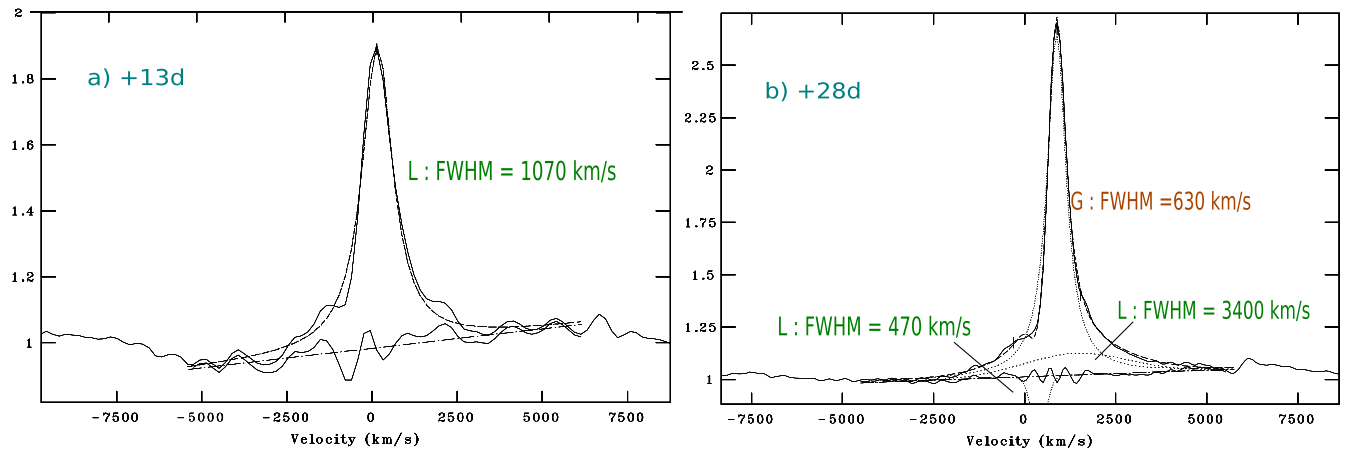}
    \caption{H$\alpha$ profile of SN 2021foa on +13\,d and +28\,d. 
    Continuum subtraction was performed for each spectrum, and they were then converted to $v=0$\,km\,s$^{-1}$ at the rest wavelength of H$\alpha$ (6562.8\,{\AA}). 
    The residuals are presented as the subtracted continuum and the normalized flux is on the y-axis.  The continuum was extracted over selected, line-free regions in the 6000\,{\AA} to 7000\,{\AA} band.}
    \label{2021foa}
    
\end{figure*}

SN 2021foa, associated with IC 0863, at $z=0.008386$ is a “transitional event" in this list, seen as a bridge between type-IIn to Ibn \citep{Gangopadhyay+2025_SN2021foa}.  It has been included here as it showed clear signs of circumstellar interaction at the epochs considered. The early (up to 3\,d before V-maximum) spectra showed characteristic narrow Balmer features from type-IIn, but post V-maximum, the Balmer lines faded away, and strong He I and broad  metal lines emerge, suggestive of a type-Ibn. 
\citet{Reguitti} interpret this as occurring either in a massive Wolf-Rayet progenitor with a substantial, but not complete, loss of its H-envelope, or in a less massive progenitor ($M < 12 M_{\odot}$) with a binary companion.  Here, we take the explosion date at reported by the ASAS-SN Supernova patrol \citep{Reguitti,2021foa} as 09 March 2021(MJD 59282), and compare its +13\,d, +28\,d and +98\,d spectra in Fig.\,\ref{2021foa} with those of SN 2017hcc at similar epochs (see Fig.\,\ref{hcc-hal1}).

\begin{figure*}
 \includegraphics[scale=0.5]{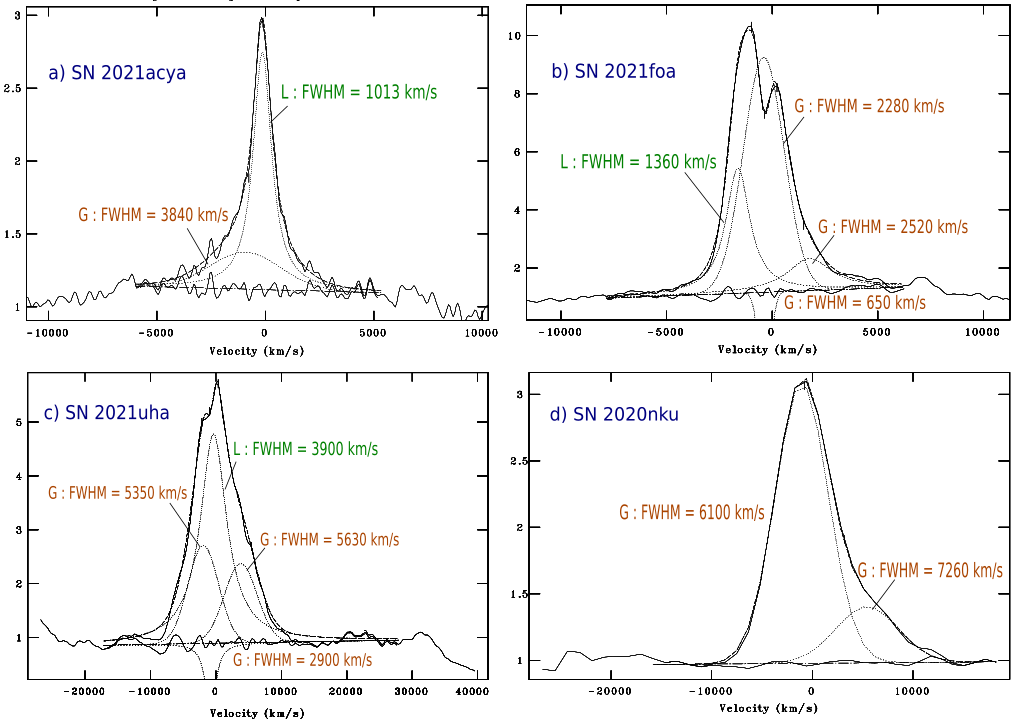}
 \caption{
The spectra containing Gaussian and Lorentzian fits on H$\alpha$ for different supernovae were compared with that of SN 2017hcc at +100\,d. Continuum subtraction was performed for each spectrum, and $v=0 $\,km\,s$^{-1}$ is set at the rest wavelength of H$\alpha$ (6562.8\,{\AA}). The residuals after line subtraction are also shown for each spectrum and the normalized flux is on the y-axis.The continuum was extracted over selected, line-free regions in the 6000\,{\AA} to 7000\,{\AA} band.}
  \label{100}
\end{figure*}

The modeling of the H$\alpha$ profile on 3 epochs (+13\,d, +28\,d and +98\,d) is shown in Fig.\,\ref{2021foa} a,b and \ref{100}.b respectively. On the earliest epoch (+13\,d), it can be fitted by single Lorentzian profile of 1070\,km\,s$^{-1}$ centered on 6566\,{\AA}. The fit residuals show absorption troughs on either side of the peak, at -630\,km\,s$^{-1}$ and +338\,km\,s$^{-1}$. This is in contrast with SN 2017hcc, which required a narrow and an intermediate component even at this early epoch.

By +28\,d, the profile can be resolved into three components, including an absorption feature. The narrow Gaussian component (FWHM 630\,km\,s$^{-1}$ centered at $\lambda$6582) now peaks redward of $\lambda$6562.8, while an absorption component of width 470\,km\,s$^{-1}$ centered at $\lambda$6571 appears. Together, they form a P-Cygni like structure with a red-shifted absorption trough at +470\,km\,s$^{-1}$. Additionally, a intermediate Lorentzian with FWHM 3400\,km\,s$^{-1}$ centered at 6594\,{\AA} also appears. If the narrow component is due to the slow moving CSM alone, the red-shifted absorption is difficult to explain in the context of a homogeneous CSM. Emission from red-shifted clumps in a highly inhomogeneous CSM, coupled with a strongly asymmetric ejecta from the explosion, may explain this profile, but that scenario highly geometry dependent. However, if we assume that the progenitor had a binary companion, with suitable line-of-sight velocity, the +23\,d spectrum may represent the epoch at which the shock hits the outer layers of the companion star. A part of the H$\alpha$ profile is then generated from shocked photosphere of the companion star itself. The emergence of He I $\lambda$5876 at this stage implies that the event had already transitioned to type-Ibn at this epoch -- we note that this He I feature diminishes greatly in the spectrum of SN 2017hcc by then and is completely removed by +100\,d.

The spectrum at +98\,d (Fig.\,\ref{100}b) resolves into 4 components, again including an absorption feature. The highest velocity Lorentzian profile is blue shifted to 6527\,{\AA}, with FWHM 1360\,km\,s$^{-1}$, followed by an absorption feature centered at 6555\,{\AA} of width 650\,km\,s$^{-1}$. The subsequent emission components include a Gaussian of width 2280\,km\,s$^{-1}$ centered at 6555\,{\AA}, and a Gaussian centered at 6600\,{\AA} of width 2520\,km\,s$^{-1}$. The fit residuals show an absorption trough centered at +65\,km\,s$^{-1}$ and an emission peak at +345\,km\,s$^{-1}$. In contrast, no absorption components were seen in SN 2017hcc on +23\,d and +86\,d.  In our observations, the first emergence is on +374\,d and is associated with the formation of the CDS in the ejecta-CSM interaction region, between the forward and reverse shocks (FS and RS respectively). 
 The CDS would  lead to a boxy profile initially, and then a stronger absorption at an intermediate velocity. Alternatively absorption in the outer region of the companion star could have led to multi component absorption and emission features.

\begin{deluxetable}{ccccccccc} 
\tablecaption{Comparison of multiple velocity components with other SN type lln on +23\,d of SN 2017hcc.\label{compare2}}
\tablehead{\colhead{Object} & \colhead{Spec.date(MJD)} & \colhead{ Telescope/}  & \colhead{ Epoch}& \colhead{ $v_1$ (km\,s$^{-1}$)} & \colhead{$v_2$(km\,s$^{-1}$)} & \colhead{$v_3$ (km\,s$^{-1}$)} &  \colhead{$v_4$ (km\,s$^{-1}$)} &  \colhead{r.m.s. of fit} \\
\colhead{(red-shift)}& \colhead{(Date)}& \colhead{Instrument}& \colhead{}& \colhead{(centroid\,{\AA})}& \colhead{(centroid\,{\AA})}& \colhead{(centroid\,{\AA})}& \colhead{(centroid\,{\AA})}& \colhead{}}
\colnumbers
\startdata
SN 2017hcc& 58049.5226& HCT /HFOSC  &+23\,d & 2030$\pm$374 &  400$\pm$30&   -& - &0.016 \\
(z=0.0168)  &(2017-10-23) & & & (6563)&(6560) &&\\
SN 2023usc&60264.3505& HCT /HFOSC  &+37\,d  & - &  440$\pm$34 &   -& -& 0.018\\
(z=0.06) &(2023-11-16 ) & & & &(6567) &&\\
SN 2023hpd& 60095.6442&LT / SPRAT &+29\,d & 1300$\pm$56& 1200$\pm$42&3090$\pm$782&-&0.032\\
(z=0.0198) & (2023-05-31)& & &(6523 )  &(6560 ) & (6582) &\\
SN 2021acya& 59543.7616& ESO-NTT / EFOSC2-NTT &+26\,d&-&1586$\pm$95&-&- &0.032\\
(z=0.06) & (2021-11-25)& & &  &(6562 ) &  &\\ 
SN 2021foa&59310.0000& NOT / ALFOSC &+28\,d &-&630$\pm$82&3400$\pm$778&470$\pm$32 &0.015\\
(z=0.008386) &(2021-04-06) & &      &&( 6582) &(6599) &(6572) \\ 
SN 2020ywx&59200.3398& ESO-NTT / EFOSC2-NTT &+46\,d &- &  1613$\pm$78& 5597$\pm$987& -&0.079\\
(z=0.0217)  &(2020-12-17) & &                  && (6571)& (6580)& \\
SN 2018khh& 58492.5386& ESO-NTT / EFOSC2-NTT &+21\,d &  -&750$\pm$32&680$\pm$87&- &0.008\\
(z=0.023) &    (2019-01-09)      &            &     &    &(6565) & (6587)& \\
SN 2016aiy& 57456.3200& ESO-NTT / EFOSC2-NTT & +22\,d & &732$\pm$85&7384$\pm$937&-&0.012\\
(z=0.01) & (2016-03-09)& &                   &&(6564)&( 6576)  & \\
SN 2015da& 57053.1669& Ekar / AFOSC &  +22\,d&-&1235$\pm$147&-&-&0.052\\
(z=0.0066) & (2015-01-31)& &              &&( 6559) & & \\
SN 2010jl &55507.6500& Keck2 / DEIMOS (UCB-SNDB) &+28\,d & 2235$\pm$340&430$\pm$18 &-&-&0.018\\
(Z=0.0107)& (2010-11-07)& & &(6560) & (6563) & & \\
\hline
\enddata
\tablecomments{A comparison of the deblended line velocities of SN 2017hcc, with other type-IIn events, at 23\,d post-explosion. $v=0$\,km\,s$^{-1}$ corresponds to H$\alpha$ rest wavelength 6562.8\,{\AA}. $v_1$,$v_2$ and $v_3$ are the velocity components into which the line can be resolved with associated uncertainties estimated from profile fitting. $v_2$ is the component whose centroid is closest to 6562.8\,{\AA}.
Additionally, the “transition" event SN 2021foa (Type-Ibn to IIn) shows an additional absorption component $v_4$. All spectra except SN 2023usc and SN 2017hcc are taken from WISeREP \citep{WISEREP}.}
\end{deluxetable}

\subsection{\texorpdfstring{SN 2020cui ($=$ ZTF20aammbvq)}{SN 2020cui (= ZTF20aammbvq)}}

SN 2020cui at $z=0.025$ was classified as type-IIn \citep{cl2020cui}.
For this event, we set  an explosion date on 09 February 2020 (MJD 58888.3562), estimated as the average of the date of last non detection (MJD 58888.2933) and first detection (MJD 58888.4190) by P48 ZTF-Cam \citep{di2020cui}.
We compare here its spectrum on +12\,d with that of SN 2017hcc on +14\,d.  The H$\alpha$ resolves into an intermediate Gaussian with FWHM 3555\,km\,s$^{-1}$ centered at 6560\,{\AA}, and a narrow, blue shifted (relative to $\lambda$6562.8) Lorentzian emission line with FWHM 1056\,km\,s$^{-1}$, centered at 6560\,{\AA}.  The fit residue (see Fig.\,\ref{14}b) shows a sharp absorption component centered at 153\,km\,s$^{-1}$ redward of the rest wavelength. 
The emission hump near $\lambda$4640 (see Fig.\,\ref{SNlln14 d}) can be explained by  the presence of N III triplet at $\lambda$$\lambda$$\lambda$ 4634,4640,4641 and  N II $\lambda$4630. A similar structure is also seen in the spectrum of SN 2017hcc, and, to a lesser extent, in SN 2018khh (Fig.\,\ref{SNlln14 d}). The N/C and N/O ratios are found to be larger than solar ratios in SN 1987A, whose progenitor was shown to undergo a blue to red to blue evolution  \citep{saio}. 
 Extensive convective dredge-up of CNO processed material is known to occur in massive stars in the red-supergiant phase. This enrichment of the stellar atmosphere is later ejected via stellar wind into the CSM. 
We note that the surface N/C ratio in core-collapse is a function of the convective overshoot parameter $f$ and is maximum for lower values of $f$, in the case of relatively low ZAMS mass progenitors (typically $\sim 13 M_{\odot}$; see  Fig.\,5, \citealt{Wagle}).
The other Balmer emission lines in the two spectra are similar in structure. However, He $\lambda5876$ emission can be fitted by a single Gaussian in SN 2020cui (FWHM $\sim 1200$\,km\,s$^{-1}$), while at the same epoch, SN 2017hcc requires 3 Gaussians, including a narrow one with FWHM $\sim 600$\,km\,s$^{-1}$. 

\subsection{\texorpdfstring{SN 2020nku ($=$ Gaia20cxs)}{SN 2020nku (= Gaia20cxs)}}

 SN 2020nku at $z=0.02$, associated with 2MASX J08210450-2230492, was classified as a type-IIn by \citep{cl2020nku}, based on a spectrum taken on 17 November 2020 (MJD 59170.4589). The explosion epoch has a large uncertainty, since the last non-detection was on 29 May 2020 (MJD 58998.8350) and the first detection was on 23 June 2020 (MJD 59023.3630) (Gaia g-band detection; \cite{di2020nku}).
 The ZTF g- and r-band photometric light curves start from 14 November 2020 (MJD 59167) -- well past the maximum, so we use here 11 June 2020(MJD 59011.59) as the average explosion date, and compare the +159\,d spectrum of SN 2020nku (see in Fig.\,\ref{SNlln100 d}) with the +100\,d spectrum of SN 2017hcc. Even at this later epoch, the H$\alpha$ profile is well fitted by two broad Gaussians: one with FWHM 6100\,km\,s$^{-1}$ centered at 6540\,{\AA} and  the other, red shifted relative to $\lambda 6562.8$ with FWHM of 7260\,km\,s$^{-1}$ centered at 6681\,{\AA} (Fig.\,\ref{100}d). This is in contrast to SN 2017hcc, which required an additional narrow component as well. The absence of a narrow component can suggest an unusually fast moving CSM, asymmetrically distributed, or an asymmetric explosion -- or both. 
The He $\lambda$7065.196 line is prominent at this stage, and the Ca II triplet $\lambda$$\lambda$$\lambda$8498.020, 8542.090, 8662.140 as well as the [Ca II]  $\lambda$$\lambda$7291.47, 7323.89 are clearly distinguishable, further confirming that this event has already reached the nebular stage.

\subsection{\texorpdfstring{SN 2020ywx ($=$ ATLAS20bewg)}{SN 2020ywx (= ATLAS20bewg)}}

SN 2020ywx ($=$ AT 2020ywx), located in the spiral arms of LEDA 213897 at $z=0.0217$ was first detected on 04 November 2020 (MJD 59157.6200) on ATLAS-MLO, o-band \citep{di2020ywx}. The last non-detection being on  21 July 2020 (MJD 59051.2700), we set the median explosion date as 12 September 2020 (MJD 59104.4500).  The event is x-ray and radio bright at late times, and a possible binary interaction has been suggested for it \citep{Baer-Way_2025}.
 The earliest available spectrum is +46\,d after the date of detection, i.e, +96\,d after the assumed date of explosion. We show the resolved components of the H$\alpha$ line in Fig.\,\ref{23}c and mention the details in Table\,\ref{compare2}. We note that at this $z$, the telluric feature at $\lambda 6730$ falls on the H$\alpha$ profile, making it difficult to fit multiple broad velocity components to it. Nevertheless, the narrow component is well-fitted by Gaussian of FWHM 1613\,km\,s$^{-1}$, and the broad component with FWHM 5597\,km\,s$^{-1}$ (see Fig.\,\ref{23}c).  The two peaks seen in the residuals at 6537\,{\AA} and 6590\,{\AA} are very close to the [NII] transition at $\lambda$$\lambda$ 6548 and 6583.  The spectrum shows a broad bump in the 8100 to 9000\,{\AA} region. The Ca II triplet $\lambda$$\lambda$$\lambda$8498.02,8542.09,8662.14 lies in this region, as do several Fe II transitions.  However, the Ca II triplet is typically associated with emission from the inner layers of the progenitors, which is very unlikely at this epoch, and hence these are likely host galaxy emission, or excited in the outer layers of a possible stripped binary companion. Alternatively, several higher order H-Pa transitions also lie in this region, and it is possible that this early epoch spectrum shows a strong NIR excess, greater than that seen in SN 2017hcc \citep{Baer-Way_2025}. 

\subsection{\texorpdfstring{SN 2018hfg ($=$ ASASSN-18xk)}{SN 2018hfg (= ASASSN-18xk)}}

SN 2018hfg in UGC4642 at $z=0.023863$ was first detected on (MJD 58400.4900) 09 October 2018 (ASASSN-Leavitt, detected in g band;\citealt{di2018hfg}) and classified as type-IIn by \citet{cl2018hfg}. Based on the last non-detection date 08 October 2018 (MJD 58395.4800), we set 06 October 2018 (MJD 58397.9850) as the average explosion epoch.

The only publicly available spectrum, taken on MJD 58412.84 (+5\,d) is compared with the +14\,d spectrum of SN 2017hcc. Like SN 2017hcc, it requires only two components: a narrow Lorentzian with FWHM 700\,km\,s$^{-1}$ centered at 6557\,{\AA} from the CSM and an intermediate  Gaussian of FWHM 2643\,km\,s$^{-1}$ centered at  6563\,{\AA}. The similarity to SN 2017hcc suggests that the progenitor of SN 2018hfg is also enveloped in a double shelled CSM, with two distinct density profiles Fig.\,\ref{14}c.
 
The other Balmer features  (H$\beta$, H$\gamma$, H$\delta$ and  H$\epsilon$) and He are as expected from a “canonical" type-IIn early in its evolution.

\subsection{\texorpdfstring{SN 2018khh ($=$ ASASSN-18abz)}{SN 2018khh (= ASASSN-18abz)}}

SN 2018khh ($=$ ASASSN-18abz), classified as a type-IIn \citep{cl2018khh}, was located in the elliptical galaxy 2MASX J22031497-5558516 at $z=0.023$.
Using the date of first detection as 20 December 2018
(MJD 58472.0400) and last non-detection (MJD 58469.0500)  as 17 December 2018
\citep{di2018khh}, we set the average explosion date as 18 December 2018 (MJD 58470.845), consistent with the value of MJD 58470.5, MJD set by \citet{Nagao-2025}.
The publicly available spectra were taken at +3\,d, +12\,d and +21\,d post explosion.  The H$\alpha$ profile of the +3\,d spectrum could be resolved into a narrow (FWHM$=384$\,km\,s$^{-1}$) and an intermediate width (FWHM $= 2591$\,km\,s$^{-1}$) Gaussian, both centered at 6563\,{\AA}. 
 This very early time spectrum also shows a strong, narrow (800\,km\,s$^{-1}$) He II $\lambda$4685.7 line, comparable at peak intensity to the H$\alpha$.  Additionally,  weak C IV emission at $\lambda$$\lambda$5801, 5811,
 and $\lambda$4658, possibly blended with N III $\lambda$$\lambda$4634, 4640 is also seen. 
High ionization lines of C and N were also seen in the early spectrum of the type-IIn SN 2008fq \citep{Taddia-2013}, who attributed it to emission from the CSM around a Wolf-Rayet (WR) star.  We note however, that WR stars are typically assumed to be progenitors of SNe Ibn, rather than canonical SNe IIn, because of their lower amount of H.  Additionally 2008fq also showed later broad absorption lines seen in type-IIL \citep{Chugai2008fq}.  
 
\begin{figure*}
 \includegraphics[scale=0.60, angle=180]{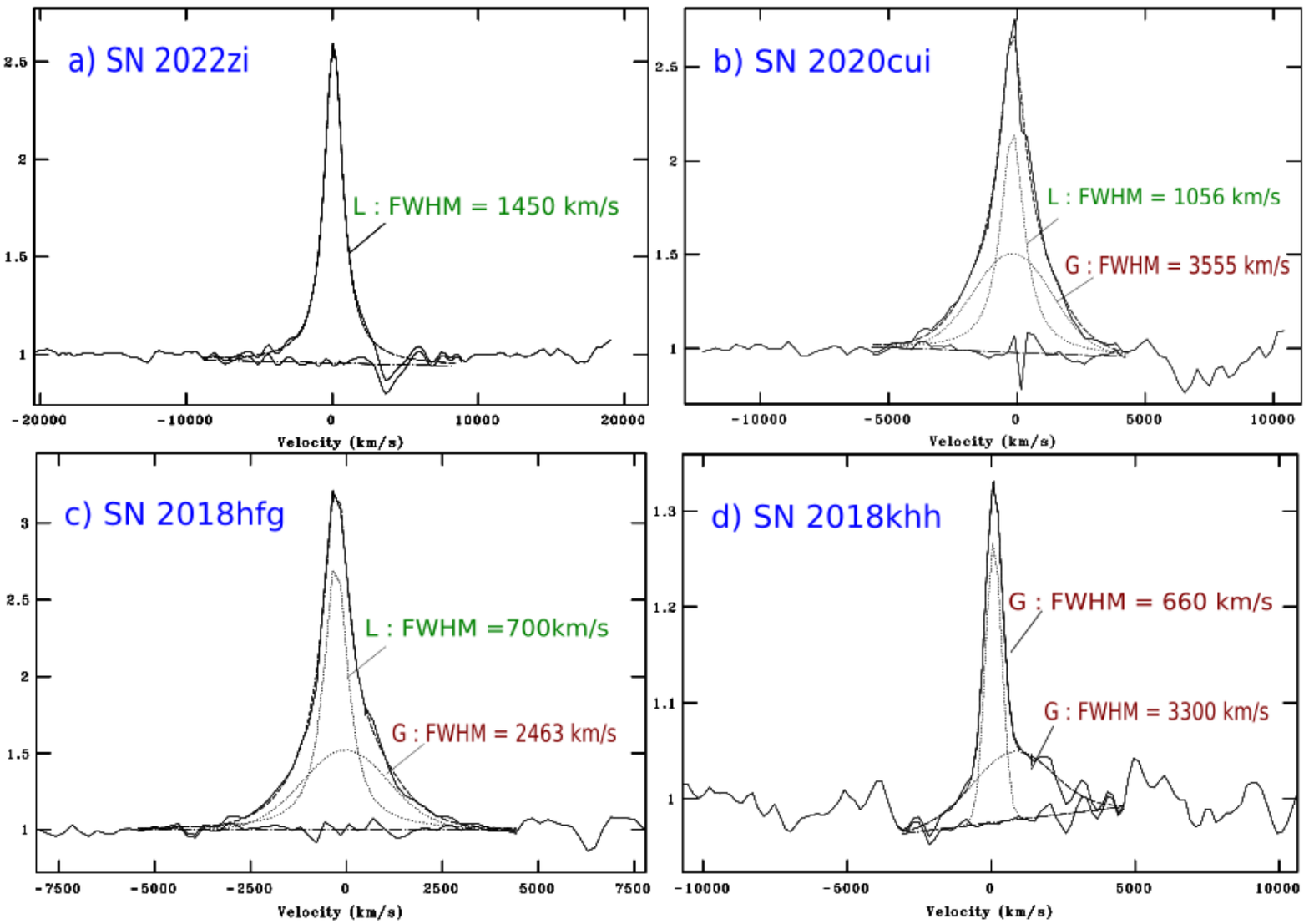}
 \caption{ Gaussian and Lorentzian fits on H$\alpha$ for early time ($\sim + 14)$d type-IIn supernovae spectra, used for comparison with +14\,d spectrum of SN 2017hcc. Continuum subtraction was performed for each spectrum, and they were then converted to $v=0$\,km\,s$^{-1}$ at the rest wavelength of H$\alpha$ (6562.8\,{\AA}). The fit residuals (observed $-$ model) are also shown and the normalized flux is on the y-axis. The continuum was extracted over selected, line-free regions in the 6000\,{\AA} to 7000\,{\AA} band.}
  \label{14}
\end{figure*}

By +12\,d, the spectrum of SN 2018khh continues to be a superposition of a narrow (FWHM 660\,km\,s$^{-1}$, centered at 6563\,{\AA}) in  Fig.\,\ref{14}d, and an intermediate (FWHM 3300\,km\,s$^{-1}$, centered on 6579\,{\AA}) Gaussian. The two component structure is similar to that of SN 2017hcc, although there the narrow component was best fitted by a Lorentzian. 
The intermediate component, emitted at higher optical depths within the CSM, shows a blue-shift expected from a CSM expanding into the line-of-sight.

The narrow component, centered almost at the rest-wavelength (6562.8\,{\AA}), are the unscattered photons emitted directly into the line-of-sight from the 
most opaque CSM. In this respect, the event behaves similarly to SN 2017hcc in a similar epoch. However, we note that by +12\,d, the He II and C IV lines are absent in 2018khh, unlike in the case of 2017hcc. If the C IV line is due to flash ionized, pre-existing dust, then by this epoch, it may have been completely sublimated.

\begin{figure*}
 \includegraphics[scale=0.60,angle=180]{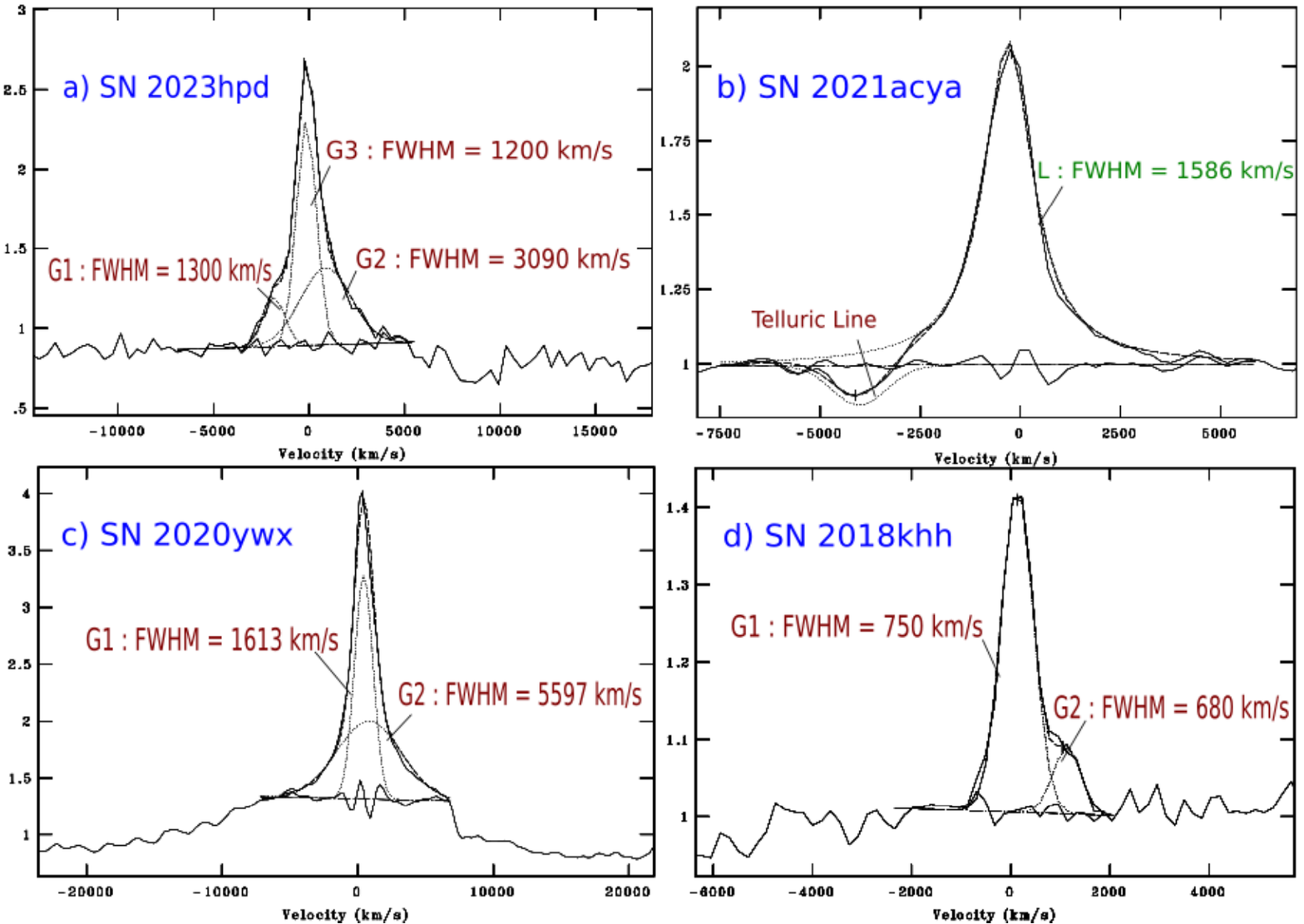}
 \caption{
   The spectra containing Gaussian and Lorentzian fits on H$\alpha$ for different supernovae were compared with the spectrum at +23\,d of SN 2017hcc. The spectra have been normalized to the continuum, and $v=0$\,km\,s$^{-1}$ at the rest wavelength of H$\alpha$ (6562.8\,{\AA}).The normalized flux is on the y-axis, and the fit residuals are also shown. The continuum was extracted over selected, line-free regions in the 6000\,{\AA} to 7000\,{\AA} band.}

  \label{23}
\end{figure*}

Comparing +21\,d of SN 2018khh  with that taken +23\,d for SN 2017hcc (see Fig.\,\ref{23}d), we find that  H$\alpha$ retains its two component Gaussian structure, but the intermediate component is now replaced by a newly emerged narrow component. The narrow feature seen at +12\,d is now slightly wider (750\,km\,s$^{-1}$, centered at 6565\,{\AA}) and also slightly red shifted. If this is exclusive to the CSM, it suggests a CSM that is very asymmetric in its distribution. The second component  (680\,km\,s$^{-1}$ centered at 6587\,{\AA}) has now drifted further redward of the broad component seen on +12\,d. 
This could be the result of a strong blue-shifted absorption of the intermediate  component, due to emission from a dense CSM at high optical depths.  We note that the 3600 ${\rm to}$  4700\,{\AA} region is now dominated by lines of Fe II, Fe III, 
in addition to N II, (similar to SN 2017hcc at +100\,d), while He II $\lambda$4685.710 feature is now blended in, and difficult to isolate at this spectral resolution in Fig.\,\ref{SNlln23 d}.

Thus, over the 3 epochs -- +3\,d, +12\,d and +21\,d -- the narrow component associated with direct emission from a low density CSM gradually widens, suggesting that the shocked CSM is rapidly becoming optically thin. At the same time, the intermediate component, associated at this stage with Thomson scattered photons from the denser regions of the CSM becomes narrower, suggesting these photons are also emitted at a greater depth. This behavior is consistent with that expected for a double layered CSM
with a sharp density gradient as described \citep{chugai}.

 \citet{Nagao-2025} also present a study of SN 2018khh, using spectra taken over 15 epochs, from +2.5\,d to +329\,d. They find that narrow lines weaken, broad emission lines emerge, and the continuum becomes redder at $\sim +100$\,d, as in the case of the long-lived type-II SN2021irp, with a disk-like CSM. Additionally, these features are also similar to those seen in some type-IIL supernovae.

\begin{deluxetable}{ccccccccc} 
\tablecaption{Comparison of multiple velocity components  with other SN type lln on +100\,d of SN 2017hcc.\label{compare3}}
\tablehead{\colhead{Object} & \colhead{Spec.date(MJD)} & \colhead{ Telescope/}  & \colhead{ Epoch}& \colhead{ $v_1$ (km\,s$^{-1}$)} & \colhead{$v_2$(km\,s$^{-1}$)} & \colhead{$v_3$ (km\,s$^{-1}$)} &  \colhead{$v_4$ (km\,s$^{-1}$)} &  \colhead{r.m.s. of fit} \\
\colhead{(red-shift)}& \colhead{(UTC)}& \colhead{Instrument}& \colhead{}& \colhead{(centroid\,{\AA})}& \colhead{(centroid\,{\AA})}& \colhead{(centroid\,{\AA})}& \colhead{(centroid\,{\AA})}& \colhead{}}
\colnumbers
\startdata
SN 2017hcc & 58112.6640 &  HCT /HFOSC&+86\,d & -&480$\pm$28&3900$\pm$100&-&0.018\\
(z=0.0168)   & (2017-12-25)& &                                & &( 6562) & (6566)& &\\ 
SN 2017hcc & 58126.6207  &  HCT /HFOSC&+100\,d &4850$\pm$1138& 410$\pm$31& 1600$\pm$155& &0.015\\
  (z=0.0168)& (2018-01-08)& &                                & (6555)&( 6562) & (6566)& &\\ 
SN 2023usc & 60289.2745&  HCT /HFOSC&+62\,d& 8030$\pm$346&470$\pm$28&-&-& 0.032\\
(z=0.06)  & (2023-12-11)& &                                & ( 6550) & (6569)& &&\\ 
SN 2023usc & 60320.1519  &  HCT /HFOSC&+93\,d &10900$\pm$1214& 440$\pm$25& -&-&0.041\\
(z=0.06)  &(2024-01-11) & &                                & (6528)&( 6570) & & &\\ 
SN 2021acya & 59617.4822&LCO2m / Spectral (Asiago)& +100\,d &3840$\pm$424& 1013$\pm$21&- &- &0.041\\
(z=0.06203)   & (2022-02-07)& &                       &(6542)&( 6560) & & &\\
SN 2021foa & 59380.5000& GTC / OSIRIS &+98\,d & 1360$\pm$48 &  2280$\pm$478&   2520$\pm$156&   650 $\pm$62 &0.143 \\(z=0.008386) &(2021-06-15) & &                             &(6535)&( 6566) & (6597)& (6553)&\\ 
SN 2021uha &59517.3403 &ESO-NTT / EFOSC2-NTT& +98\,d &5350$\pm$654& 3900$\pm$821& 5630$\pm$451&2900$\pm$132 &0.059\\
(z=0.0079)   &(2021-10-30) & &                       &(6518)&( 6572) & (6655)& (6570)&\\
SN 2020nku & 59170.4589& P60 / SEDM (ZTF) &+159\,d &6100$\pm$1023 &  -& 7260$\pm$964&  - &0.021\\
(z=0.02)   &(2020-11-17) & &        &( 6540)& & (6681)& &\\
SN 2015da & 57141.1179&  NOT / ALFOSC & +100\,d& 3340$\pm$452& 1400$\pm$70 & 1875$\pm$85&-&0.024\\
(z=0.02) &(2015-04-29) & &                                &(6524)&( 6558) & (6591)& &\\ 
SN 2010jl &	55571.8201 &  BAO-2.16m / Cassegrainz&+92\,d & 3484$\pm$190&652$\pm$32 &-&-&0.02\\
(Z=0.0107)&(2011-01-10) & & &(6556) & (6563) & & \\
\hline
\enddata
\tablecomments{A comparison of the deblended line velocities of SN 2017hcc, with other type-IIn events, at 86 and 100\,d post-explosion. Here $v=0$\,km\,s$^{-1}$ corresponds to H$\alpha$ rest wavelength 6562.8\,{\AA}. The profiles are resolved into atmost 3 emission components -- $v_1$, $v_2$, $v_3$ -- and an absorption component $v_4$, which is seen only in “transition" event SN 2021foa (Type-Ibn to IIn) and SN 2021uha (Type-IIn). $v_2$ is the component whose centroid is closest to 6562.8\,{\AA}. Uncertainties are estimated from profile fitting}
\end{deluxetable}

\begin{figure*}
    \centering
    \begin{minipage}{0.45\textwidth}
        \centering
       \includegraphics[scale=0.48, trim=2cm 0cm 2cm 2.5cm]{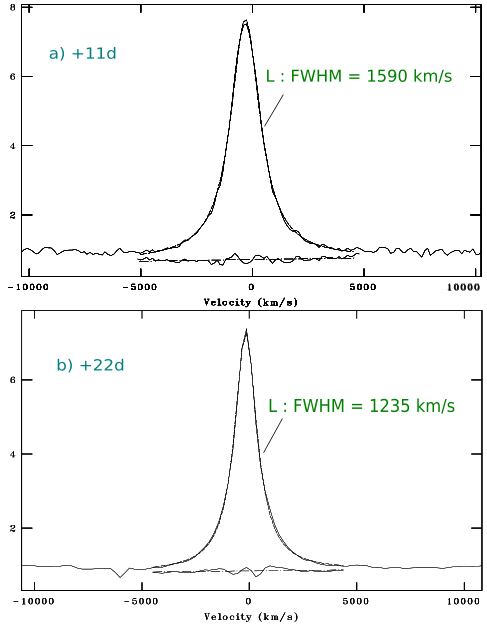}
    \includegraphics[scale=0.48]{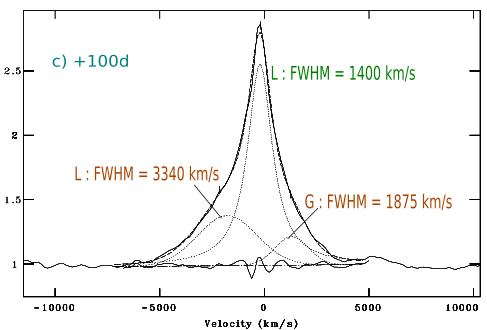}
        \caption{H$\alpha$ profile of SN 2015da on earlier epochs. Continuum subtraction was performed for each spectrum, and $v=0$\,km\,s$^{-1}$ was set at the rest wavelength of H$\alpha$ (6562.8\,{\AA}). The residuals are presented as the subtracted continuum and the normalized flux is on the y-axis. The continuum was extracted over selected, line-free regions in the 6000\,{\AA} to 7000\,{\AA} band.}
    \label{2015 da}
    \end{minipage}
    \hfill
    \begin{minipage}{0.45\textwidth}
        \centering
    \includegraphics[scale=0.4, trim=2cm 0cm 2cm 0cm]{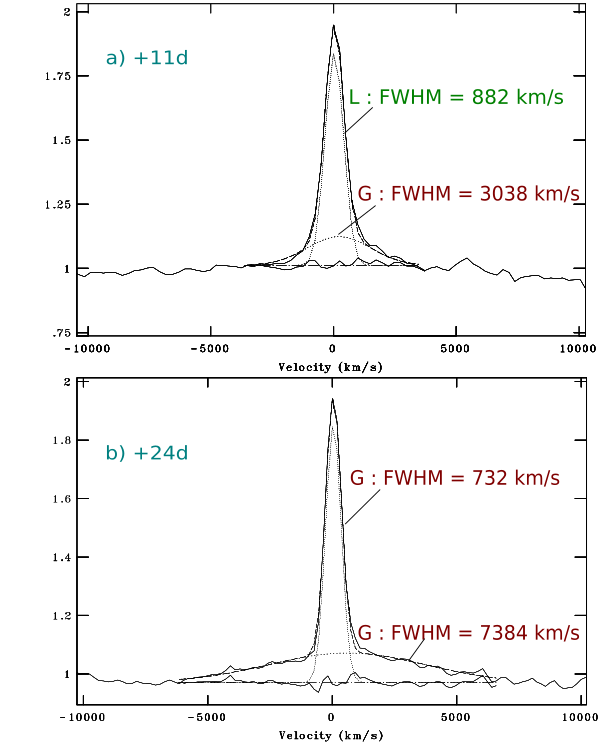}
    \includegraphics[scale=0.4, trim=2cm 0cm 2cm 0cm]{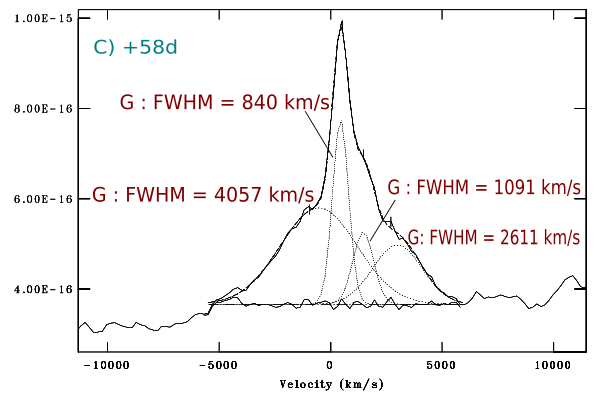}
        \caption{H$\alpha$ profile of SN 2016aiy on +11, +24 and +58\,d. $v=0$\,km\,s$^{-1}$ was set at the rest wavelength of H$\alpha$ (6562.8\,{\AA}). For the spectra on +11 and +24\,d, a local continuum was extracted over a selected, line-free regions in the 6000\,{\AA} to 7000\,{\AA} band and the continuum subtracted, normalized spectra used to model the line profile. On +58\,d, it is difficult to define a continuum because of strong absorption features in 6200-6400\,{\AA} band.  The fit range (dashed line) and superposed fit residuals are also shown. The normalized flux is on the y-axis.}
    \label{2016aiy}
    \end{minipage}
\end{figure*}

\subsection{\texorpdfstring{SN 2016aiy ($=$ ASASSN-16bw)}{SN 2016aiy (= ASASSN-16bw)}}

SN 2016aiy exploded in the host galaxy ESO 323-G084  at $z=0.01$.
The date of the last non-detection was on 11 February 2016 (MJD 57429.3600) while that of the first detection (ASASSN-Cassius, V-band) was on (MJD 57435.3700) 17 February 2016 \citep{di2016aiy}. No light curves being publicly available, we take the average of these two epochs as the explosion date 14 February 2016 (MJD 57432.3650) $\pm$ 3.

The +11\,d spectrum of SN 2016aiy is compared with that of SN 2017hcc at +14\,d. The H$\alpha$ profile is well fitted by a narrow Lorentzian of FWHM 882\,km\,s$^{-1}$ centered at 6563\,{\AA} and an intermediate Gaussian with FWHM 3038\,km\,s$^{-1}$ centered at 6567\,{\AA} (Fig.\,\ref{2016aiy}a). We note that unlike SN 2017hcc, no blue shift is seen in the peak of the narrow component in this early spectrum.

The He I  $\lambda5875.624$ emission line shows a characteristic similar to a P-Cygni with an absorption trough at $\lambda5821$, which is absent in SN 2017hcc. The H$\beta$ shows a possible “Cachito" \citep{cachito} at $\lambda$4823\,{\AA}, although it could also be an absorption feature due to the Fe-group lines. If real, the “Cachito" -- an absorption notch blueward of the rest wavelength -- would be typically associated with high velocity features and seen more commonly in type-II supernovae.

Comparing +24\,d spectrum of SN 2016aiy with the +23\,d spectrum of SN 2017hcc (Fig.\,\ref{2016aiy}b), we find that the H$\alpha$ of SN 2016aiy can be resolved into two Gaussian components -- a narrow one with 732\,km\,s$^{-1}$ centered at 6564\,{\AA}, and a broad one with FWHM 7384\,km\,s$^{-1}$ centered at 6576\,{\AA}. However, the fit residual shows a clear blue shifted P-Cygni profile with an absorption trough at -176\,km\,s$^{-1}$.  The absorption feature in the He I and H$\beta$ emission line, discussed above was still present but was not detected in SN 2017hcc.

No spectra of SN 2016aiy at around +100\,d post-explosion is publicly available. However, examining the +58\,d spectrum, we find that the H$\alpha$ profile of SN 2016aiy is now fitted by four Gaussian components while three were enough for SN 2017hcc. The narrow component with FWHM 840\,km\,s$^{-1}$ is red-shifted to $\lambda$6572. The intermediate components include: 
(a) a blue-shifted one with FWHM 4057\,km\,s$^{-1}$ centered at 6551\,{\AA}, (b) a red shifted one with FWHM 2611\,km\,s$^{-1}$, centered at $\lambda6628$, and (c) one further red-shifted with FWHM 1091 km/s centered at $\lambda6594$. The large, blue-shifted absorption trough seen in the 6140\,{\AA} to 6450\,{\AA} range, enhanced by Si II and O I / [O I] absorption features (Fig.\,\ref{2016aiy}c), may be responsible for the perceived red-shift of all components in this profile fit.

\subsection{\texorpdfstring{SN 2015da ($=$ PSN J13522411+3941286)}{SN 2015 da (= PSN J13522411+3941286)}}

SN 2015da, in host galaxy NGC 5337 \citep{di2015da}, is a well studied type-IIn  with a long interaction time scale, and an extended, massive CSM, similar to SN 2017hcc \citep{daTartaglia}. 
{\citet{Smith-2024}} set a median explosion date of  2015-01-08 (MJD 57030.4450)  for this event, which we use here.
Like SN 2017hcc, whose $V-K_s$ color continued to redden well up to 500\,d post-explosion (see Fig.\,4, \citealt{moran}) SN 2015da also shows NIR excess, albeit on a shorter timescale (see Table.\,2, \citealt{daTartaglia}). IR excess in type-IIn is most likely due to either pre-existing dust (clearly inferred in the case of SN 2017hcc) or due to newly formed dust at a later stage, as in the case of SN 2015da, which showed a re-brightening in the K-band from +400\,d onward, lasting for $\sim 240$\,d. Dust formation, or pre-existing dust, can cause a preferential scattering of the shorter wavelengths, thus enhancing the blue-shifted absorption seen in the line profiles. 

The H$\alpha$ profile of the +11\,d spectrum of SN 2015da is fitted well by a single Lorentzian with  FWHM of 1590\,km\,s$^{-1}$ centered at 6556\,{\AA}. In contrast, SN 2017hcc required a narrow as well as intermediate Gaussian component Fig.\,\ref{2015 da}a. He I is present in both spectra (although more prominent in 2017hcc), while He II is not discernible in either spectrum. Additionally, narrow [N II] $\lambda$ 5755 emission, possibly emitted from the overlying ISM or a CSM, lost very early in the progenitor's evolution, is also seen in SN 2015da, but absent SN 2017hcc. Finally, a broad O I $\lambda8446.3$ line is present in SN 2015da but absent in SN 2017hcc.

Next, we compare the +22\,d spectrum of SN 2015da with SN 2017hcc at +23\,d. The H$\alpha$ emission line was fitted by a single Lorentzian of velocity 1235\,km\,s$^{-1}$ centered at 6559\,{\AA} (Fig.\,\ref{2015 da}b), in contrast to SN 2017hcc, which again required two components.

On comparing +100\,d spectra of SN 2015da and SN 2017hcc, we note the emergence of two additional, intermediate velocity, Gaussian components -- a blue shifted one centered at 6524\,{\AA} with FWHM of 3340\,km\,s$^{-1}$, and a  red shifted one centered at 6591\,{\AA} with FWHM of 1875\,km\,s$^{-1}$ -- in addition to the component centered at 6558\,{\AA} having 1400\,km\,s$^{-1}$. The fit residual  (Fig.\,\ref{2015 da}c) shows a distinct, blue shifted P-Cygni of -400\,km\,s$^{-1}$. He I is present in both SN 2017hcc and SN 2015da and many Fe absorption lines begin to appear in the  4500\,{\AA} to 5500\,{\AA} band. Our H$\alpha$ line deconstruction is consistent with that of \citet{daTartaglia}, who also find that an additional Gaussian component is needed to fit the profile at $> +79 $\,d, which again vanishes by +138\,d, when a single Lorentzian component is sufficient to model it.

\subsection{\texorpdfstring{SN 2010jl ($=$ PTF 10aaxf)}{SN 2010jl (= PTF 10aaxf)}} 

 SN 2010jl, which exploded in the host galaxy UGC 5189A, is classified as a SLSN-type-IIn (like SN 2017hcc), with a progenitor with a dense CSM, and showed evidence of high-velocity ejecta in the NIR spectra. \citet{Stoll-2011} (in their Table\,1) reported the first photometry of this event on HJD 2455479.14. It has been well-studied by several groups in the optical \citep{2010fransson,2010jencson,2010jlzhang} and in the IR \citep{2010borish} bands. Additionally, late time optical/IR observations revealing dust formation have been presented by \citet{2010bevan}.

\begin{figure}
    \centering
    \includegraphics[scale=0.45]{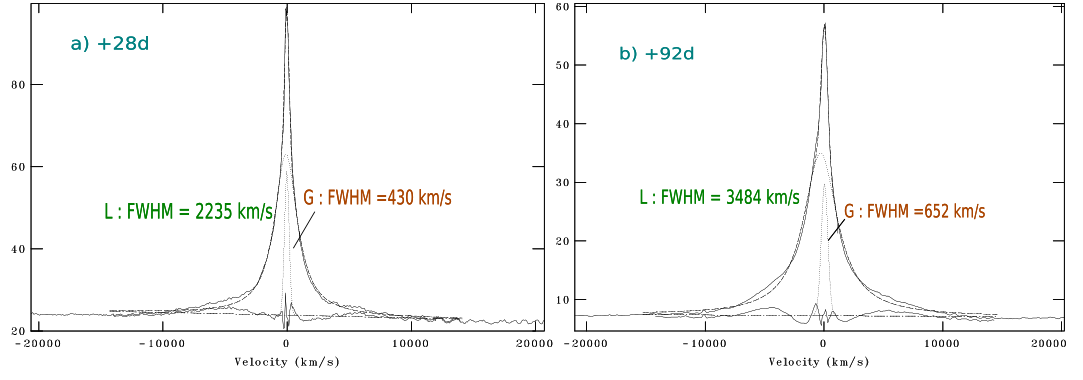}
    \caption{H$\alpha$ profile of SN 2010jl at early epochs. $v=0$\,km\,s$^{-1}$ is set 6562.8\,{\AA} for both epochs and the normalized flux is on the y-axis. The continuum was extracted over selected, line-free regions in the 6000\,{\AA} to 7000\,{\AA} band.}
    \label{2010jl}
\end{figure}

We note that optical spectra of 2010jl and 2017hcc at similar epochs in our study show similar structures.
Comparing the +28\,d spectrum of SN 2010jl \citep{2010jlsmith} with +23\,d of SN 2017hcc, we find that the H$\alpha$ profile is well fitted by a narrow Gaussian (most likely from the CSM), with FWHM 430 km$s^{-1}$, centered at 6563\,{\AA}, and a blue shifted intermediate Lorentzian component with FWHM of 2235\,km\,s$^{-1}$ centered at 6560\,{\AA}(Fig.\,\ref{2010jl}a). We note that the center of the H$\alpha$ is very nearly at the rest wavelength (accounting for instrumental resolution), while that of SN 2017hcc was blue-shifted by 3\,{\AA}, suggesting some red-shifted absorption even at this stage. The fit residual for SN 2010jl shows sharp absorption features at -284\,km\,s$^{-1}$ and 107\,km\,s$^{-1}$, unlike SN 2017hcc, wherein these features are less pronounced. Strong absorption in the early spectrum suggests that the CSM in SN 2010jl would have been significantly optically thicker, and hence more dense, than in SN 2017hcc.

Next, we compared the +92\,d 
epoch of SN 2010jl with +86\,d of SN 2017hcc. The H$\alpha$ resolves into a narrow Gaussian component, associated with the CSM, centered at 6563\,{\AA} with FWHM 652\,km\,s$^{-1}$ and a blue shifted intermediate Lorentzian with FWHM of 3484\,km\,s$^{-1}$ centered at 6556\,{\AA}. 
The fit residual has absorption components at -245\,km\,s$^{-1}$ and 313\,km\,s$^{-1}$, which sit on the narrow component. Additionally, there are also absorption features in the fit residuals at -1586\,km\,s$^{-1}$ and at 1788\,km\,s$^{-1}$, suggesting there is blue- and red-shifted absorption in both the narrow and intermediate velocity components, in the case of SN 2010jl, unlike in the case of SN 2017hcc. 
Examination of the +96\,d spectrum of SN 2010jl (nearest archival spectrum to +100\,d of SN 2017hcc) shows that H$\alpha$ is fitted by a narrow Lorentzian profile of FWHM 320\,km\,s$^{-1}$centered at 6563\,{\AA}, along with an intermediate blue shifted Gaussian component 3146\,km\,s$^{-1}$ at 6555\,{\AA}.
In contrast, SN 2017hcc required 3 components, including an intermediate and a broad component. If the intermediate component is indeed from the shock-CSM interaction region, then this stage is reached much later in the case of SN 2010jl. This may suggest that the CSM of SN 2010jl was located further from the progenitor (and hence formed earlier), than in the case of SN 2017hcc.

\newpage

\section{Discussion}\label{Discussion}

The ``classical'' model  of type-IIn supernova consists of a progenitor (M $\sim 25$ M$_{\odot}$), still with an extensive H-envelope, surrounded by a cooler, dense, slow-moving CSM, created during its pre-supernova evolution. The mass loss of the progenitor is expected to be very large -- ($10^{-4}$ to $10^{-2}$ M$_{\odot}$ yr$^{-1}$) has been proposed, based on a study of 5 type-IIn supernovae, by \citet{Taddia-2013}. The morphology and composition of the CSM is expected to depend on multiple factors, including the progenitor's evolutionary state when it was formed, and its binarity. In general, a shell of continuous CSM can be formed by a steady mass-loss wind from a progenitor during its post-AGB phase. However, evidence also exists for clumpy CSM clouds, formed via LBV-like episodic mass ejection in a few decades to several centuries before the terminal explosion. \citet{Ofek-2014-precursor} examined PTF observations of the fields of 16 type-IIn SNe, and identified precursor activity among 5 of them, with $ > 50 \%$ of type-IIn progenitors showing at least one eruption brighter than $10^7 L_{\odot}$ within $\sim 4$ months prior to the supernova explosion. Additionally, \citet{Strotjohann-2021-precursor} analyzed ZTF forced photometry light curves of 196 interacting SNe (including  131 type-IIn SNe,  12 Type Ibn SNe, and 7 SNe Ia-CSM), and detected precursor eruptions in the progenitors of 18 type-IIn and 1 type Ibn supernovae, although they do not find any clear correlation between precursor activity and luminosity of the final supernova explosion.
\citet{Bonanos-2024} proposed candidates for episodic mass-loss based on a spectroscopic study of 185 massive stars in ten galaxies using the Very Large Telescope (VLT). More importantly, \citet{Reguitti-2024} detect 7 such pre-cursor eruptions in  archival, pre-explosion images of 27 type-IIn events, with earliest precursor eruption occurring 10 yr before the explosion of SN 2019bxq. Most interestingly, pre-cursor activity is detected in SN 2016aiy (1070 to 249\,d pre-explosion) and in the transition event IIn to Ibn 2021foa, in the 20\,d immediately prior to explosion.  Likewise, \citet{Hiramatsu-2024-SN2021qqp} find a 100\,d long precursor activity in the light curve of the type IIn SN 2021qqp, prior to the terminal supernova explosion.  However, no such precursor activity in the case of the long-lived  SN 2010jl  or SN 2015da  \citep{Reguitti-2024} -- both of which may be thought of as analogs to SN 2017hcc.

\begin{figure}
    \centering
    \includegraphics[scale=.3]{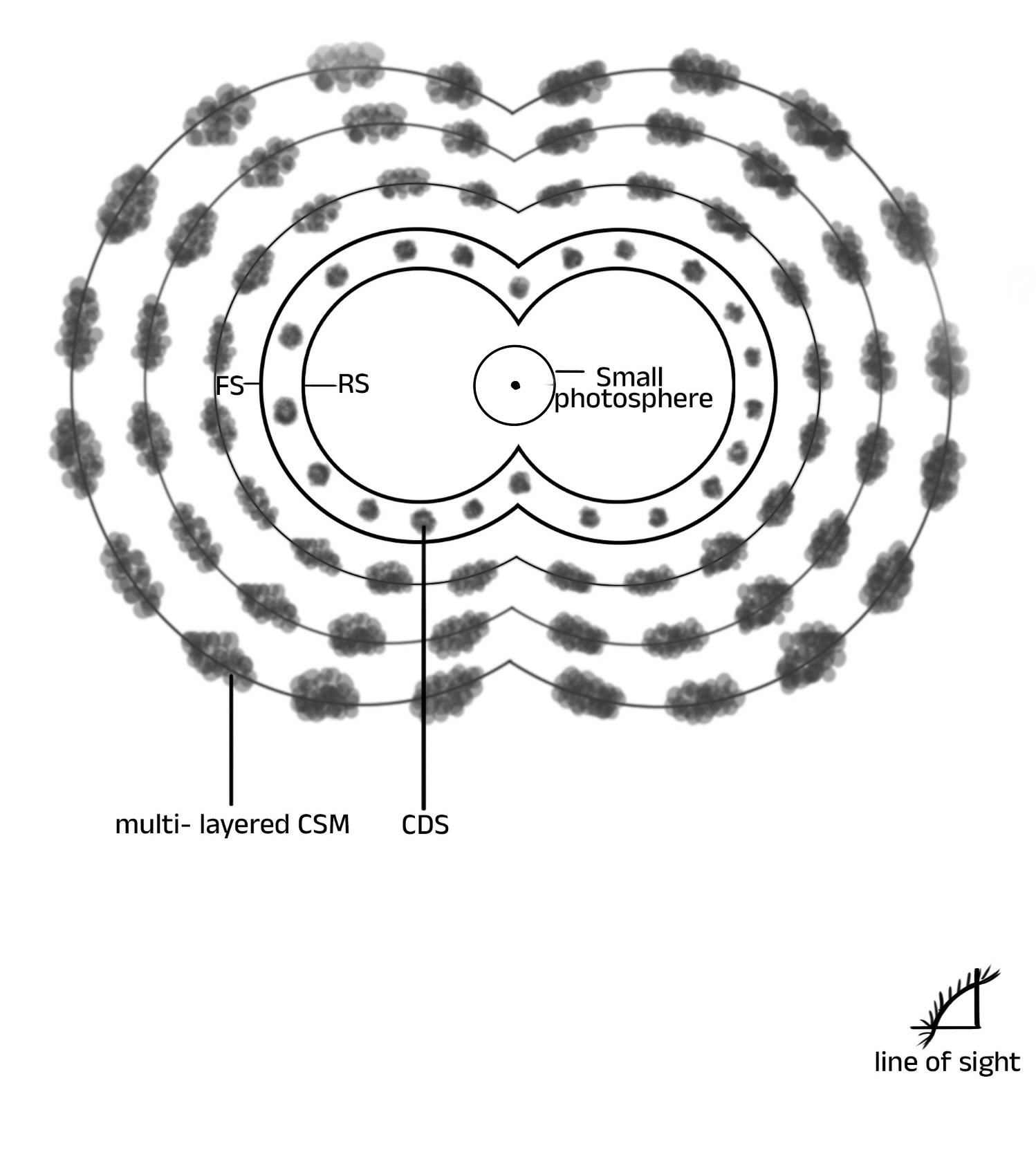}
    \caption{A schematic of early stages of a supernova explosion in a progenitor with a clumpy CDS, formed at different evolutionary epochs, resulting in multiple shells. Also shown is a cool dense shell (CDS) generated between the forward reverse shocks in this morphology. In this schematic, the photosphere of the expanding ejecta is still enclosed well within the CSM shell.}
    \label{fig:csmbi}
\end{figure}

\begin{figure*}
    \centering
    \begin{minipage}{0.45\textwidth}
        \centering
      \includegraphics[scale=0.25]{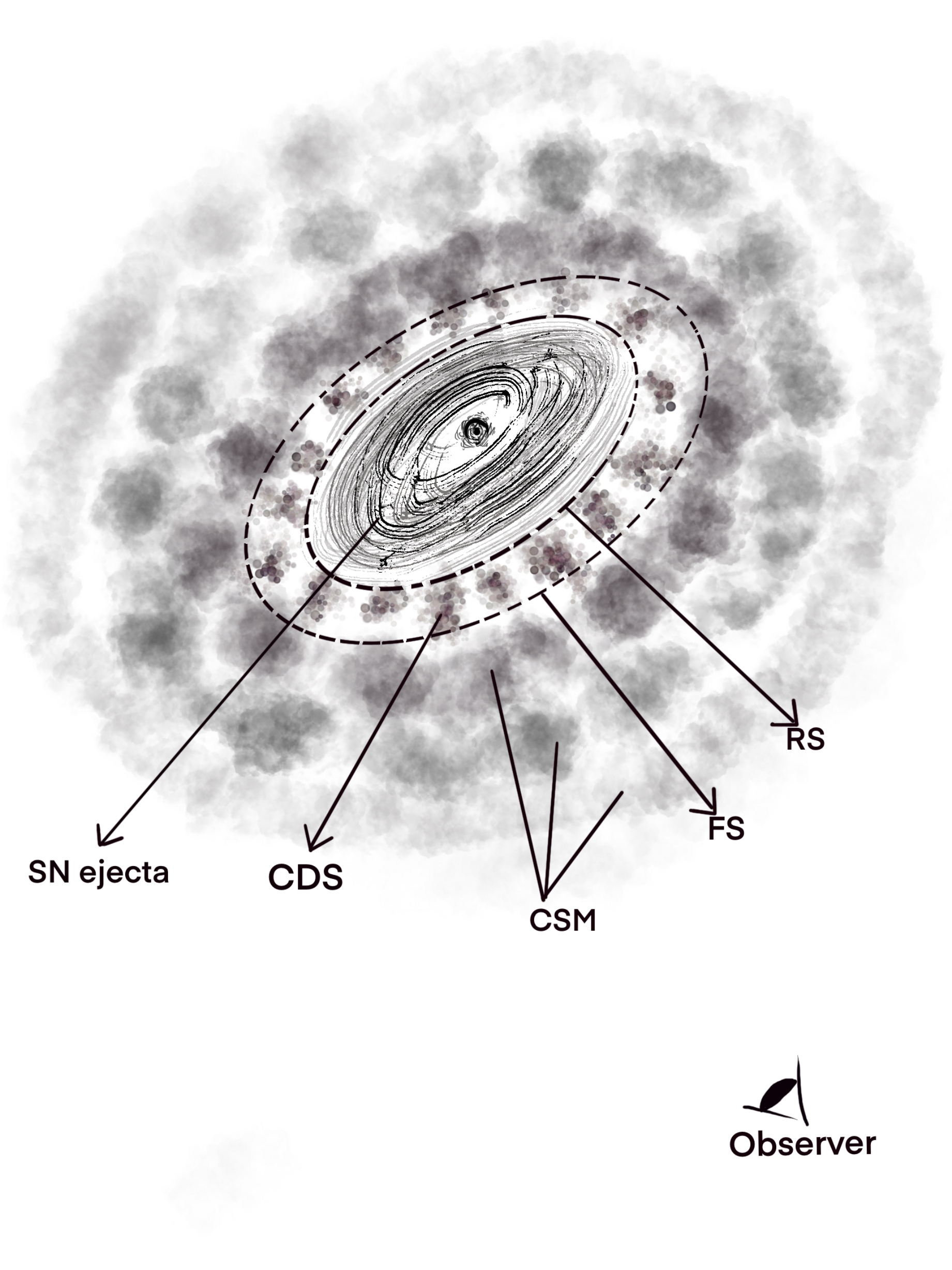 }
    \caption{An interacting supernova in a progenitor whose CSM morphology consists of clumpy, broken, rings. At the epoch shown, the ejecta is still enclosed within the CSM. Thus the observer sees a direct narrow emission line from the outer, least dense region of the CSM, superposed on an intermediate component from the inner, denser layers, as well as broad component from the ejecta itself. A morphology with clumpy, nested, shells would have a similar structure.
    }
    \label{fig:late}
    \end{minipage}   
    \hfill
    \begin{minipage}{0.45\textwidth}
        \centering
      \includegraphics[scale=0.25]{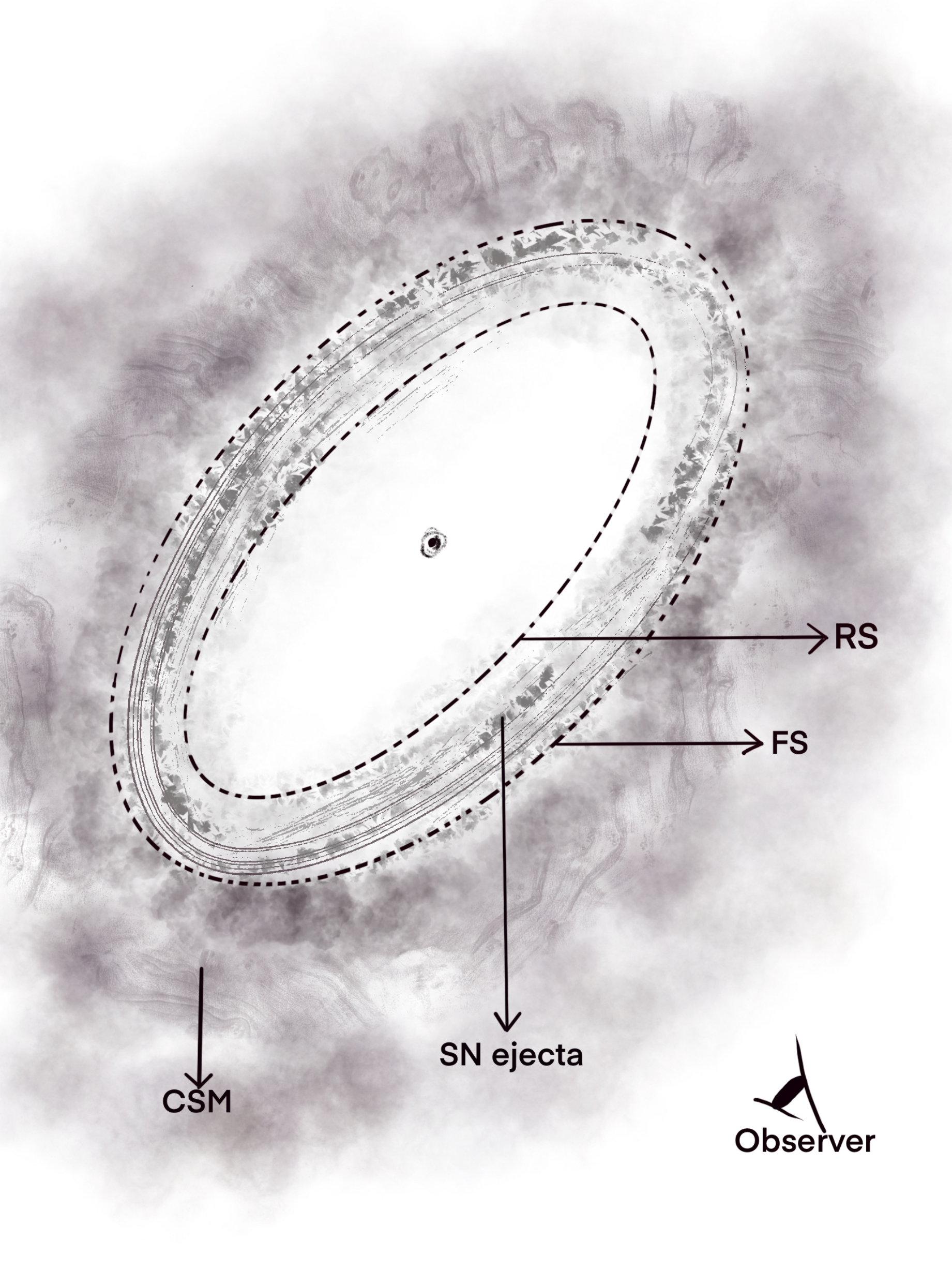}
    \caption{The schematic shows interaction of supernova ejecta with CSM, in a progenitor with an extended, shell-like CSM with a smooth, radially dependent density profile. At this late epoch, the ejecta is interacting with a CDS formed between the forward and reverse shocks.}
    \label{fig:early}
    \end{minipage}
\end{figure*}

Additionally, more exotic geometries of CSM formation during stellar lifetimes have also been seen, e.g., in the multiple, nested dusty shells in the  WR 140 binary system \citep{Lau-2022}, the double-lobed CSM of the $\eta$-Carina system, and even in the central ring of SN 1987A. A schematic of such a bipolar, multi-shelled CSM is depicted in Fig.\,\ref{fig:csmbi}, and schematics of ring-like, or toroidal CSMs are shown in Figs.\,\ref{fig:early} and\,\ref{fig:late}. Emission from both bipolar and toroidal morphologies would be strongly polarized, especially at early times due to extreme asymmetry. \citet{Liu} proposed the existence of multi-shell CSM, to explain the multi-humped light curves of the super luminous, but H-poor iPTF15esb and iPF13 dcc. However, the absence of multiple episodes of rebrightening in the light-curve (LC) of SN 2017hcc (see \citealt{moran}; Fig.\,2 and Fig.\,5) does not necessarily preclude the existence of multi-shell CSM   -- we also note that due to conjunction with the sun, there is a gap of almost 100\,d  in its LC. Finally, \citet{vanmarle} carried out 2D and 3D simulations of the CSM formation around evolved massive ($M>30 M_{\odot}$) stars, and found that the structure of the CSM is strongly dependent on how the fast-moving, but low-density, wind from the late, LBV stages of the massive progenitor, interact with the denser, but slow-moving material ejected in its AGB stage. The non-detection of precursor activity (if it not a selection effect due to sporadic photometric sampling of the host galaxy) in the case of SN 2010jl and SN 2015da suggests that the last LBV-like activity from the progenitors of these long-interacting (SL-SN IIn) occurred several decades to centuries before the terminal explosion. The various generations of CSM have already interacted and mixed, forming regions of Rayleigh-Taylor instabilities (filaments or ``fingers"), as seen in the simulations of \citet{vanmarle}.

The early emergence of a narrow line in SN 2017hcc in the classification spectrum ($\sim +10$\,d post explosion, \citealt{Dong-2017}) with velocity $v = 500$\,km\,s$^{-1}$ (+14\,d), which continues to hover around $\sim 450$\,km\,s$^{-1}$ throughout its entire evolution, (up to +100\,d in our study) confirms the presence of a dense CSM in the vicinity of the progenitor. The persistence of the narrow lines for up to +400\,d suggests that the shocked ejecta had not completely overtaken CSM by then. Moreover,  MIKE echelle \citep{Bernstein} spectra, with $R=19000$ (red) and $R=25000$ (blue)  taken on +24\,d \citep{smith}, shows a very narrow line (FWHM $\sim$ 50\,km\,s$^{-1}$) with a superposed P-Cygni, with an absorption at ($\sim$ 51\,km\,s$^{-1}$) -- not seen at lower resolutions. If this slow moving component was created by stellar winds ejected from  progenitor's AGB phase, then the component at $\sim 450$\,km\,s$^{-1}$ can be associated with fast moving winds emitted later in progenitor's evolution.

{Observations of SN~2017hcc point to intense precursor activity, with multiwavelength data indicating enhanced mass loss over the decade before explosion at rates of about $0.1\,M_{\odot}\,{\rm yr}^{-1}$ \citep{chandra,kumar}. This dense CSM strongly influenced the supernova’s early behavior, driving powerful ejecta--CSM interaction, early IR emission, and marked polarization within the first weeks \citep{kumar,Reguitti-2024}. The large CSM mass and high mass-loss rates are best explained by eruptive episodes, similar to LBV-like or super-Eddington outbursts, rather than steady stellar winds \citep{chandra,kumar,mauerhan}. Persistent polarization and asymmetric spectral features further suggest that binary interaction or equatorial ejections shaped the CSM \citep{Reguitti-2024,mauerhan}, while nuclear burning instabilities or wave-driven outflows remain viable additional triggers \citep{dickinsonASSASN,bilinski2014}. Overall, SN~2017hcc represents the long-lived, extended class of Type~IIn precursors, where years of eruptive and asymmetric mass loss build up the dense surroundings that dominate the explosion’s evolution.
}

We find (Fig.\,\ref{shift}) that the centroid of the narrow velocity components of both the H$\alpha$ and H$\beta$ lines drifts from a blue-shift (+14\,d and +23\,d) to a red-shift (+365\,d, +411\,d) as the supernova evolves -- a finding first noted by \citet{moran} who tracked both H$\alpha$ and H$\beta$ line centroids up to 1765\,d post explosion.
In a spherical CSM model, this shift of the narrow component suggests that it may be emitted in a clumpy CSM, with the blue shifted line initially emitted from forward moving clouds.
The red-shifted emission emerges later, as the line-of-sight CSM expands and becomes optically thin, and photons from the receding side of the CSM become visible. Alternatively, an additional redshifted component may have been superposed on the spectrum, possibly emitted as light-echo from the receding clouds of dust in the CSM glimpsed through the transparent regions between the clumps.


 We now examine models for the CSM morphology in long-lived type-II SNe. 
In order to explain the emergence of the fast SN ejecta by +75\,d (see Fig.\,\ref{hcc-hal2} and Table\,\ref{hcc}); \citet{smith} proposed a non-spherical, bi-polar geometry for the CSM for SN 2017hcc (see Fig.\,14, \citealt{smith}). In this geometry, the CSM shell is at some distance from the epicenter of the explosion, and the solid angle subtended by the ejecta to the observer plays an important role in determining the width of the various velocity components. The viewer only “sees" the part of CSM that is in the direct line of sight to the supernova ejecta. Thus, at relatively early ($\sim +86 $\,d and earlier) epochs, when the ejecta is still well within the CSM ``bubble", the viewer sees a narrow component from the shocked CSM, and a component from the photosphere of the ejecta alone, highly attenuated by the CSM. The later ($\sim$ +100\,d) emergence of multiple, high velocity components, and the widening of the narrow component, is then attributed to ejecta filling the CSM bubble (a broad component), and (at $\sim$ +200\,d) interacting with the CSM itself (an intermediate velocity component). A problem with bipolar CSM distribution is that a binary companion seems to be needed to create the double lobed structure in massive star system.  While no search for a binary companion of the progenitor of SN 2017hcc has been reported, we note that no binary companion has been detected in the HST images of the long interacting type-IIn SN 2010jl \citep{Ni2010jl}. 
The absence of a binary companion may not pose a big problem if the binary system merged, leading to an LBV core-collapse as in the case of the  long-lived SN 2015bh \citep{Goranskij}. It is to be noted that the progenitor of SN 2015bh has also shown precursor activity up to $\sim 1$ yr prior to the SN explosion  \citep{Elias2015bh}, \citet{Thone-2017_SN2015da} suggesting that the initial LBV core-collapse lead to a faint SN because of fall back. The secondary brightening in the light curve of SN 2015bh is then attributed to ejecta-CSM interaction.  We note here that no precursor activity has been detected in the archival search of CHASE and PTF images of the field of SN 2010jl by \citet{Reguitti}.


We now turn our attention to the other type-IIn, whose spectra have been compared with that of SN 2017hcc at similar epochs. 

At +14\,d (SN 2017hcc) we only have one near-contemporaneous dataset each for SN 2022zi, SN 2020cui, and SN 2018hfg (see Table.\,\ref{compare1}). The H$\alpha$ feature in the early spectra of SN 2020cui and SN 2018hfg  possess a two-component structure, similar to SN 2017hcc. However, SN 2020cui appears to have an intermediate ($\sim$ 1000 km $s^{-1}$),  a broad ($\sim 3500$ km s${-1}$) component and the narrow component is missing. This suggests that initially, the SN ejecta of SN 2018hfg, like that of SN 2017hcc, is enveloped in a CSM bubble, but that in the case of the SN 2020cui, the CSM is much closer to the progenitor, and the ejecta-CSM interaction begins early. Additionally, survival of pre-existing dust being unlikely in this early stage, the blue shifted centroid of SN 2020cui ($\lambda$6560) is most likely due to red-shifted absorption from a dense, optically thick, shell consisting of decelerating ejecta and swept-up CSM \citep{dessart-2015}. In contrast we note that the H$\alpha$ profile of SN 2022zi and SN 2015da at similar epochs require only a single, intermediate velocity, Lorentzian to produce a good fit. This may suggest that for these events, the CSM is still very dense and opaque, resulting in near-complete absorption of Thomson scattered photons along the line-of-sight -- the narrow component would, in this case, be emitted from the optically thin, outer layers of the shocked CSM (Fig.\,\ref{fig:csmbi}).    

In the early epochs of SN 2023hpd (see Fig.\,\ref{23}a), multiple Gaussian profiles were required to resolve the blue and red wings of the H$\alpha$ profile, exhibiting similar velocities. This characteristic suggests the presence of a clumpy CSM. Notably, the absence of a narrow ($ \ll 1000$ km $s^{-1}$) peak at the rest wavelength of H$\alpha$ suggests that even the outermost layers of the CSM are still very optically dense, and each broad component may be from Thomson scattered photons streaming out from within individual clumps.

The presence of an intermediate component in the H$\alpha$ profile of SN 2020ywx can be attributed to the still-dense CSM, in which the photons are undergoing Thomson scattering  or from the CDS \citep{chugai}.

SN 2018khh exhibited an interesting profile, particularly during the early epochs following the explosion. The spectrum obtained one day post-explosion displayed a typical Type IIn spectrum, but with strong He II emission, surpassing the flux or magnitude of the Balmer emission lines. However, by +12\,d, these features had disappeared. More intriguingly, the profile at +21\,d  resembled that of SN 2017hcc at +100\,d, rather than at +23\,d. The presence of narrow component in H$\alpha$ on +12\,d was clearly from the outer layers of the dense CSM and intermediate component from Thomson scattering deep within CSM \citep{chugai}. 
In this model, the CSM consists of two components -- a dense core surrounded by a low density halo. The later emergence of two narrow components on +21\,d possibly indicates the emission is occurring from clumpy CSM, which by then had become optically thin. 

Our spectra of SN 2023usc is relatively noisier, since the event was rather faint, and the H$\alpha$ line was contaminated by telluric features, which were subsequently corrected for. Even so, it showed a narrow emission line in the early days, which later evolved into a narrow Gaussian feature sitting on top of a broad H$\alpha$ emission line originating from fast-moving ejecta. The evolution of the H$\alpha$ profile is almost a “text-book" case for an explosion in a CSM shell. The narrow line hovers around ($\sim 450$\,km\,s$^{-1}$) for up to +93\,d, while the intermediate ($\sim 2000 $\,km\,s$^{-1}$ velocity feature seen on +12\,d, vanishes  by +37\,d, and is replaced by a high velocity feature ($>$ 8000\,km\,s$^{-1}$) by +62\,d. The intermediate velocity feature being associated with Thomson scattered photons deep within the CSM, this suggests that the CSM has become optically thin by +37\,d, and emission from the ejecta  is now visible (+62\,d). We note that the narrow component “widens" ($\sim 700$\,km\,s$^{-1}$), which may be due to the fast moving ejecta running into the CSM leading to the formation of a CDS. Additionally, the presence of [N II] indicates that the supernova is embedded deep within the galaxy. 
However, the other Balmer lines, except for H$\alpha$, were not prominent. 
Thus, it is likely that the explosion occurred in an asymmetric, possibly clumpy, optically thick shell. Assuming that the shocked ejecta has a velocity $v_s$ when it rams into the CSM moving with velocity $v_{CSM}$, at rest frame time $t_{r,int} = t_{int}*\sqrt{(1-z^2)}$, the last epoch of creation of the CSM can be estimated as $t= ({1\over v_{CSM}} - {1 \over v_s})v_s t_{r,int}$. In the case of SN 2023usc, if we assume that this occurred by +62\,d, then this “look back" time can be estimated at $t=1257$\,d pre-explosion.


 In addition to the dense CSM, we detected Fe-group elements from the ejecta,  in multiple excited states, in the late-phase spectra ($\sim$100\,d) of several SNe. The emergence of an ``Fe bump'' in interacting supernovae at similar phases has also been reported by \citet{Nagao-2025}.
We also note that the spectra of several type-IIn SNe appear to show boxy H$\alpha$ lines emission from the CDS at around $\sim 100$\,d as well.
The emergence of both features at the same epoch can be reconciled if the CSM is highly aspherical, or even clumpy.
In such a scenario, the part of the CSM that is less extended (or less dense) becomes optically thin earlier, allowing emission from the ejecta to emerge, even as ``boxy'' H$\alpha$ lines emerge from the CDS in the inner regions of the extended (or denser) regions of the CSM.

The emission from the supernova’s own optical photosphere has various velocity components and contributes to the overall spectrum. The typical photospheric velocities are few to several $\times 1000$\,km\,s$^{-1}$. They exhibit both blue-shifted  absorption and red-shifted emission components, resulting in a P-Cygni profile. However, in the early epochs, the emission from the circumstellar medium -- both direct, and rescattered within the optically dense segment --  dominates over the P-Cygni profile of the photospheric components, resulting in a narrow emission profile with a broad base in the hydrogen emission spectrum.

As the shocked CSM expands and becomes optically thin, the emission from the shocked ejecta starts to dominate. As a result, the photons emitted by the ejecta's photosphere can reach the observer, giving rise to a P-Cygni profile that becomes more pronounced than the narrow H$\alpha$ line from the CSM. Additionally, metals in shocked ejecta produce absorption lines, creating distinctive notches in the spectral profile and broadening. These effects are particularly evident in the later epochs of the supernova. By 100 days, shocked SN ejecta, which
was hitherto expanding freely, catches up the now optically thin CSM, and the following phenomena occur:\\
i. The freely expanding ejecta experiences a slight deceleration as it interacts with the matter at the inner edge of the CSM.  \\
ii. The optical photosphere continues to recede further into the ejecta

These results contribute to the increasing prominence of the P-Cygni profile, particularly in conjunction with the H$\alpha$ line from the CSM. The interaction between the ejecta and the CSM gives rise to line profiles consisting of different velocity components creating distinct notches in the spectral profile and broadening  (e.g. SN 2021foa, SN 2021uha in Fig.\,\ref{SNlln100 d}).

After 100 days, the H$\alpha$ line shows intermediate-width components with velocities ranging up to a few thousands of km/s, suggesting that they originate from the cool dense shell (CDS), as it undergoes ionization and reionization.
We note that SN 2017hcc was detected by  Chandra/ACIS-S only
on +727\,d, at a relative high column density $N_H=1.13 \times 10^{22}$ cm$^{-2}$ \citep{chandra}. This appears to be consistent with a CDS.
The CDS could have a similar structure to the CSM, meaning it may exhibit a clumpy, discontinuous morphology, as shown in Fig.\,\ref{fig:csmbi}. Eventually, it thins out over time. The presence of these intermediate components in some supernovae and their absence in others suggests that, unlike the continuous dense CSM model proposed by \citet{smith}, a clumpy, discontinuous CSM/CDS is formed.

The presence of strong and broad Ca II emission lines by $\sim $+100\,d in SN 2021uha, SN 2020nku and SN 2020ywx suggests that the SN ejecta itself may have already reached the nebular stage. Also notable  are the multiple ionization states of nitrogen such as N III, N II in SN 2020cui and SN 2018hfg.

\section{Conclusions}
  In this paper we have carried out a comparative study of spectra of type-IIn supernovae at various evolutionary epochs, using, as cases in point,  the well-studied, long-lived SN 2017hcc, and the more recent, short duration type-IIn SN 2023usc. Our immediate purpose is to search for common features in H$\alpha$ profile (typically, the brightest feature in any type II SN spectra), at similar epochs in the events' light curves. The ultimate aim of this study is to link the individual components of a complex H$\alpha$ line profile to the gradually unveiling structure of the progenitor's extended CSM, and its  interaction with the SN shock and ejecta. 
  {Previous studies of type-IIn supernovae \citep{Ransome+2021_IIn-reclassification, Salamaso-2025} have shown the tremendous range of photometric and spectral diversity in this class of objects, suggesting that type-IIn SNe are a phenomena linked not only to the mass of the progenitor, but to its immediate environment and, by extension, to its evolutionary history. \citet{Ransome+2021_IIn-reclassification} re-examined the classification of 115 SNe, previously identified as type-IIn, and redistributed them in a “gold class" and a “silver class" depending on whether the strong, narrow, spectral features characteristic of CSM interaction are present throughout the evolution of the event or not. In this scheme SN 2010jl and SN 2015da are assigned to the gold class, while SN 2017hcc was assigned to the silver class, although this may have been because they did not have access to the full dataset \citep{moran} for this object. The other objects in our sample are more recent and thus not included in their sample. }
 
In the comparative analysis section of our study, we have compared the early spectral characteristics of SN 2017hcc and SN 2023usc with several type-IIn SNe (and one transition event) at 3 epochs -- +14\,d, +24\,d and +100\,d -- with the selected epochs being limited by data availability. At early epochs (+14 and +24\,d, see tables\,\ref{compare1},\,\ref{compare2} respectively), we find that $ 50 - 60 \%  $ of the IIn-SNe (including SN 2021foa, the one type IIn-Ibn event) have at least one H$\alpha$ component with $v < 1000$\,km\,s$^{-1}$. However, at +100\,d (see Table\,\ref{compare3}), we find that true narrow lines ($v < 1000$\,km\,s$^{-1}$), as seen in 2017hcc and 2023usc, persist only in SN 2010jl, while SN 2015da shows persistent intermediate width lines at this epoch. 2017hh, 2010jl and 2015da are long-interacting events, and therefore this suggests that an extended CSM halo may be present only in the  progenitors of some long-interaction events.

The combined spectro-polarimetric and spectroscopic diagnostics for SN 2017hcc strongly support the presence of a highly extended, asymmetric, and clumpy circumstellar medium (CSM) surrounding long-lived Type IIn supernovae. The unusually high continuum polarization measured during the first month \citep{mauerhan}, together with its pronounced blue-rising slope and subsequent decline as the transient approached peak luminosity, implies that both dust and electron scattering contributed significantly to the polarized flux. Crucially, the near-constancy of the polarization position angle across both the dust-dominated and electron-scattering phases indicates that the underlying scattering material shared a common geometric structure that persisted throughout the SN–CSM interaction. This alone argues against a spherically symmetric wind, as such a configuration would suppress net polarization, and instead points to an intrinsically aspherical CSM-likely bipolar, equatorial, or fragmented into discrete clumps.

Additional spectroscopic evidence independently corroborates this interpretation. The evolution from narrow to intermediate-width components in the H$\alpha$ and H$\beta$ profiles, the blue-to-red centroid drift of the narrow emission as the CSM became optically thin, and the emergence of multiple Gaussian subcomponents at later times all suggest a radially stratified, clumpy CSM shaped by episodic mass-loss events prior to core collapse. The persistence of narrow lines for $\gtrsim  100$\,d  indicates that the shocked ejecta encountered dense CSM at large radii, while the later emergence of broad components implies that the freely expanding SN ejecta eventually overtook these partial shells as the interaction weakened. Similar signatures have been reported for other interacting transients (e.g., SN 2009ip, SN 2010jl), where asymmetric line profiles and high early-time polarization have been interpreted as evidence for equatorial disks (SN 2010jl; \citealt{fransson}) or bipolar lobes (SN 2009ip, SN 1998S), possibly shaped by binary interactions or LBV-like eruptions \citep{smith,mauerhan}.

 One of the aims of this paper is to determine if there are any defining
characteristics of true-to-type type-IIn supernovae in their very early spectra. Our comparison epochs ($\sim$ +14\,d, +23\,d and +86/+100\,d) are limited by the publicly available data, but they represent shock-cooling phase ( $\sim$ +14\,d, +23\,d), and late cooling phase ($\sim$ 100\,d) of the light curves. In the LSST era, when it may not be possible to characterize events via temporally dense spectroscopic-photometric follow-up of (especially) faint events at large z, the ability to distinguish a true type-IIn from other interacting supernovae can constrain, among other factors, the distribution of massive progenitors with dense CSM at high z.


\section{Acknowledgements:}
 We thank the anonymous referee for helpful comments and suggestions. We thank the staff of IAO, Hanle, and CREST, Hosakote, who made the observations of SN 2017hcc and SN 2023usc possible. The facilities at IAO and CREST, operated by the Indian Institute of Astrophysics, Bangalore. We acknowledge the use of the Weizmann Interactive Supernova Data Repository (WISeREP\href{https://www.wiserep.org}; \cite{WISEREP}) and the Transient Name Server (TNS \href{https://www.wis-tns.org/}; \citealt{TNS}).
We acknowledge Ms. Fathimath Faseela and Mr. Akshay.M.K for the schematic diagram.

\bibliography{SN2017hcc}{}
\bibliographystyle{aasjournal}



\end{document}